\newcommand\submitms{y} 

\newcommand\psj{\ref@jnl{PSJ}}

\documentclass[apj]{emulateapj}

\usepackage{apjfonts}
\usepackage{ifthen}
\usepackage{natbib}
\usepackage{amssymb, amsmath}
\usepackage{appendix}
\usepackage{etoolbox}
\usepackage{graphicx}
\usepackage{rotating}
\usepackage{subfigure}
\bibpunct[, ]{(}{)}{,}{a}{}{,}

\usepackage[
bookmarks=true,           
bookmarksnumbered=true,   
colorlinks=true,          
citecolor=blue,           
linkcolor=blue,           
menucolor=blue,           
urlcolor=blue,            
linkbordercolor={0 0 1},  
pdfborder={0 0 1},
frenchlinks=true]{hyperref}

\providecommand{\adsurl}[1]{\href{#1}{ADS}}

\makeatletter
\patchcmd{\NAT@citex}
  {\@citea\NAT@hyper@{%
     \NAT@nmfmt{\NAT@nm}%
     \hyper@natlinkbreak{\NAT@aysep\NAT@spacechar}{\@citeb\@extra@b@citeb}%
     \NAT@date}}
  {\@citea\NAT@nmfmt{\NAT@nm}%
   \NAT@aysep\NAT@spacechar\NAT@hyper@{\NAT@date}}{}{}

\patchcmd{\NAT@citex}
  {\@citea\NAT@hyper@{%
     \NAT@nmfmt{\NAT@nm}%
     \hyper@natlinkbreak{\NAT@spacechar\NAT@@open\if*#1*\else#1\NAT@spacechar\fi}%
       {\@citeb\@extra@b@citeb}%
     \NAT@date}}
  {\@citea\NAT@nmfmt{\NAT@nm}%
   \NAT@spacechar\NAT@@open\if*#1*\else#1\NAT@spacechar\fi\NAT@hyper@{\NAT@date}}
  {}{}
\makeatother

\makeatletter
\DeclareRobustCommand{\lowcase}[1]{\@lowcase#1\@nil}
\def\@lowcase#1\@nil{\if\relax#1\relax\else\MakeLowercase{#1}\fi}
\makeatother

\newcommand\chisq{\ifmmode{\chi\sp{2}}\else\math{\chi\sp{2}}\fi}
\newcommand\redchisq{\ifmmode{ \chi\sp{2}\sb{\rm red}}
                    \else\math{\chi\sp{2}\sb{\rm red}}\fi}

\DeclareSymbolFont{UPM}{U}{eur}{m}{n}
\DeclareMathSymbol{\umu}{0}{UPM}{"16}
\let\oldumu=\umu
\renewcommand\umu{\ifmmode\oldumu\else\math{\oldumu}\fi}
\newcommand\micro{\umu}
\renewcommand\micron{\micro m}
\newcommand\microns{\micron}

\let\oldsim=\sim
\renewcommand\sim{\ifmmode\oldsim\else\math{\oldsim}\fi}
\let\oldpm=\pm
\renewcommand\pm{\ifmmode\oldpm\else\math{\oldpm}\fi}
\newcommand\by{\ifmmode\times\else\math{\times}\fi}

\newcommand\tablebox[1]{\begin{tabular}[t]{@{}l@{}}#1\end{tabular}}
\newbox{\wdbox}
\renewcommand\c{\setbox\wdbox=\hbox{,}\hspace{\wd\wdbox}}
\renewcommand\i{\setbox\wdbox=\hbox{i}\hspace{\wd\wdbox}}
\newcommand\n{\hspace{0.5em}}

\newcount\timect
\newcount\hourct
\newcount\minct
\newcommand\now{\timect=\time \divide\timect by 60
         \hourct=\timect \multiply\hourct by 60
         \minct=\time \advance\minct by -\hourct
         \number\timect:\ifnum \minct < 10 0\fi\number\minct}

\catcode`@=11

\newcommand\comment[1]{}

\newcommand\commenton{\catcode`\%=14}
\newcommand\commentoff{\catcode`\%=12}

\renewcommand\math[1]{$#1$}
\newcommand\mathshifton{\catcode`\$=3}
\newcommand\mathshiftoff{\catcode`\$=12}

\comment{alignment tab}
\let\atab=&
\newcommand\atabon{\catcode`\&=4}
\newcommand\ataboff{\catcode`\&=12}

\let\oldmsp=\sp
\let\oldmsb=\sb
\def\sp#1{\ifmmode
           \oldmsp{#1}%
         \else\strut\raise.85ex\hbox{\scriptsize #1}\fi}
\def\sb#1{\ifmmode
           \oldmsb{#1}%
         \else\strut\raise-.54ex\hbox{\scriptsize #1}\fi}
\newbox\@sp
\newbox\@sb
\def\sbp#1#2{\ifmmode%
           \oldmsb{#1}\oldmsp{#2}%
         \else
           \setbox\@sb=\hbox{\sb{#1}}%
           \setbox\@sp=\hbox{\sp{#2}}%
           \rlap{\copy\@sb}\copy\@sp
           \ifdim \wd\@sb >\wd\@sp
             \hskip -\wd\@sp \hskip \wd\@sb
           \fi
        \fi}
\def\msp#1{\ifmmode
           \oldmsp{#1}
         \else \math{\oldmsp{#1}}\fi}
\def\msb#1{\ifmmode
           \oldmsb{#1}
         \else \math{\oldmsb{#1}}\fi}
\def\supon{\catcode`\^=7}
\def\supoff{\catcode`\^=12}
\def\subon{\catcode`\_=8}
\def\suboff{\catcode`\_=12}
\def\supsubon{\supon \subon}
\def\supsuboff{\supoff \suboff}

\newcommand\actcharon{\catcode`\~=13}
\newcommand\actcharoff{\catcode`\~=12}

\newcommand\paramon{\catcode`\#=6}
\newcommand\paramoff{\catcode`\#=12}

\comment{And now to turn us totally on and off...}

\newcommand\reservedcharson{ \commenton  \mathshifton  \atabon  \supsubon 
                             \actcharon  \paramon}

\newcommand\reservedcharsoff{\commentoff \mathshiftoff \ataboff \supsuboff 
                             \actcharoff \paramoff}

\catcode`@=12
\reservedcharsoff

\reservedcharson

\comment{ Must have ONLY ONE of these... trust these macros, they work

}
\reservedcharsoff

\actcharon
\if\submitms y

\else

\comment{
\received{}
\revised{}
\accepted{}
}
\fi

\if\submitms y
\slugcomment{\tablebox{
Revised on {\today} \now. Submitted to {\em PSJ}}.}
\else
\slugcomment{Revised on {\today} \now. Submitted to {\em PSJ}}.}
\fi

\reservedcharson

\shorttitle{BART III: Initialization, Post-processing Routines, and WASP-43 b}
\shortauthors{Blecic {\em et al.}}

\comment{Paper-specific macros}

\begin{document}

\title {AN OPEN-SOURCE BAYESIAN ATMOSPHERIC RADIATIVE TRANSFER (\textsc{BART}) CODE: \newline III. INITIALIZATION, ATMOSPHERIC PROFILE GENERATOR, POST-PROCESSING ROUTINES 
\newline AND APPLICATION TO EXOPLANET WASP-43\lowercase{b}}

\author{Jasmina Blecic\altaffilmark{1, 2, 3}, Joseph
  Harrington\altaffilmark{3, 4}, Patricio E. Cubillos\altaffilmark{3, 5}, M. Oliver Bowman\altaffilmark{3}, Patricio Rojo\altaffilmark{6}, Madison Stemm \altaffilmark{3}, Ryan C.\ Challener\altaffilmark{3}, Michael~D.~Himes\altaffilmark{3}, Austin J.\ Foster\altaffilmark{3}, Ian Dobbs-Dixon\altaffilmark{1,2,7}, Andrew S.\ D.\ Foster\altaffilmark{3, 8}, Nathaniel B.\ Lust\altaffilmark{2, 9}, Sarah D.\ Blumenthal\altaffilmark{3, 10}, Dylan Bruce\altaffilmark{3}, Thomas~J.~Loredo\altaffilmark{11}}

\affil{\sp{1} Department of Physics, New York University Abu Dhabi, PO Box 129188 Abu Dhabi, UAE.}
\affiliation{\sp{2} Center for Astro, Particle and Planetary Physics (CAP$^3$), New York University Abu Dhabi, PO Box 129188, Abu Dhabi, UAE}
\affil{\sp{3} Planetary Sciences Group, Department of Physics,
  University of Central Florida, Orlando, FL 32816-2385, USA}
\affil{\sp{4}Florida Space Institute, University of Central Florida, Orlando, FL 32826-0650, USA}
\affil{\sp{5} Space Research Institute, Austrian Academy of Sciences, Schmiedlstrasse 6, A-8042 Graz, Austria}
\affil{\sp{6} Department of Astronomy, Universidad de Chile,
       Santiago de Chile, Chile}
\affiliation{\sp{7} Center for Space Science, NYUAD Institute, New York University Abu Dhabi, PO Box 129188, Abu Dhabi, UAE}
\affil{\sp{8} Center for Radiophysics and Space Research, Space Sciences
       Building, Cornell University, Ithaca, NY 14853-6801}
\affil{\sp{9} Department of Astrophysical Sciences, Princeton University, Princeton, NJ 08544, USA}
\affil{\sp{10} Department of Physics, University of Oxford, Oxford OX1 3PU, United Kingdom}
\affil{\sp{11} Center for Astrophysics and Planetary Science, Space Sciences
Building, Cornell University, Ithaca, NY 14853-6801, USA}

\email{jasmina@nyu.edu}

\begin{abstract}

This and companion papers by \citeauthor{HarringtonEtal2021-BART_I}\,\,and \citeauthor{CubillosEtal2021-BART_II}\,\,describe an open-source retrieval framework, Bayesian Atmospheric Radiative Transfer (\textsc{BART}), available to the community under the reproducible-research license via \href{https://github.com/exosports/BART}{https://github.com/exosports/BART}. \textsc{BART} is a radiative-transfer code (\href{https://github.com/exosports/transit}{\tt transit}, \citeauthor{Rojo2009-Transit}), initialized by the Thermochemical Equilibrium Abundances (\href{https://github.com/dzesmin/TEA}{\textsc{TEA}}) code (\citeauthor{BlecicEtal2016-TEA}\,), and driven through the parameter phase space by a differential-evolution Markov-chain Monte Carlo (\href{https://github.com/pcubillos/mc3}{\textsc{MC3}}) sampler (\citeauthor{CubillosEtal2017apjRednoise}\,). In this paper we give a brief description of the framework, and its modules that can be used separately for other scientific purposes; outline the retrieval analysis flow; present the initialization routines, describing in detail the atmospheric profile generator and the temperature and species parameterizations; and specify the post-processing routines and outputs, concentrating on the spectrum band integrator, the best-fit model selection, and the contribution functions. We also present an atmospheric analysis of WASP-43b secondary eclipse data obtained from space- and ground-based observations. We compare our results with the results from the literature, and investigate how the inclusion of additional opacity sources influence the best-fit model.
 
\end{abstract}

\section{Introduction}
\label{intro}

The rapid increase of detected extrasolar planets in the last decade (4375 confirmed and 2614 candidates as of April 2021; \href{https://exoplanetarchive.ipac.caltech.edu}{https://exoplanetarchive.ipac.caltech.edu}) and the number of novel techniques employed to analyze their data \citep[e.g.,][]{SwainEtal2008ApJhd209458bspec, CarterWinn2010, KnutsonEatl2009ApJ-PhaseVariationHD149026b, StevensonEtal2012apjHD149026b, DemingEtal2013arXivHD209458b-Xo1b-WCF3, deWitEtal2012aapFacemap, DemingEtal2015-PixelDecor, MayStevenson2020, ChallenerEtal2021PSJ}, have prompted theorists to develop different approaches to model their atmospheres. To get insights into their thermal structures and chemical compositions, initial methods started from first principles in a one-dimensional (1D) scheme \citep[e.g.,][]{FortneyEtal2005apjlhjmodels} and later developed into complex three-dimensional (3D) models that studied the atmospheric dynamics and circulations \citep[e.g.,][]{ShowmanEtal2009-3Dcirc, Dobbs-DixonAgol2013-3DRT}.

\begin{figure*}[t!]
\centerline{
\includegraphics[width=0.99\textwidth, clip=True]{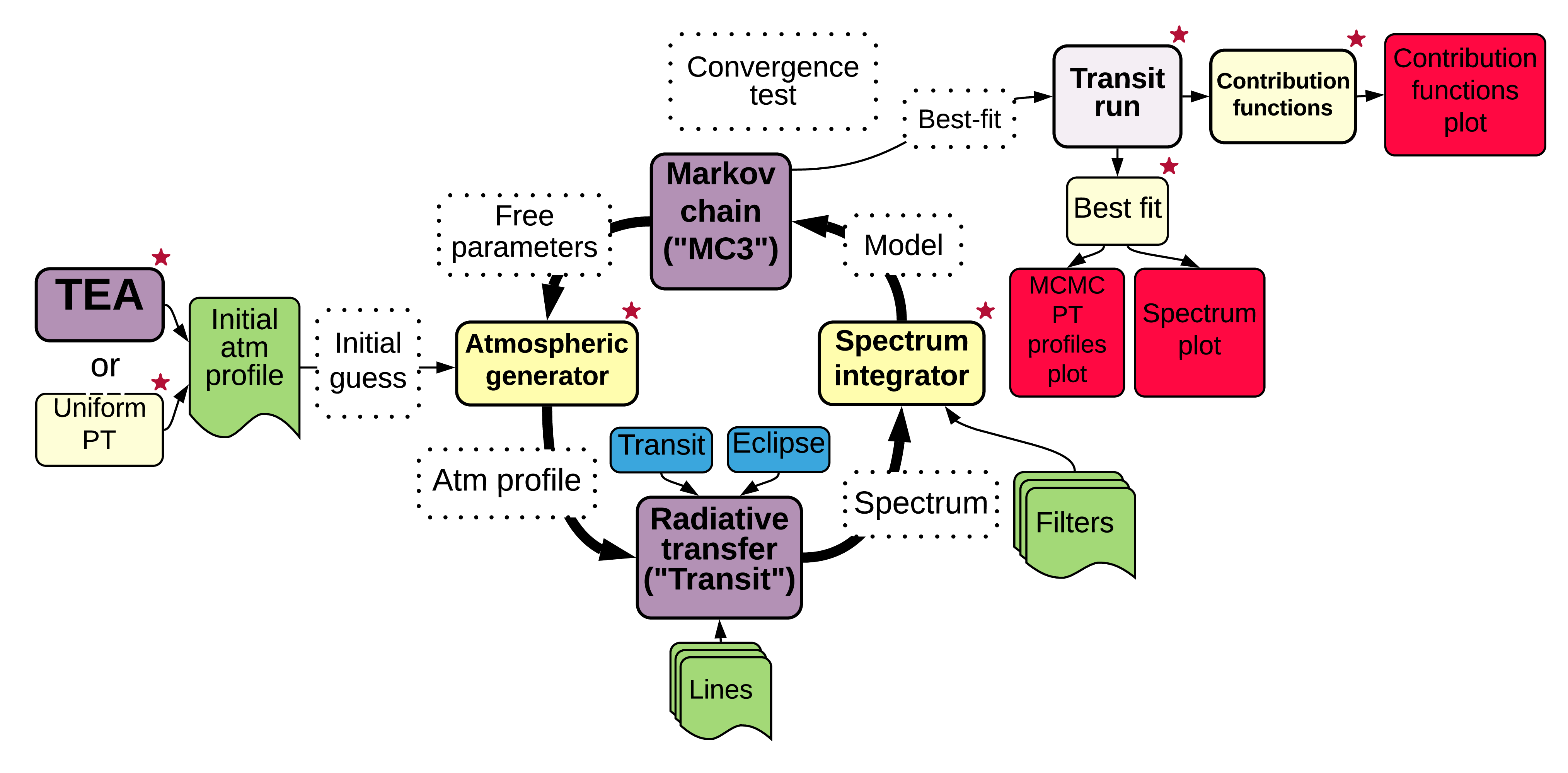}}
\caption{\footnotesize
Simplified \textsc{BART} flow chart. In purple are the three major \textsc{BART} modules. \textsc{TEA} and \textsc{MC3} are written in Python, while {\tt transit} is written in C. Other supporting modules are written in Python and given in yellow. In white are the inputs/outputs of each module. In green are the input files. In blue are transit and eclipse raypath-solution submodules. Stars denote modules and packages that will be described in more detail in this paper.}
\label{fig:BARTChart}
\end{figure*}

Today, we have several major approaches to atmospheric modeling. One applies a direct, forward modeling technique that provides a set of parameters to generate the observed spectra \citep[e.g.,][]{SingEtal2016natHotJupiterTransmission, Fortney2008, BurrowsEtal2008apjSpectra}; the other utilizes observations to determine the model's best-fit parameters and their uncertainties; and the most recent one applies machine learning methods that use a precomputed grid of atmospheric models to retrieve the full posterior distribution of parameters \citep{Marquez-NeilaEtal2018-machineLearning, ZingalesWaldmann2018-machineLearning, CobbEtal2019-machineLearning, OreshenkoEtal2020}. 

The inverse, retrieval approach, such as the one implemented in the framework we present here, determines the properties of the planetary atmosphere based on the available observations. It uses a statistical algorithm, usually a Markov-chain Monte Carlo method, MCMC, or nested sampling \citep[e.g.,][and references within]{Ford2005, Skilling2006-nestedSampling} to estimate the posterior distribution of the model given the data. By searching for the regions of space that fit data the best, this approach also determines the uncertainties in the model parameters. 

Today, we have more than a handful of retrieval frameworks used in the exoplanetary field that implement various complexity of physical phenomena, statistical criteria and optimization techniques. In the study of exoplanetary atmospheres, the retrieval approach was first introduced by \citet{MadhusudhanSeager2009ApJ-AbundanceMethod}. They used a multidimensional grid-optimization scheme within a radiative-transfer model to explore the phase space of model parameters. In their following paper, \citet{MadhusudhanSeager2010} utilized the first application of an MCMC algorithm. Soon after \citet{LeeEtal2012-CF} introduced a non-linear optimal estimation algorithm {\em NEMESIS} \citep{irwin2008nemesis, Rodgers2000-Retrieval} that applied the correlated-\math{k} technique, allowing for a rapid integration of the model spectra and an order of magnitude faster exploration of the phase space than previous line-by-line algorithms. Their method also required fewer model evaluations, attributed to the assumption that the parameter error distributions are Gaussians, and offered a single best-fit solution calculated using the Levenberg-Marquardt algorithm \citep{Levenberg1944, Marquardt1963}. The same year, \citet{BennekeSeager2012-Retrieval} released another Bayesian-based retrieval algorithm, {\em SCARLET}. This framework introduced the {\em parameter credible regions}, considered non-Gaussian uncertainties of the model parameters, employed a radiative-convective model to calculate the temperature profile, and allowed for the presence of a cloud deck or solid surface. The following year, \citet{BennekeSeager2013-Retrieval} go a step further and introduce, for the first time, the nested-sampling algorithm for efficient parameter exploration \citep{Skilling2004-nestedSampling, FerozHobson2008-nestedSampling} and a Bayesian model comparison via Bayesian evidence \citep{Gregory2007-modelComparison, Trotta2008-BayesianEvidence}. In 2013, \citet{LineEtal2013-Retrieval-II} presented {\em CHIMERA} and tested three retrieval exploration algorithms: optimal estimation, differential evolution Markov chain Monte Carlo (DEMC), and bootstrap Monte Carlo. 

Several years later, \citet{Waldmann2015-TAU} introduced the \math{\tau}{\em-REx} code that utilized the molecular line lists from the  ExoMol project and a custom built software that identifies likely absorbers/emitters in the spectra. In 2017, \citet{LavieEtal2017-HELIOS} released {\em HELIOS-R} open-source code that uses the {\tt PyMultiNest} package \citep{BuchnerEtal2014} and can apply free or self-consistent equilibrium chemistry retrieval. The same year, \citet{Wakeford2017Sci-ATMO} and \citet{EvansEtal2017-ATMOretrieval} published a Bayesian framework based on the one-dimensional forward model {\em ATMO} \citep{TremblinEtal2015ApJ-ATMO}, and \citet{MacDonaldMadhusudhan2017-POSEIDON} introduced {\em POSEIDON}, a two-dimensional atmospheric retrieval algorithm that includes inhomogeneous cloud prescription. A year after, the same group \citep{GandhiMadhusudhan2018MNRAS-HYDRA} presented a new disequilibrium retrieval framework, {\em HyDRA}, that introduced a self-consistent model to constrain layer-by-layer deviations from chemical and radiative-convective equilibrium. Since then two open-source codes have been published: {\em petitRADTRANS} that uses correlated-k method and implements various cloud parametrization \citep{MolliereEtal2019petitRADTRANS}, and {\em PLATON} \citep{ZhangEtal2019}, which implements a two-parameter, equilibrium chemistry model, and includes a Mie-scattering cloud model and unocculted starspot corrections. Also, most of the previous frameworks got upgraded: the {\em NEMESIS} framework \citep{BarstowEtal2020-retrievalComparison} now includes the {\tt PyMultiNest} algorithm \citep{BuchnerEtal2014, Krissansen-TottonEtal2018} and multiple parametrized cloud models \citep{Barstow2020-modelSelection}, and the {\em HELIOS-R} code implemented the first hybrid CPU-GPU approach \citep{KitzmannEtal2020A}. Very recently \citet{MinEtal2020} presented the {\em ARCiS} retrieval framework that combines a self-consistent physical modelling with a parametrized approach.

Here, we present another open-source framework, the Bayesian Atmospheric Radiative Transfer (\textsc{BART}) code for forward and retrieval modeling. This paper is one of three papers that describe the architecture, individual modules and packages, and theory implemented in this code. Figure \ref{fig:BARTChart} shows a simplified flow of the \textsc{BART} code. The underlying theory and the code structure of the modules and packages marked by the stars will be covered in more detail in this paper. The radiative-transfer module is described in the collaborative paper by \citet{CubillosEtal2021-BART_II}, submitted. The overall design of the code, the validation tests, performance, and optimization are described in the collaborative paper by \citet{HarringtonEtal2021-BART_I}, submitted. In addition, each paper presents an atmospheric analysis of a hot exoplanet using \textsc{BART}, showing its diverse features.

In Section \ref{sec:BART}, we present \textsc{BART} and describe the implementation of the initialization routines, atmospheric profile generator, spectrum integrator, best-fit routines, and contribution function module. We also briefly outline the characteristics of the three major modules: \textsc{TEA}, {\tt transit} and \textsc{MC3}, which are described in detail in \citet{BlecicEtal2016-TEA, CubillosEtal2021-BART_II} and \citet{CubillosEtal2017apjRednoise}, respectively. In Section \ref{sec:analWASP43b}, we present an atmospheric analysis of WASP-43b secondary eclipse data using \textsc{BART}. In Section \ref{sec:conc}, we give a short summary and our conclusions.

\section{The Bayesian Atmospheric Radiative  Transfer Code}
\label{sec:BART}

\textsc{BART} is an open-source Bayesian, thermochemical, radiative-transfer code written in Python and C, and available to the community under the Reproducible-Research Software License via \href{https://github.com/exosports/BART}{https://github.com/exosports/BART}.  It consists of three major parts (Figure \ref{fig:BARTChart}, purple boxes): the Thermochemical Equilibrium Abundances module (\textsc{TEA}, see Section \ref{sec:TEA}), a radiative-transfer module ({\tt transit}, see Section \ref{sec:transit}), and the Multi-core Markov-chain Monte Carlo statistical module (\textsc{MC3}, see Section \ref{sec:MC3}). Each of the modules works independently and can be used for other scientific purposes. \textsc{TEA} is an open-source thermochemical equilibrium abundances code that calculates the volume mixing fraction of gaseous molecular species \citep[][\href{https://github.com/dzesmin/TEA}{https://github.com/dzesmin/TEA}]{BlecicEtal2016-TEA}. {\tt Transit} is an open-source radiative-transfer code that applies a 1D opacity-sampling computation to generate model spectra (see the collaborative paper by \citealt{CubillosEtal2021-BART_II}, \href{https://github.com/exosports/transit}{https://github.com/exosports/transit}). \textsc{MC3} is an open-source Bayesian framework that estimates a posterior distribution and the best-fit model \citep[][\href{https://github.com/pcubillos/mc3}{https://github.com/pcubillos/mc3}]{CubillosEtal2017apjRednoise}.

To start off retrieval (Figure \ref{fig:BARTChart}), \textsc{BART} generates an initial atmospheric model (see Section \ref{sec:Init}). Given a range of pressures, \textsc{BART} evaluates the temperature profile by using one of several parameterization schemes (see Section \ref{sec:atmGenerator}). Then, for given elemental and molecular species, the species volume mixing ratios are calculated using the  \textsc{TEA} module (see Section \ref{sec:TEA}) or a routine that produces vertically uniform abundance profiles. Arbitrary abundance profiles may also be used. The terms of the temperature-pressure, \math{T(p)}, profile and species abundances are free parameters. 

Given an initial guess, the atmospheric generator produces the atmospheric model, passes it to {\tt transit} (see Section \ref{sec:transit}) to calculate the transit or eclipse spectrum, and integrates the spectrum over the observational bandpasses (see Section \ref{sec:bandInteg}). The band-integrated values are sent to \textsc{MC3} to compare them against the observations and calculate \math{\chi\sp{2}}. Then, \textsc{MC3} generates a new set of free parameters, repeating the process until the phase space of all parameters is fully explored, and two convergence criteria, the Gelman and Rubin convergence test \citep{GelmanRubin1992} and the desired number of effectively independent samples (ESS), are satisfied \citep[see][Section 5 and Appendix C for more details on the ESS criterion]{HarringtonEtal2021-BART_I}. 

The best-fit parameters are then used to run the {\tt transit} module once more to reproduce the spectrum of the best-fit model atmosphere (see Section \ref{sec:bestFit}). In this run, {\tt transit} generates a file with the optical depth values for each atmospheric layer and wavelength, which is used to calculate the contribution functions of each observation (see Section \ref{sec:cf}). 

Finally, \textsc{BART} returns the best-fit parameters' values and credible regions, and plots the best-fit spectrum, the \math{T(p)} and abundance profiles, and the contribution functions. In addition, \textsc{MC3} plots the parameters' traces, pairwise posterior distributions, and marginal posterior histograms.

\begin{figure}[hb!]
\centerline{
\includegraphics[width=7cm, trim=0 0 0 20, clip=True]{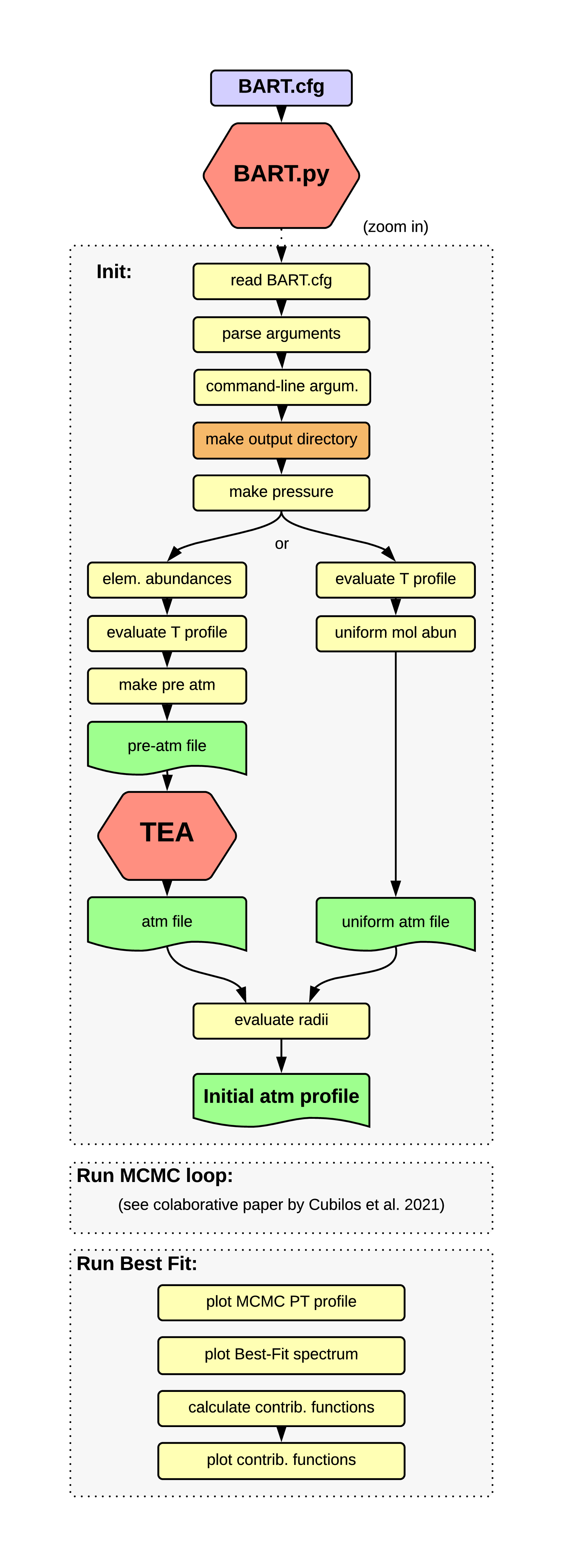}}
\vspace{-20pt}
\caption{\footnotesize
Flow chart of the \textsc{BART} driver, {\tt BART.py}. The gray sections show the initialization routines, the {\tt MCMC} loop (described in the collaborative paper by \citealt{CubillosEtal2021-BART_II}), and the best-fit routines (see Section \ref{sec:bestFit}). All routines are called inside {\tt BART.py}.}
\label{fig:BARTpy}
\end{figure}

In the following sections, we describe in more detail the implementation of the initialization routines, Section \ref{sec:Init}; the atmospheric profile generator, Section \ref{sec:atmGenerator}; the spectrum integrator, Section \ref{sec:bandInteg}; the best-fit routines, Section \ref{sec:bestFit}; and the contribution function module, Section \ref{sec:cf}. In Sections \ref{sec:TEA}, \ref{sec:MC3}, and \ref{sec:transit} we briefly describe \textsc{TEA}, \textsc{MC3}, and {\tt transit} modules for completeness, but refer the reader to \citet{BlecicEtal2016-TEA, CubillosEtal2017apjRednoise} and \citet{CubillosEtal2021-BART_II}, respectively, where we describe these codes in detail. 

\subsection{\textsc{BART} Initialization}
\label{sec:Init}

\textsc{BART} configures an initial model atmosphere following the user's educated guess about the planetary temperature, pressure, and species volume mixing ratios (abundances) at each atmospheric layer. The choice of this initial model atmosphere does not influence the final result, in the case when, as assumed in \textsc{BART}, one retrieves constant-with-altitude abundances profiles.The \math{T(p)} profile can be generated either using parameterized or isothermal schemes, while the chemical species can be calculated using either the \textsc{TEA} code or assuming vertically uniform profiles. The initial configuration parameters are set in an ASCII file, {\tt BART.cfg}, that carries, in addition, the full set of other parameters to run retrieval. \textsc{BART} driver, {\tt BART.py}, accepts this file and executes all modules and subroutines. It communicates the initial parameters to \textsc{MC3} for reference, but also runs the initialization instead of the Atmospheric Profile Generator on the first iteration. Figure \ref{fig:BARTpy} shows a simplified execution order of the routines called in {\tt BART.py}. Section {\tt Init} outlines steps executed to generate the initial atmospheric profile based on the user's choice of the elemental and molecular species, elemental volume mixing ratios, C/O ratio, metallicity, and whether chemical equilibrium or uniform abundances are assumed.  Section {\tt Run MCMC Loop} is described in the collaborative paper by \citet{CubillosEtal2021-BART_II}, and Section {\tt Run Best Fit} is described in Section \ref{sec:bestFit} of this paper.

\subsection{Atmospheric Profile Generator}
\label{sec:atmGenerator}

Once the initial atmospheric model is generated and compared to the observations using a radiative-transfer algorithm ({\tt transit}), any subsequent atmospheric models are generated using the atmospheric profile generator (see Figure \ref{fig:BARTChart}). 

This submodule generates models based on the parameters passed by a statistical algorithm (\textsc{MC3}). To calculate the abundances of free chemical species \textsc{BART} changes the initial abundance profiles using one scaling factor per species. To create a \math{T(p)} profile, \textsc{BART} can currently use one of four parameterized temperature-profile schemes: isothermal, adiabatic, one originally developed by \citet{Guillot2010A-LinePTprofile}, and one based on the parameterization described in \citet{MadhusudhanSeager2009ApJ-AbundanceMethod}.  

The isothermal parametrization scheme, often utilized in transmission geometry, uses one free parameter, \math{T\sb{iso}}, assuming the same temperature across all altitudes. 

The adiabatic parametrization scheme,  originally developed for brown-dwarf atmospheres, assumes that an adiabat is driven by the relationship \math{p\sp{1-\gamma}T\sp{\gamma} = \textrm{constant}}, where \math{\gamma} is the ratio of specific heats at constant pressure and constant volume. The temperature as a function of pressure \math{p} is then given as \math{T = {T_0}/\left[{1 + \frac{\gamma - 1}{\gamma}\textrm{ln}\left({p_0}/{p}\right)}\right]}, and has three free parameters: reference temperature \math{T\sb{0}}, reference pressure \math{p\sb{0}}, and \math{\gamma}.

The last two parametrization schemes, utilized most often in the eclipse geometry, allow for more complex temperature profiles to be explored in retrieval. We denote them as T-Parameterization schemes I and II, and in the following Sections \ref{sec:Line} and \ref{sec:Madhu} we outline their basic equations implemented in \textsc{BART}. More details about the parameters and physics behind these models can be found in \citet{LineEtal2013-Retrieval-II} and \citet{MadhusudhanSeager2009ApJ-AbundanceMethod}, respectively, while details on the limitations of these approaches are discussed in \citet{HengEtal2014} and \citet{BlecicEtal2017}.

\subsubsection{T-Parameterization Scheme I}
\label{sec:Line}

This parameterization scheme was originally formulated by \citet{Guillot2010A-LinePTprofile} and subsequently modified by \citet{HengEtal2012LinePTprofile, LineEtal2013-Retrieval-II} and \citet{ParmentierGuillot2014-LinesPTprofile}. In this approach the planet's temperature is given as

\begin{equation}
T^4(\tau) = \frac{3T_{\rm{int}}^4}{4} \big(\frac{2}{3} + \tau\big) + \frac{3T_{\rm{irr}}^4}{4}(1 -\alpha)\,\xi_{\gamma_{\rm 1}}(\tau) + \frac{3T_{\rm{irr}}^4}{4}\,\alpha\,\xi_{\gamma_{\rm 2}}(\tau)\, ,
\label{Line}
\end{equation}

\noindent where \math{\xi\sb{\gamma_{\rm i}}} is

\begin{equation}
\xi_{\gamma_{\rm i}} = \frac{2}{3} + \frac{2}{3\gamma_{\rm i}}\big[1 + \big(\frac{\gamma_{\rm i}\tau}{2} - 1\big)\,e^{-\gamma_{\rm i}\tau}\big] + \frac{2\gamma_{\rm i}}{3} \big(1 - \frac{\tau^2}{2}\big)\,E_2(\gamma_{\rm i}\tau)\, .
\end{equation}

\noindent Parameters \math{\gamma\sb{\rm 1}} = \math{\kappa\sb{\upsilon\sb{1}}/\kappa\sb{IR}}  and \math{\gamma\sb{\rm 2}} = \math{\kappa\sb{\upsilon\sb{2}}/\kappa\sb{IR}} are ratios of the mean opacities in the visible to the infrared. The parameter \math{\alpha} ranges between 0 and 1 and describes the partition between the two visible streams, \math{\kappa\sb{\rm{\upsilon{1}}}} and \math{\kappa\sb{\rm{\upsilon{2}}}}. \math{E\sb{2}(\gamma\tau)} is the exponential integral function. The planet internal flux  is given as

\begin{equation}
T\sb{\rm{int}} = \beta\,\big(\frac{R_{*}}{2a})^{1/2}\, T_{*} \, 
\end{equation}

\noindent where \math{R\sb{*}} and \math{T\sb{*}} are the stellar radius and temperature, and \math{a} is the semimajor axis. The incident solar flux is denoted as \math{T\sb{\rm{irr}}}. Both \math{T\sb{\rm{int}}} and \math{T\sb{\rm{irr}}} variables have fixed values. The parameter \math{\beta} accounts for albedo, emissivity, and day-night redistribution, and has a value around 1 for zero albedo and unit emissivity. The parameter \math{\tau} is the optical depth and it is calculated using the mean infrared opacity \math{\kappa\sb{\rm{IR}}}, pressure \math{P}, and the planet surface gravity \math{g} at the 1 bar level

\begin{equation}
\tau = \frac{\kappa_{\rm{IR}}\,P}{g} \, .
\end{equation}

\begin{figure}[h!]
    \includegraphics[height=5.2cm, width=0.45\textwidth]{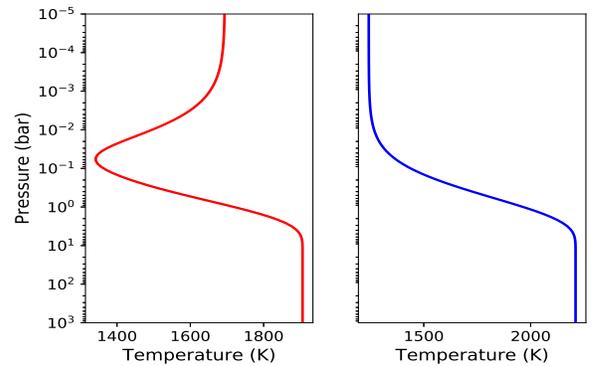}
\caption{Parametric T(p) profiles for inverted (left) and non-inverted (right) atmosphere based on \citet{Guillot2010A-LinePTprofile} and \citet{LineEtal2013-Retrieval-II}, generated using Equation \ref{Line}.}
\label{fig:LinePT}
\end{figure}

This parameterized approach has five free parameters: \math{\kappa\sb{\rm{IR}}}, \math{\kappa\sb{\rm{\upsilon{1}}}}, \math{\kappa\sb{\rm{\upsilon{2}}}}, \math{\alpha}, and \math{\beta}. The energy balance at the top of the atmosphere is accounted for through the parameter \math{\beta}. The existence of a temperature inversion is allowed through the parameters \math{\kappa\sb{\rm{\upsilon{1}}}} and \math{\kappa\sb{\rm{\upsilon{2}}}}. More details on these parameters can be found in \citet{Guillot2010A-LinePTprofile}. For boundaries imposed on these parameters see \citet{LineEtal2014-Retrieval-I}, Section 3.2. Figure \ref{fig:LinePT} shows one inverted and one non-inverted temperature profile generated using Equation \ref{Line}. Figure 21 of \citet{BlecicEtal2017} shows an exploration of all parameters.

\subsubsection{T-Parameterization Scheme II}
\label{sec:Madhu}

Our second option for the temperature-profile generator is a parameterization scheme similar to the one developed by \citet{MadhusudhanSeager2009ApJ-AbundanceMethod}. We made some minor changes to this method, which we describe below.

\begin{figure}[h]
\vspace{-10pt}
    \includegraphics[clip,width=0.50\textwidth]{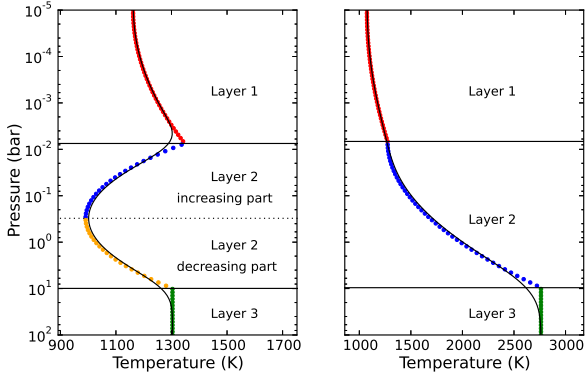}
\caption{Parametric T(p) profiles for inverted and non-inverted atmosphere based on \citet{MadhusudhanSeager2009ApJ-AbundanceMethod}. {\bf Left:} For the inverted atmosphere, Layer 2 consists of two parts: one where temperature decreases with height, and the other where temperature increases with height (thermal inversion occurs). Different colors depict partial profiles for each layer generated using Equation (\ref{invert}). The dots display the number of levels in the atmosphere, which is equally spaced in log-pressure space. The thin black line shows the smoothed profile. {\bf Right:} Non-inverted atmospheric profile. The different colors depict partial profiles for each layer generated using Equation (\ref{NONinvert}). The thin black line shows the smoothed profile.}
\label{fig:MadhuPT}
\end{figure}

In this scheme, the profiles are generated for inverted and non-inverted atmospheres separately. The atmosphere is divided into three layers based on the physical constraints expected in hot Jupiters (Figure \ref{fig:MadhuPT}). For more details on each of those layers see Section 2.1 in \citet{MadhusudhanSeager2009ApJ-AbundanceMethod}. The following set of equations mimics the behaviour in each atmospheric layer

\begin{align}
\label{MadhuPT}
  P_0 &< P < P_1& &P = P_0e^{\alpha_1(T - T_0)^{\beta_1}} &\hspace{0.2cm} \rm{layer\,\,1} \nonumber \\
  P_1 &< P < P_3& &P = P_2e^{\alpha_2(T - T_2)^{\beta_2}} &\hspace{0.2cm} \rm{layer\,\,2}  \\
  P~&> P_3& &T = T_3 &\hspace{0.2cm} \rm{layer\,\,3}  \nonumber
\end{align}

\noindent The set contains 12 variables: \math{P}\sb{0}, \math{P}\sb{1}, \math{P}\sb{2},  \math{P}\sb{3}, \math{T}\sb{0}, \math{T}\sb{1}, \math{T}\sb{2}, \math{T}\sb{3}, \math{\alpha\sb{1}}, \math{\alpha\sb{2}}, \math{\beta\sb{1}}, and \math{\beta\sb{2}}. To decrease the number of free parameters, we first set \math{P}\sb{0} to the pressure at the top of the atmosphere. The parameters \math{\beta\sb{1}} and \math{\beta\sb{2}} are empirically determined to be \math{\beta\sb{1}} = \math{\beta}\sb{2} = 0.5 \citep[see Section 2.3 in ][]{MadhusudhanSeager2009ApJ-AbundanceMethod}.Two of the parameters can be eliminated based on the two constraints of continuity at the two layer boundaries. The temperature \math{T\sb{3}} is estimated based on the effective (surface) temperature of the planet. When a planet total emissivity in the observed wavelength band is less than one, due to the presence of an atmosphere, its emissivity is less than that of a black body and the actual temperature of the object is higher than the effective temperature. Thus, to account for the presence of an atmosphere and spectral features, we use a scaling factor of 1 to 1.5 to constrain the maximum range of the \math{T\sb{3}} temperature.

The effective temperature of the planet is calculated based on the energy balance equation as

\begin{equation}
\label{Teff}
T_{\rm eff}^4 = f\,\, T_{\rm star}^{4}\, \Big(\frac{R}{a}\Big)^2\,(1-A)\, ,
\end{equation}

\noindent where factor \math{f} describes the energy redistribution from the day to the night side, and \math{T\sb{\rm star}} is the temperature of the star. \math{f} = 1/4 defines the uniform redistribution of energy between the day and the night side of the planet. In the case where the energy received is uniformly
redistributed on the planet's dayside, and none of the energy is
transferred to the night side, \math{f = 1/2}. For zero albedo, Equation (\ref{Teff}) becomes

\begin{equation}
\label{PlanetTeff}
T_{\rm eff}^4 = \frac{1}{2}\,T_{\rm star}^{4}\, \Big(\frac{R}{a}\Big)^2\, .
\end{equation}

To remove one more free parameter, we rewrite Equation (\ref{MadhuPT}) to distinguish the increasing and decreasing part of the Layer 2 curve

\begin{align}
\label{eqn:JB-PT}
  P_0 &< P < P_1& &P = P_0e^{\alpha_1(T - T_0)^{1/2}} &\hspace{0.2cm} \nonumber \\
  P_1 &< P < P_2& &P = P_2e^{\alpha_2(T - T_2)^{1/2}} &\hspace{0.2cm} \\
  P_2 &< P < P_3& &P = P_2e^{-\alpha_2(T - T_2)^{1/2}} &\hspace{0.2cm} \nonumber \\
  P~&> P_3& &T = T_3 &\hspace{0.2cm}\nonumber
\end{align}

\vspace{8pt}
\hspace{45pt} 2.2.2.1 \hspace{2pt} {\small \em Inverted \math{T(p)} Profile} 
\vspace{6pt}

The parametric profile for the inverted atmosphere has six free parameters: \math{P}\sb{1}, \math{P}\sb{2},  \math{P}\sb{3}, \math{T}\sb{3}, \math{\alpha\sb{1}}, and \math{\alpha\sb{2}}. We calculate the \math{T\sb{0}}, \math{T\sb{1}}, and \math{T\sb{2}} temperatures as

\begin{align}
\label{invert}
  T_2 = T_3 - \big(\frac{\log(P_3/P_2)}{\alpha_2}\big)^2   \nonumber \\
  T_0 = T_2 - \big(\frac{\log(P_1/P_0) }{\alpha_1}\big)^2 + \big(\frac{\log(P_1/P_2)}{-\alpha_2}\big)^2  &\hspace{0.4cm} \\
  T_1 = T_0 + \big(\frac{\log(P_1/P_0)}{\alpha_1}\big)^2 &\hspace{0.2cm} \nonumber
\end{align}

An example of an inverted \math{T(p)} profile is shown in Figure \ref{fig:MadhuPT}, left panel. To remove sharp kinks on the layer boundaries (between Layers 1-2 and Layers 2-3) we again followed the \citet{MadhusudhanSeager2009ApJ-AbundanceMethod} approach and used a SciPy smoothing function with the nearest neighbor and the lowest standard deviation settings, smoothing only a few points around the boundaries. Such an approach does not cause any correlation between distant atmospheric layers and mimics the smooth transitions observed in temperature profiles of Solar system planets \citep[see Figure 2 in][]{MadhusudhanSeager2009ApJ-AbundanceMethod}.

\vspace{8pt}
\hspace{42pt} 2.2.2.2 \hspace{2pt} {\small \em Non-Inverted \math{T(p)} Profile}
\vspace{6pt}

For the non-inverted atmosphere, we assume that the Layer 2 follows an adiabatic temperature profile and exclude \math{P\sb{2}} as a free parameter. Thus, the parametric profile for the inverted atmosphere has five free parameters: \math{P}\sb{1}, \math{P}\sb{3}, \math{T}\sb{3}, \math{\alpha\sb{1}}, and \math{\alpha\sb{2}}.  We calculate \math{T\sb{0}} and \math{T\sb{1}} as

\begin{align}
\label{NONinvert}
  T_1 = T_3 - \big(\frac{\log(P_3/P_1)}{\alpha_2}\big)^2   \\
  T_0 = T_1 - \big(\frac{\log(P_1/P_0)}{\alpha_1}\big)^2  &\hspace{0.1cm} \nonumber
\end{align}

An example of a non-inverted \math{T(p)} profile is shown in Figure \ref{fig:MadhuPT}, right panel. 
Figure 19 of \citet{BlecicEtal2017} shows an exploration of all parameters.

\subsubsection{Species Factors}
\label{sec:specsFac}

In addition to the temperature profile parameters, the atmospheric generator accepts free parameters for the molecular species scaling factors. For each selected free species, we modify its initial volume mixing ratio profile by multiplying it with a scaling factor. As commonly assumed in the literature, for retrieval we adopt constant-with-altitude initial abundances profiles. In this way, the choice of the initial model atmosphere does not influence the final retrieval results. We allow the species volume mixing ratios between a maximum value of 1 and a minimum value low enough such that the species does not affect the absorption spectrum (e.g., \math{10\sp{-12}}). Such an approach allows for non-equilibrium conditions in the planetary atmosphere to occur. Additionally, one can impose a maximum value for the sum of the fitting volume mixing ratios (e.g., to ensure a H\sb{2}/He-dominated atmosphere). 

We have as many free parameters as species we want to fit in our model. Our scaling factors apply to the entire initial profile for a species, and we are retrieving {\em log} of the species abundances to prevent negative and physically implausible mixing ratios and to allow them to vary over several orders of magnitude. This code can easily be modified to use abundance profiles with multiple free parameters, as with \math{T(p)}.

\subsection{\textsc{TEA} Module}
\label{sec:TEA}

To calculate the equilibrium abundances for the species of interest, we use the Thermochemical Equilibrium Abundances code, \textsc{TEA}, \citet{BlecicEtal2016-TEA}. This calculation allow us to estimate the initial constant-with-altitude species abundances for retrieval. \textsc{TEA} calculates the volume mixing fractions of gaseous molecular species following the methodology of \citet{WhiteJohnsonDantzig1958JGibbs} and \citet{Eriksson1971}. Given a \math{T(p)} profile and elemental abundances, \textsc{TEA} determines the volume mixing fractions of the desired molecular species by minimizing the total Gibbs free energy of the system \citep{ZeleznikGordon:1960}. The minimization is done using an iterative Lagrangian steepest-descent method that minimizes a multi-variate function under constraint. In addition, to guarantee physically plausable results, i.e., positive volume mixing fractions, \textsc{TEA} implements the Lambda Correction algorithm \citep{WhiteJohnsonDantzig1958JGibbs}. 

This approach requires a knowledge of the free energy of the species as a function of temperature. These are obtained from the JANAF (Joint Army Navy Air Force) tables \citep[\href{http://kinetics.nist.gov/janaf/}{http://kinetics.nist.gov/janaf/},][]{ChaseEtal1982JPhJANAFtables, ChaseEtal1986bookJANAFtables}. Thus, \textsc{TEA} has an access to 84 elemental species and the thermodynamical data for more than 600 gaseous molecular species (valid between 100 and 6000 K for the species of our interest here). We use the reference table containing elemental solar abundances given in \citet[][Table 1]{AsplundEtal2009-SunAbundances}. 

\textsc{TEA} is tested against the original method developed by \citet{WhiteJohnsonDantzig1958JGibbs}, the analytic method developed by \citet{BurrowsSharp1999apjchemeq}, and the Newton-Raphson method implemented in the free Chemical Equilibrium with Applications code (CEA, \href{http://www.grc.nasa.gov/WWW/CEAWeb/}{http://www.grc.nasa.gov/WWW/CEAWeb/}). Using the free energies listed in \citet[][]{WhiteJohnsonDantzig1958JGibbs}, their Table 1, and derived free energies based on the thermodynamic data provided in CEA's {\tt
  thermo.inp} file, \textsc{TEA} produces identical final
abundances for both approaches, but with a higher numerical precision. \textsc{TEA} is also benchmarked against the analytical codes developed by \citet{HengTsai2016ApJ, HengLyons2016ApJ, CubillosBlecicDobbs-Dixon2019ApJ}, and the numerical code developed by \citet{WoitkeEtal2018}, {\em GGchem}. 

The thermochemical equilibrium abundances obtained with \textsc{TEA} can be used in static atmospheres as well as a starting point in models of gaseous chemical kinetics and abundance retrievals. \textsc{TEA} is written in Python in a modular way, it is documented (the start guide, the user manual, the code document, and the theory paper are provided with the code), actively maintained, and available to the community via the open-source development site \href{https://github.com/dzesmin/TEA}{https://github.com/dzesmin/TEA}.

The thermochemical equilibrium abundances of the desired chemical species to be used in \textsc{BART} could also be calculated using any other available equilibrium abundances code like {\em GGChem} \citep{WoitkeEtal2018}, {\em FastChem} \citep{StockEtal2018-FastChem}, and {\em CEA} \citep{GordonMcBride:1994} providing the same input format as \textsc{BART} requires.

\subsection{Statistical Module}
\label{sec:MC3}

\textsc{BART} explores the parameter space of thermal profiles and species abundances using the Multi-core Markov-chain Monte Carlo module \citep[MC3,][]{CubillosEtal2017apjRednoise}. MC3 is an open-source general-purpose statistical package for model fitting. Using Bayesian Inference through a MCMC algorithm, MC3 provides three routines to sample the parameters' posterior distributions: Differential-Evolution \citep[DEMC, ][]{Braak2006DifferentialEvolution}, Snooker DEMC \citep{Braak2008SnookerDEMC}, and Metropolis Random Walk (using multivariate Gaussian proposals). It handles Bayesian priors (uniform, Jeffrey's, or informative), and implements the Gelman-Rubin convergence test \citep{GelmanRubin1992} together with ESS (see Section \ref{sec:BART}) that defines the accuracy of credible regions. It utilizes single-CPU and multi-core computation, supported through Messaging Passing Interface, MPI. The code, written in Python with several C-routines, is documented and available to the community via \href{https://github.com/pcubillos/}{https://github.com/pcubillos/mc3}. 

There are two versions of MC3, both written in Python 3. The main repo uses Python's {\tt multiprocessing} module, while a branch uses MPI. \textsc{BART} uses the MPI branch of MC3 for concurrent processing. The MC3 code consists of a central routine that coordinates the MCMC run and multiple worker routines that evaluate the {\tt Transit} model. On initialization, the central routine creates one worker for each MCMC chain, which remain instantiated throughout the run for efficiency. The central routine communicates with the workers via MPI by sending sets of model parameters and receiving back the corresponding $\chi^2$ for each iteration. These instances use shared memory to store the large opacity table and other objects used in the calculation, vastly reducing memory use. 

\subsection{Transit}
\label{sec:transit}

The {\tt transit} module solves the radiative-transfer equation to generate planetary spectra. It is an open-source  radiative-transfer code originally developed by \citet{Rojo2009-Transit} that applies a 1D opacity-sampling computation in local thermodynamic equilibrium to generate model spectra. This code, written in C, was built specifically to attempt to detect water in the extrasolar planet HD 209458b using transit spectroscopy. Since then, {\tt transit} has been significantly improved to handle eclipse geometry, multiple line-list and collision-induced absorption (CIA) sources, and to perform opacity grid and Voigt profile calculations (see the collaborative paper by \citealt{CubillosEtal2021-BART_II}). The \textsc{BART} project also added user and programmer documentation to the package. The code is available via \href{https://github.com/exosports/transit}{https://github.com/exosports/transit}.

\subsection{Spectrum Band Integrator}
\label{sec:bandInteg}

To compare the model spectra to the data, \textsc{BART} integrates the spectra over the spectral response curve for each observing band. For transit geometry, the observed transit depths are directly
compared to the band-integrated transmission spectra.  For eclipse geometry, the observed eclipse depths correspond to the planet-to-star flux ratio

\begin{equation}
  F\sb{p}/F\sb{star} = \frac{F\sp{\rm sf}\sb{\rm p}}{F\sp{\rm sf}\sb{\rm star}}\left(\frac{R\sb{p}}{R\sb{star}}\right)\sp{2},
\end{equation}
where \math{F\sp{\rm sf}\sb{\rm p}} is the surface flux spectrum of the planet, while \math{F\sp{\rm sf}\sb{\rm star}} is the surface flux spectrum of the star.
\textsc{BART} incorporates the Kurucz models for the stellar spectra
\citep{CastelliKurucz-2004new}.

\subsection{Best Fit}
\label{sec:bestFit}

Upon running the required number of MCMC iterations to satisfy both convergence criteria (see Section \ref{sec:BART}), \textsc{MC3} generates the best-fit parameters file that we use to run the {\tt transit} module one more time. This produces the final \textsc{BART} outputs, the best-fit atmospheric file, and the best-fit spectrum file. In addition, the \textsc{MC3} module generates a file with parameters' best-fit values and 68\math{\%} credible regions, trace plots showing the sequence of parameters' values for each MCMC iteration, 1D plots showing the parameters' marginalized posterior probability
distributions, and 2D plots showing pairwise posterior marginalizations for all the combinations of free-parameter pairs. These plots help identify possible non-convergence, multi-modal posteriors, correlations, or incorrect priors. Using this information \textsc{BART} plots the best-fit spectrum, the \math{T(p)}-profile posteriors, the abundances profiles, and the contribution functions. 

\subsection{Contribution Functions}
\label{sec:cf}

The contribution functions provide information on where the emission measured by the telescope originates \citep{Chamberlain1978-PlanAtmosp,Griffith1998-CF, KnutsonEtal2009ApJ-redistribution}. To assess the contribution from a certain layer to the observed intensity, we calculate two quantities: the transmission weighting function and the Planck function at a given temperature. The weighting function, which describes how transmission is changing with altitude, is the kernel of the radiative-transfer integral. It weights the contribution to the intensity from the Planck functions at different log-pressure altitudes. The contribution function, the product of the weighting function and the Planck function, is the integrand of the radiative-transfer integral.

The weighting function, \math{W}, is defined as a derivative of the transmission function, \math{\mathcal{T}} (\math{\mathcal{T} = e^{-\tau/\mu}}), as \math{W = \partial{\mathcal{T}}/\partial{z}}, where \math{\tau} is the optical depth, \math{z} is the altitude and \math{\mu} is the cosine of the ray path's angle to the normal. The altitude where the peak of the weighting function is found depends on the opacity at that wavelength. In other words, it is sensitive to the atmospheric thermal structure and composition (volume mixing fractions). The contribution function assesses vertical sensitivity of the emission spectrum by giving the pressure level at which thermal emission from the atmosphere contributes most to the intensity observed at the top of the atmosphere in each wavelength. 

To investigate the contribution from a certain atmospheric layer to the observed intensity, we start with the equation that defines the intensity at the top of the atmosphere

\begin{eqnarray}
I_\nu\,(\tau=0) & = & \int_{0}^{\tau}\frac{B_\nu}{\mu}\,e^{-\tau^{'}/\mu} d\tau^{'}\, , 
\label{final0}
\end{eqnarray}

\noindent and by using the hydrostatic balance equation and the equation of state rewrite it in terms of pressure
\begin{eqnarray}
I_\nu\,(p) =  \int_{p_{t}}^{p_{b}}{B_\nu}\,\frac{d\,\mathcal{T}}{d(\log\,p)}\, d(\log\,p) \,,
\label{final5}
\end{eqnarray}

\noindent where \math{p\sb{b}} and \math{p\sb{t}} are pressures at the top and the bottom of the atmosphere. The weighting function is then given as

\begin{eqnarray}
\label{weightP}
W(p) = \frac{d\mathcal{T}}{d(\log\,p)}\, ,
\end{eqnarray}

\noindent and the contribution function as

\begin{eqnarray}
\label{CF}
CF(p) = {B_\nu}\,\frac{d\mathcal{T}}{d(\log\,p)}\, ,
\end{eqnarray}

Since the weighting function is the convolution of the rising transmission and falling density, it will have a roughly Gaussian shape for all wavelengths that do not sense the surface. When the atmosphere is more transparent, the peak of the weighting function moves towards larger pressures (lower altitudes). For wavelengths where the atmosphere is completely transparent, the weighting function is below the surface and the shape becomes exponential. Above the peak, the telescope does not sense the atmosphere that well due to the low atmospheric density and few emitting molecules. Below the peak, the emitting radiation is mostly absorbed by the atmosphere above.

The weighting and contribution functions are key quantities in temperature retrievals and they strongly depend on the best-fitting models. \textsc{BART} calculates the contribution functions after \textsc{MC3} has determined the best-fit parameters. Using the best-fit parameters, we run the {\tt transit} module again and reproduce the optical-depth array of the best-fit model. The optical depths are used to calculate contribution functions at each wavelength across the planet's spectrum. The band-averaged contribution functions are obtained by integrating the calculated contribution functions across the filter bandpasses of our observations (the transmission response functions) at every pressure layer.

\section{Atmospheric Analysis of WASP-43\lowercase{b}}
\label{sec:analWASP43b}

WASP-43b \citep{HellierEtal2011aaWASP43bdisc} is one of the closest-orbiting hot Jupiters, revolving around one of the coldest stars (4400 {\pm} 200 K) that hosts hot Jupiters.  Attributed to the small radius of the host star \citep[0.667 {\pm} 0.011 \math{R\sb{\rm \sun}}, ][]{GillonEtal2012AA-WASP-43b}, its cool temperature, and small semi-major axis (0.01526 {\pm} 0.00018 AU), the system produces significant dips both in transit and eclipse. This makes WASP-43b one of the most favorable targets today for space and ground observations and a perfect exoplanet for atmospheric characterization. 

Previous WASP-43b atmospheric analyses suggest a carbon-rich composition \citep{ZhouEtal2014-WASP-43b} with C to O ratio larger than solar but smaller than 1 \citep{LineEtal2014-Retrieval-I, benneke2015strict}, water abundance consistent with solar composition \citep{StevensonEtal2014-WASP43b, kataria2014atmospheric, benneke2015strict}, super-solar metallicity \citep{StevensonEtal2014-WASP43b, kataria2014atmospheric, Stevenson2017AJ-WASP43phaseCurve} and possible existence of high altitude clouds or hazes \citep{ChenEtal2014-WASP43b, Stevenson2017AJ-WASP43phaseCurve}.
The presence of thermal inversion was initially ruled out by \citet{GillonEtal2012AA-WASP-43b, LineEtal2014-Retrieval-I} and \citet{BlecicEtal2014-WASP43b}, but found to be localized on the dayside of the planet by \citet{kataria2014atmospheric}. The planet energy budget and redistribution were analyzed by many groups, initially suggesting a poor day-to-night redistribution \citep{GillonEtal2012AA-WASP-43b, BlecicEtal2014-WASP43b, StevensonEtal2014-WASP43b, Stevenson2017AJ-WASP43phaseCurve}. However,  the inclusion of the reflected light in models by \citet{KeatingCowan2017ApJ-WASP-43b} revealed a much hotter night side. This result was subsequently confirmed by \citet{MendonMalikDemoryHeng2018AJ-WASP43b} utilizing their improved data reduction technique and assuming clouds on the planet's night side. A cloudy night side was also hypothesised by \citet{VenotEtal2020-JWST-WASP-43b}, who performed phase-curve retrieval analysis of WASP-43b synthetic {\em James Webb Space Telescope}/Mid-Infrared Instrument (JWST/MIRI) observations. Very recently,  \citet{Irwin2020MNRAS} performed a multidimensional '2.5D retrieval' of all WASP-43b orbital phases simultaneously, using an optimal-estimation algorithm, confirming again thick clouds on the night side of WASP-43b.

In this paper, we performed an atmospheric analysis of WASP-43b secondary eclipse data from space- and ground-based observations using \textsc{BART} (see Table \ref{table:eclDep}). Our goal was to compare our results with the results from the literature \citep{LineEtal2014-Retrieval-I, KreidbergEtal2014-WASP43b} and in this way validate our framework, and to investigate several cases previously unexplored for WASP-43b. 

{\renewcommand{\arraystretch}{1.0}
\begin{table}[hb!]
\caption{\label{table:eclDep} Eclipse Depths}
\atabon\strut\hfill\begin{tabular}{lcc}
    \hline
    \hline
    Source and Instrument                                           & Wavelength      & Eclipse Depth               \\
                                                      & \micron        &     (\%)                     \\
    \hline
    \citet{GillonEtal2012AA-WASP-43b}, {\em VLT}/HAWK I\tablenotemark{a}   & 1.19            & 0.079 {\pm} 0.032             \\ 
                                                      & 2.09            & 0.156 {\pm} 0.014             \\
    \hline
    \citet{WangEtal2013-WASP43b}, {\em CFHT}/WIRCam\tablenotemark{b}   & 1.65            & 0.103 {\pm} 0.017             \\
                                                      & 2.19            & 0.194 {\pm} 0.029             \\
    \hline  
    \citet{ChenEtal2014-WASP43b}, {\em MPG/ESO}/GROND\tablenotemark{c}
 & 0.806           & 0.037 {\pm} 0.022             \\
                                                      & 2.19            & 0.197 {\pm} 0.042             \\
    \hline
    \citet{BlecicEtal2014-WASP43b}, {\em Spitzer}    & 3.6             & 0.347 {\pm} 0.013             \\
                                                      & 4.5             & 0.382 {\pm} 0.015             \\
    \hline
    \citet{ZhouEtal2014-WASP-43b}, {\em IRIS2}\tablenotemark{d}       & 2.15            & 0.181 {\pm} 0.027             \\
    \hline
    \citet{StevensonEtal2014-WASP43b}, {\em HST}\tablenotemark{e}     & 1.1425          & 0.0365 {\pm} 0.0045           \\
                                                      & 1.1775          & 0.0431 {\pm} 0.0039           \\
                                                      & 1.2125          & 0.0414 {\pm} 0.0038           \\
                                                      & 1.2475          & 0.0482 {\pm} 0.0036           \\
                                                      & 1.2825          & 0.0460 {\pm} 0.0037           \\
                                                      & 1.3175          & 0.0473 {\pm} 0.0033           \\
                                                      & 1.3525          & 0.0353 {\pm} 0.0034           \\
                                                      & 1.3875          & 0.0313 {\pm} 0.0030           \\
                                                      & 1.4225          & 0.0320 {\pm} 0.0036           \\
                                                      & 1.4575          & 0.0394 {\pm} 0.0036           \\
                                                      & 1.4925          & 0.0439 {\pm} 0.0033           \\
                                                      & 1.5275          & 0.0458 {\pm} 0.0035           \\
                                                      & 1.5625          & 0.0595 {\pm} 0.0036           \\
                                                      & 1.5975          & 0.0614 {\pm} 0.0037           \\
                                                      & 1.6325          & 0.0732 {\pm} 0.0042           \\
    \hline
    \citet{StevensonEtal2017AJ}, {\em Spitzer}        & 3.6             & 0.3300 {\pm} 0.0089           \\
                                                      & 4.5             & 0.3827 {\pm} 0.0084           \\
    \hline
\end{tabular}\hfill\strut\ataboff
\tiny{
\begin{minipage}[t]{0.86\linewidth}
\tablenotetext{1}{Cryogenic near-IR imager HAWK at the Very Large Telescope}
\tablenotetext{2}{WIRCam on the Canada-France-Hawaii Telescope}
\tablenotetext{3}{Gamma Ray Burst Optical and Near-Infrared Detector on the MPG/ESO 2.2m telescope at La Silla Observatory, Chile}
\tablenotetext{4}{Very Large Telescope Infrared Image Sensor}
\tablenotetext{5}{Hubble Space Telescope}
\end{minipage}}
\end{table}

\subsection{\textsc{BART} Setup}
\label{sec:BARTsetup}

The pressure range for all of our models was constrained between 10\sp{2} and 10\sp{-5} bars, and sampled 100 times uniformly in log space. We used the parameterization from Section \ref{sec:Line} to generate our \math{T(p)} profile. We performed several trial runs including all five \math{T(p)} profile parameters and concluded that we can fix \math{\gamma\sb{2}} and \math{\alpha} to zero. Parameter \math{\gamma\sb{2}} is redundant when atmosphere is not inverted, as all our initial tests revealed, and parameter \math{\alpha} can then be set to zero. Fixing these parameters also leads to simpler model and preferred BIC. Thus, the three free parameters of our \math{T(p)} profile were \math{\kappa\sb{\rm IR}}, \math{\gamma\sb{1}}, and \math{\beta}. In addition, we constrained the temperature range between 300 and 3000 K to prevent \textsc{MC3}'s random walk from stepping outside of the plausible temperature range (this range is bounded by the HITRAN/HITEMP databases' partition functions). The remaining free parameters of our model were scaling factors of the species abundances. 

We choose our initial \math{T(p)} profiles by taking the parameter values from the literature. To estimate the appropriate values for the vertically uniform abundances, we used \textsc{TEA} to calculate the mixing ratios at every atmospheric level and then we  utilized the 0.1 bar level (approximate photosphere level) values as our initial guesses for species abundances.

When comparing our results with the results from \citet{LineEtal2014-Retrieval-I} and \citet{KreidbergEtal2014-WASP43b}, we included the same, four most spectroscopically active species (CO, CO\sb{2}, CH\sb{4}, and H\sb{2}O) and their opacities in our model. To test additional chemical scenarios, we also included other spectroscopically active species relevant for hot-Jupiter atmospheres known to have spectral features in the near-infrared spectral region (HCN, C\sb{2}H\sb{2}, C\sb{2}H\sb{4}, NH\sb{3}, HS, H\sb{2}S, TiO, and VO). \citet{MacDonaldMadhusudhan2017-Nytrogen} show that the inclusion of nitrogen species is crucial when dealing with the {\em HST} observations \citep[see also ][]{KilpatrickEtal2018apjWASP63bWFC3}, thus we included HCN and NH\sb{3} opacities in some of our models. The inclusion of C\sb{2}H\sb{2}, C\sb{2}H\sb{4}, and HCN opacities is relevant for the atmospheres with C/O ratio larger than one, as these species become even more abundant than H\sb{2}O under those conditions \citep[e.g, ][]{MosesEtal2013-COratio, BlecicEtal2016-TEA}. We also included TiO, VO and H\sb{2}S in some of our models, as TiO and VO are shown to cause thermal inversions in hot-Jupiter atmospheres \citep{Spiegel2009apjInversion, Hubeny2003, FortneyEtal2008-TiO}, while \citet{Zahnle09-SulfurPhotoch} argues that H\sb{2}S is also a potential inversion agent. To be consistent with previous studies, we neglected the effect of clouds and scattering in all of our models.

We used the ExoMol \footnote{\href{http://www.exomol.com/data/data-types/xsec}
{http://exomol.com/data/data-types/xsec}} database for H\sb{2}O, CO\sb{2}, NH\sb{3}, C\sb{2}H\sb{2}, C\sb{2}H\sb{4}, H\sb{2}S, TiO, and VO; and HITEMP \footnote{\href{https://www.cfa.harvard.edu/hitran/HITEMP.html}{https://www.cfa.harvard.edu/hitran/HITEMP.html}} database for CO and CH\sb{4} species as our sources for the molecular line-list data. For H\sb{2}O we used the molecular line-list data from \citet{PolyanskyEtal2018mnrasPOKAZATELexomolH2O}; for CO\sb{2} the line-list from \citet{rothman2010-hitemp}; for HCN from \citet{HarrisEtal2006mnrasHCNlineList, HarrisEtal2008mnrasExomolHCN}; for NH\sb{3} -- \citet{YurchenkoEtal2011mnrasNH3opacities, Yurchenko2015jqsrtBYTe15exomolNH3}; for C\sb{2}H\sb{2} -- \citet{WILZEWSKI-c2h2, GordonEtal2017jqsrtHITRAN2016}; for C\sb{2}H\sb{4} -- \citet{MantEtal2018-ExomolC2H4}; for TiO -- \citet{McKemmishEtal2019-ExomolTiO}, H\sb{2}S -- \citet{AzzamEtal2016-ExomolH2S}; and for VO -- \citet{McKemmishEtal2016-ExomolVO}. For CH\sb{4} we used the newly available HITEMP line-list from \citet{HargreavesEtal2020apjsHitempCH4} shown to provide more accurate linelist information than ExoMol, and for CO we used the HITEMP linelist from \citet{LiEtal2015apjsCOlineList}.  Since many ExoMol databases consist of billions of line transitions, we used the REPACK package \citep{Cubillos2017apjCompress} to extract only the strongest line transitions that dominate the opacity spectrum. The partition functions for the HITEMP opacity sources were calculated based on \citet{LaraiaEtal2011TIPS}, and for ExoMol we used the tabulated values provided with the line lists. In addition to the molecular line lists, we included the H\sb{2}-H\sb{2} collision induced opacities from \citet{BorysowEtal2001-H2H2highT} and \citet{Borysow2002-H2H2lowT}, and H\sb{2}-He collision induced opacities from \citet{RichardEtal2012-HITRAN-CIA}.

To assess how species opacities influence the WASP-43b spectrum and which ones have dominant effects, we generated a custom atmospheric forward model and its emission spectrum for each species in the 0.6--5.5 {\microns} wavelength range. We assumed a non-inverted temperature profile, solar elemental composition, and thermochemical equilibrium species abundances. Figure \ref{fig:opacs} shows the individual opacity cross sections for each of the 11 molecular species used in our analysis, at the temperature of 1500 K and pressure of 1 bar. We also show the corresponding emission spectra (using the same colors as the species opacities), showing the spectral regions where each species impact the WASP-43b spectrum.

The system parameters (planetary mass and radius, star's metallicity, effective temperature, mass, radius, gravity, and the semi-major axis) were taken from \citet{GillonEtal2012AA-WASP-43b}. These parameters were used in our parameterized temperature model and to generate the stellar spectrum by interpolating the stellar grid models from \citet{CastelliKurucz-2004new}. 

The response functions of IRIS2/AAT observations were provided by \citet{ZhouEtal2014-WASP-43b}, while the transmission response functions for Very Large Telescope/High Acuity Wide field K-band Imager (VLT/HAWK I), GROND, and WIRCam observations were provided by \citet{ChenEtal2014-WASP43b}. The {\em Spitzer} response functions for the channel 1 and 2 subarray observations were found on the {\em Spitzer} website. For each of the {\em HST} observations, we used the top-hat response functions.

We used uniform priors for \math{\beta} (the temperature-profile parameter, see Section \ref{sec:Line}) and log-uniform priors for all other temperature-profile parameters and molecular species, with boundary limits set wide enough to allow \textsc{MC3} to explore the parameter phase space thoroughly. On each iteration, we rescaled the mixing ratios of H\sb{2} and He, preserving their original ratio such that the total sum of all species' abundances is unity. To be able to compare our results with \citet{LineEtal2014-Retrieval-I}, we impose the same constraint as they do, that the sum of the fitted molecular species abundances must not exceed 15\%. Prior to proceeding with this approach we performed tests confirming that our conclusions are not changed by imposing this limit. This constraint is bounding the atmospheric compositions to a hydrogen-dominated atmosphere, roughly restricting the metallicity to less than 200 solar. Similar approach has been applied recently by \citet{Tsiaras2018AJ}. 

We generated the line-list data files for the species and the wavelength range of interest and from them we generated the opacity tables with the opacity grid between 300 to 3000 K in 100 K intervals, and between 0.6 - 5.5 {\microns} in 1 cm\sp{-1} intervals in the wavenumber space. The maximum optical depth was set to ten for all models ({\tt transit} stops the extinction calculation at each wavelength when the optical depth reaches the user-defined value \math{\tau\sb{\rm max}}, see collaborative paper by \citealp{CubillosEtal2021-BART_II}).

We ran ten independent chains and enough iterations until the Gelman and Rubin convergence test for all free parameters drops below 1\% \citep{GelmanRubin1992}, and until we reach a satisfactory ESS value \citep[see Section 5 and Appendix C of][]{HarringtonEtal2021-BART_I}. 

\begin{figure*}[ht!]
\centering
\includegraphics[width=.26\textwidth, height=3.5cm]{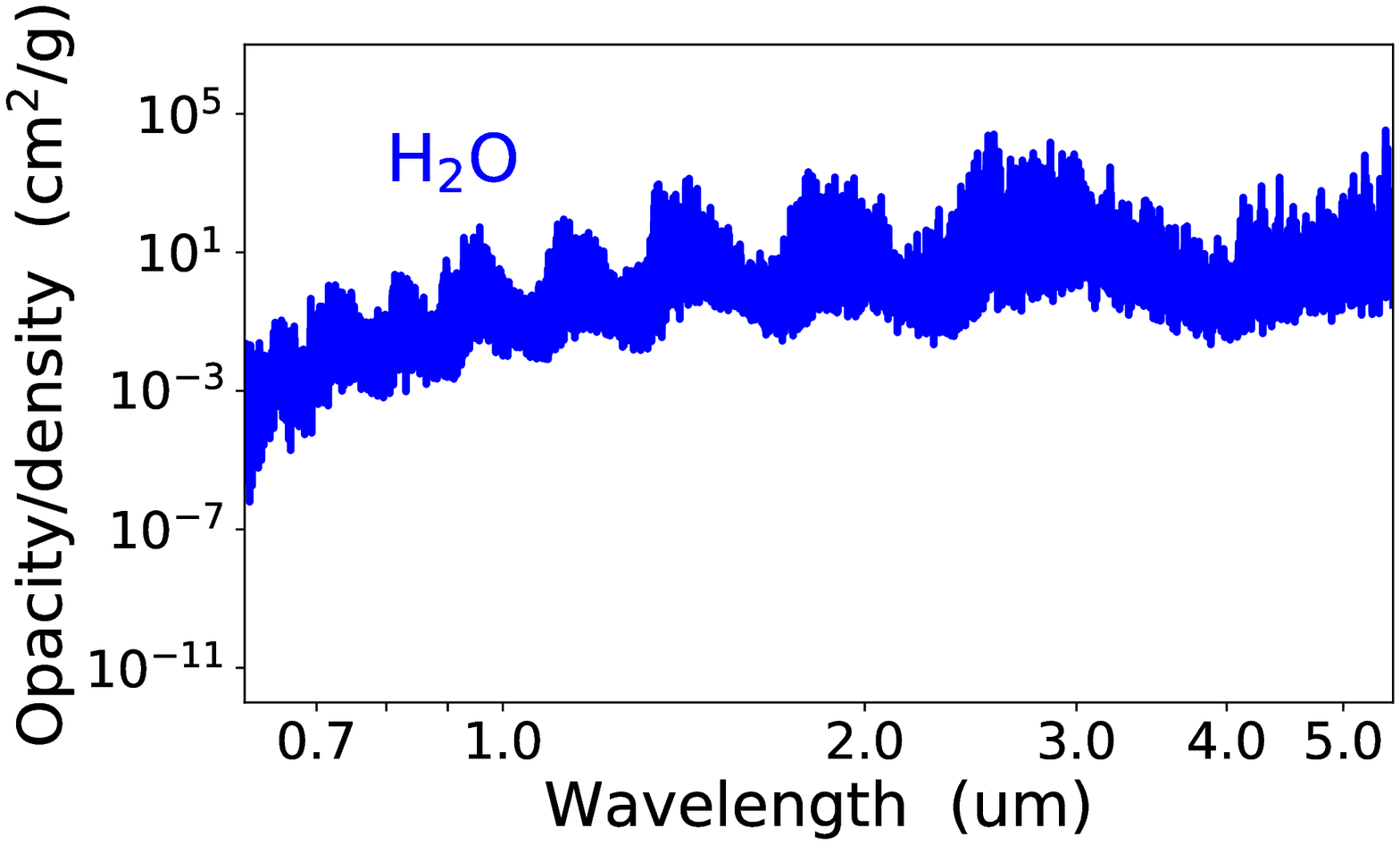}\hspace{-13pt}
\includegraphics[width=.26\textwidth, height=3.5cm]{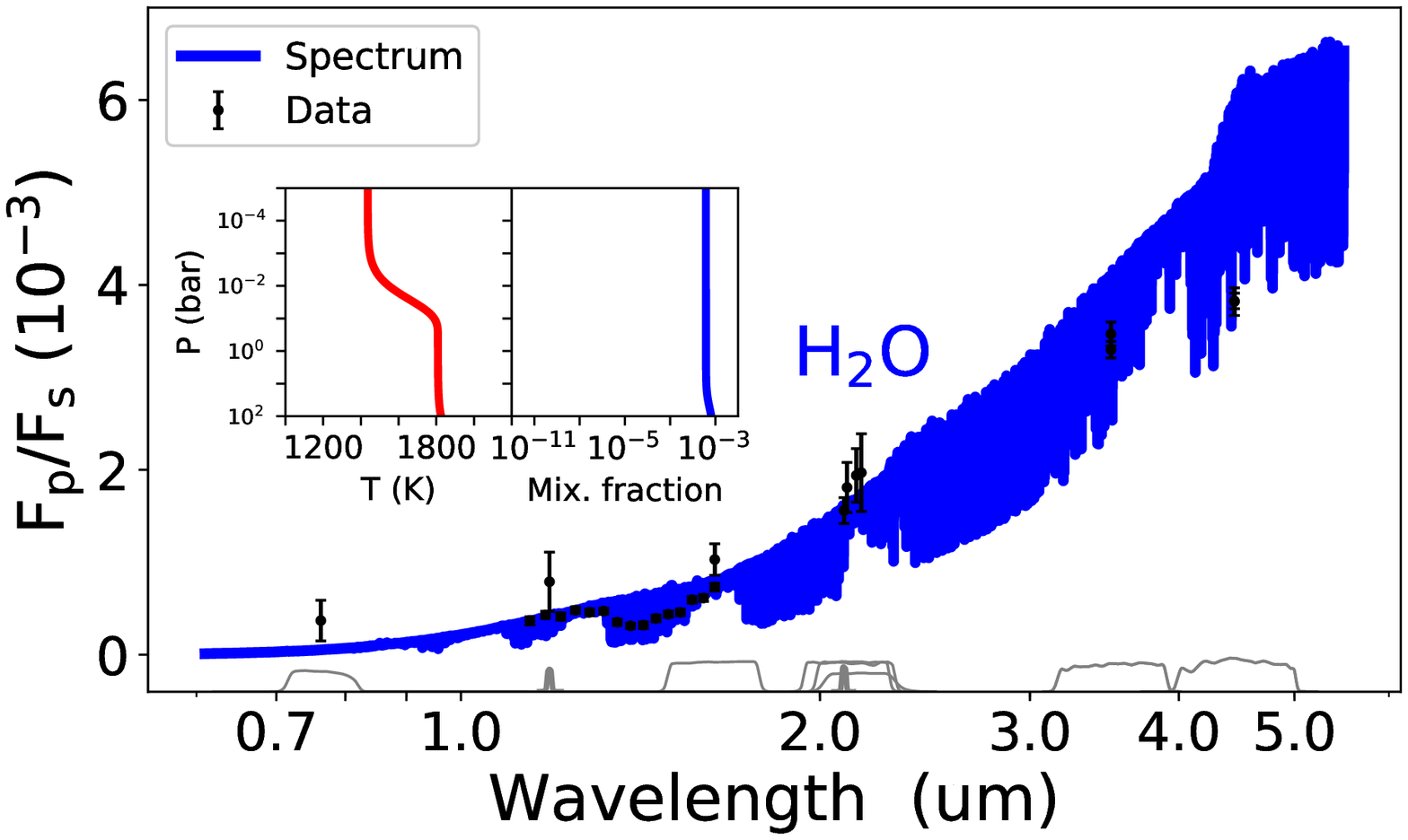}\hspace{-13pt}
\includegraphics[width=.26\textwidth, height=3.5cm]{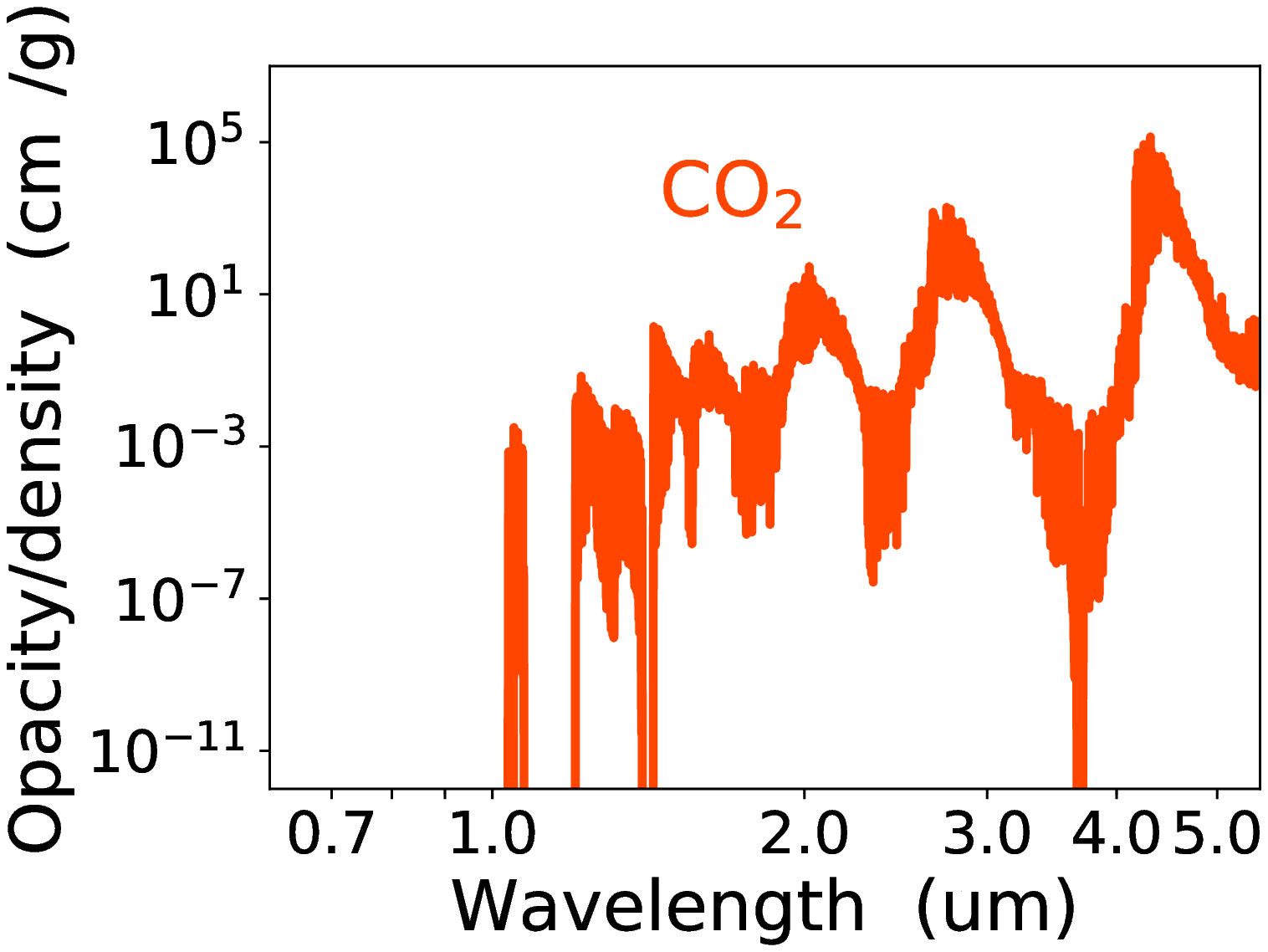}\hspace{-13pt}
\includegraphics[width=.26\textwidth, height=3.5cm]{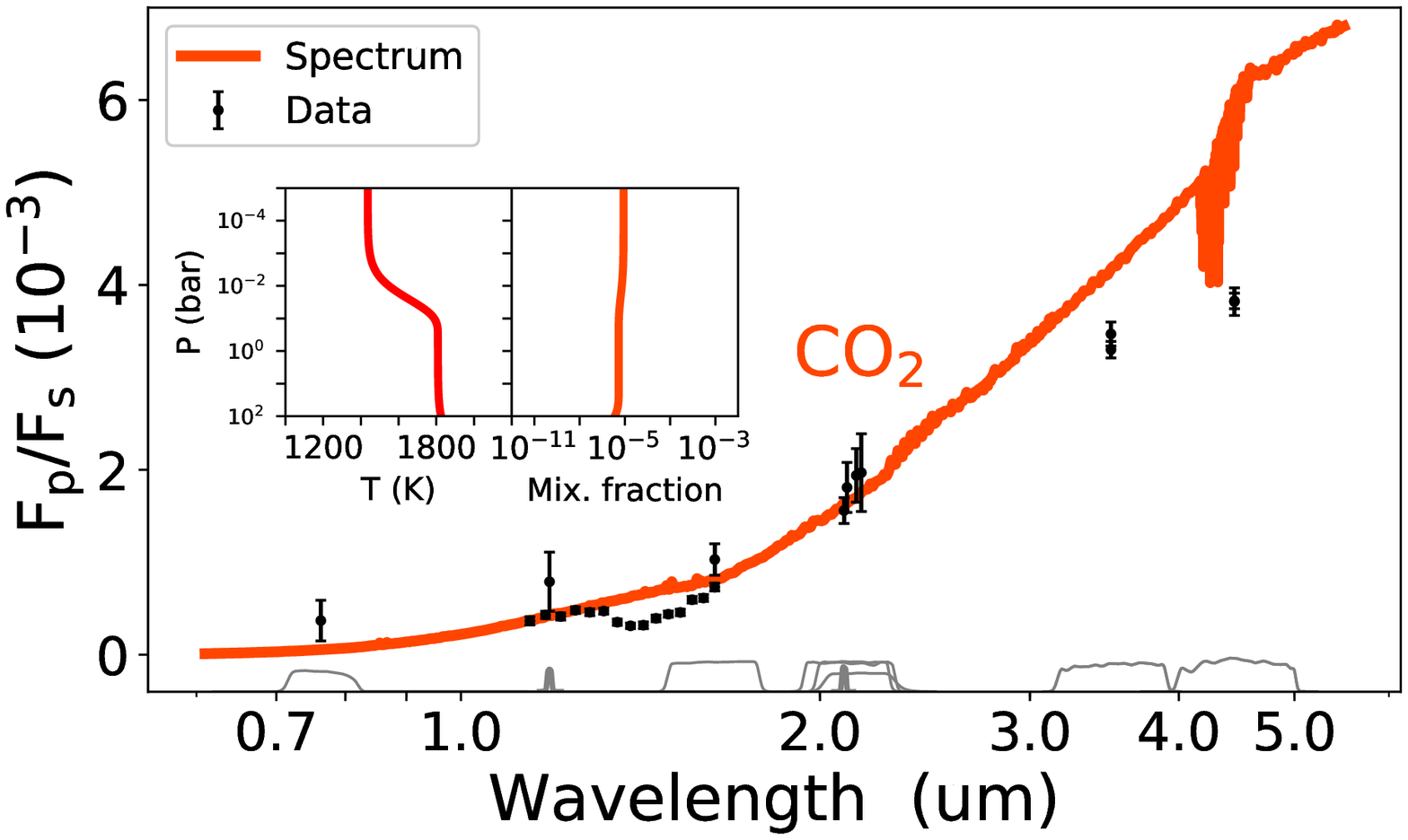}
\vspace{-2pt}

\includegraphics[width=.26\textwidth, height=3.5cm]{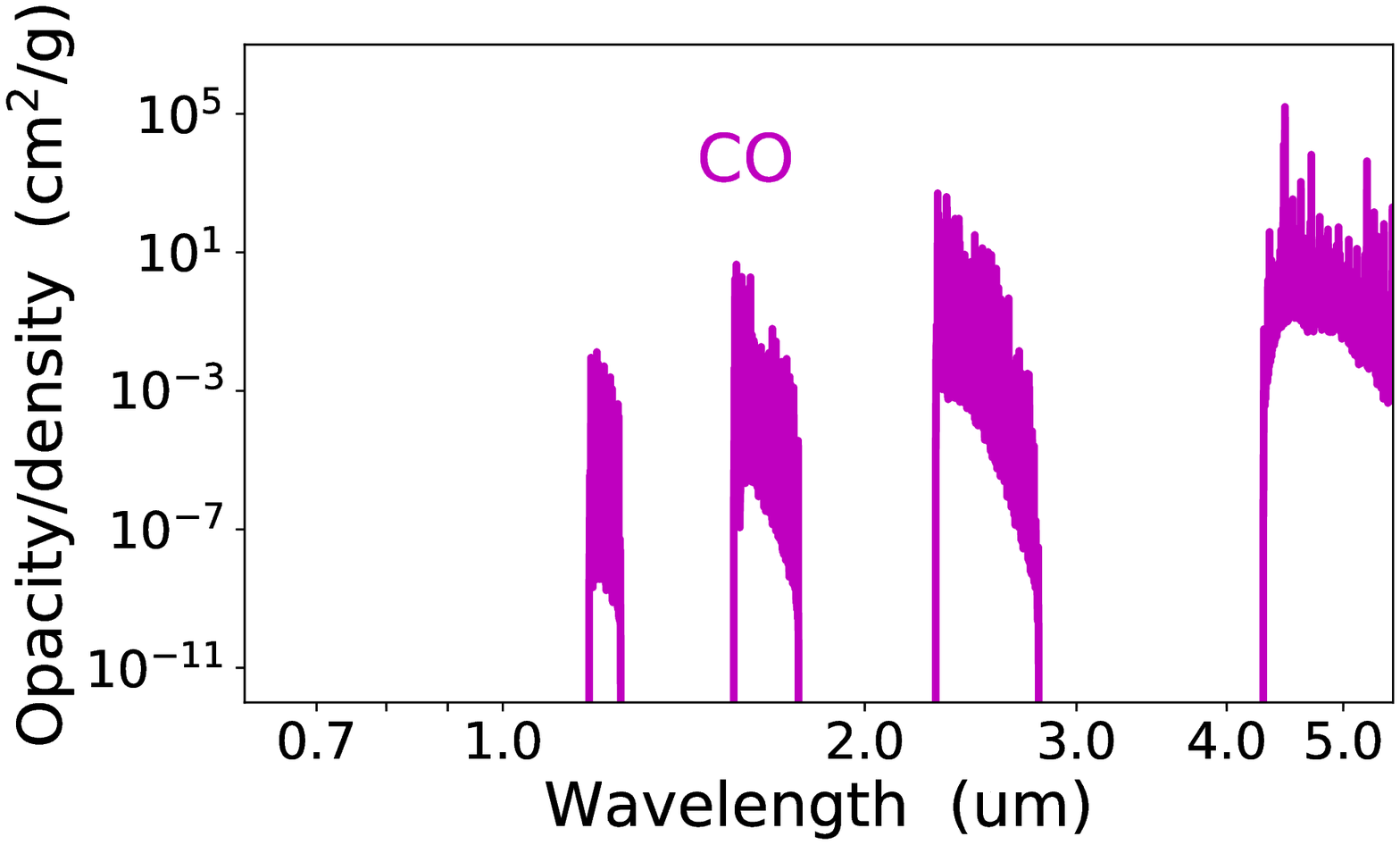}\hspace{-13pt}
\includegraphics[width=.26\textwidth, height=3.5cm]{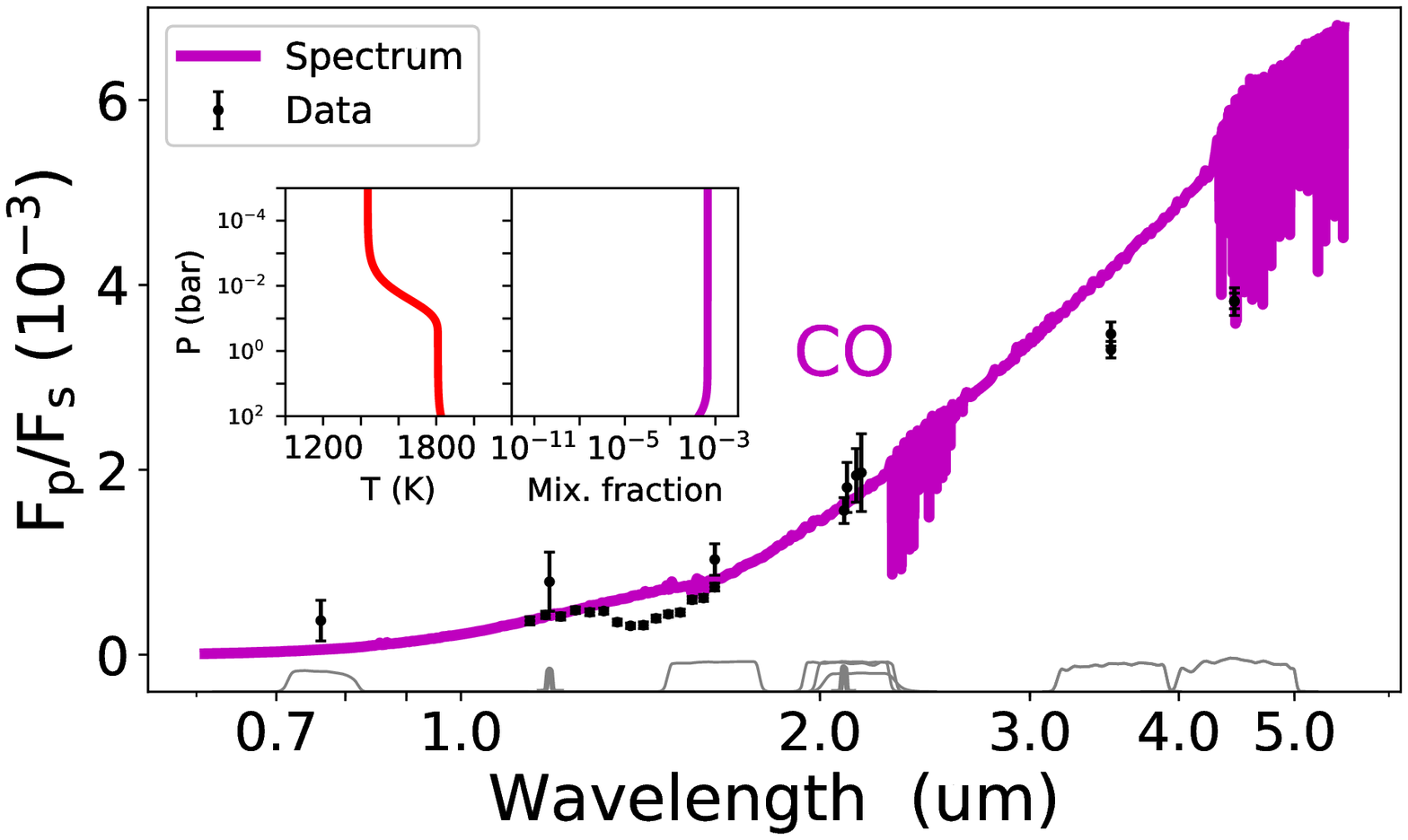}\hspace{-13pt}
\includegraphics[width=.26\textwidth, height=3.5cm]{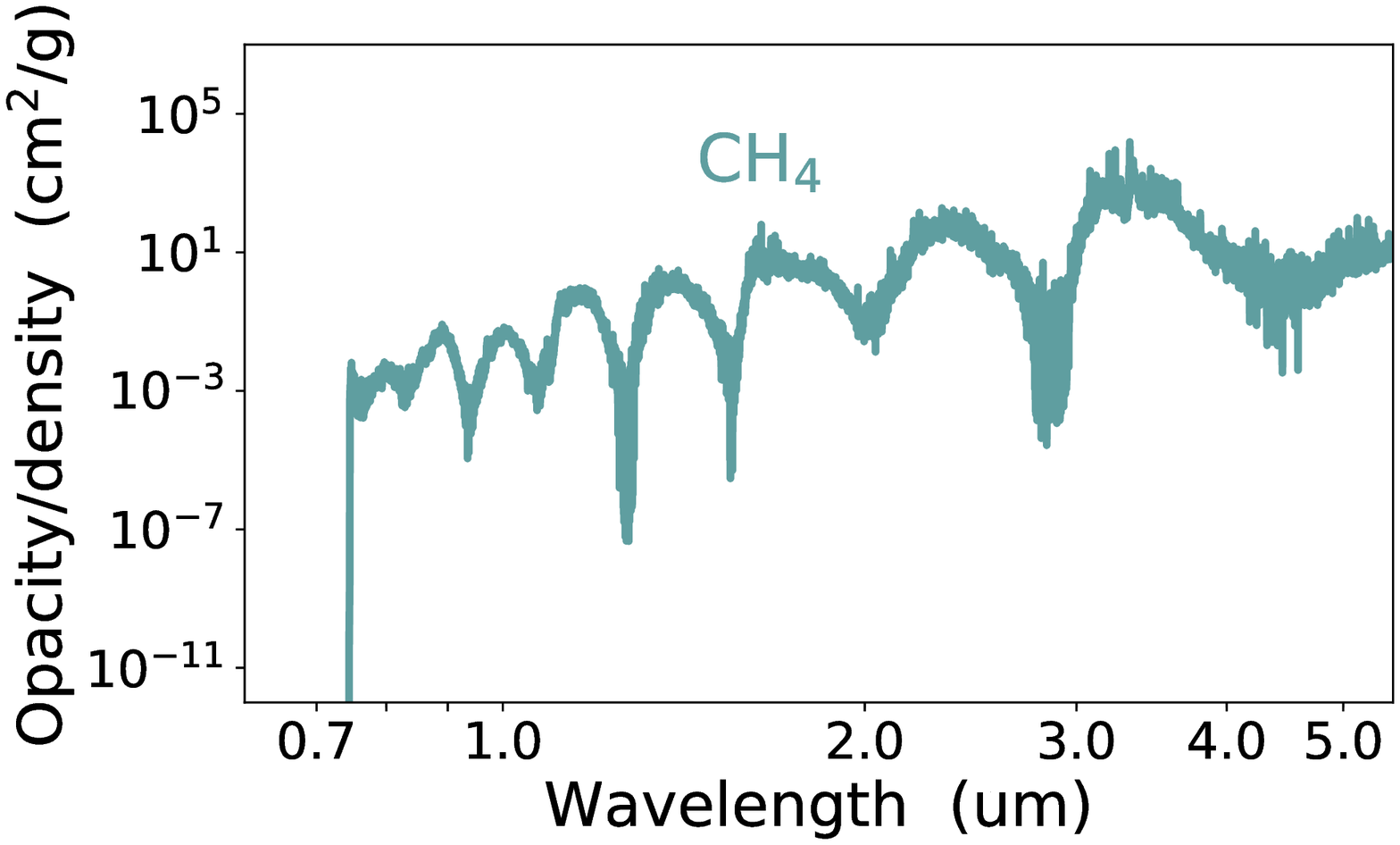}\hspace{-13pt}
\includegraphics[width=.26\textwidth, height=3.5cm]{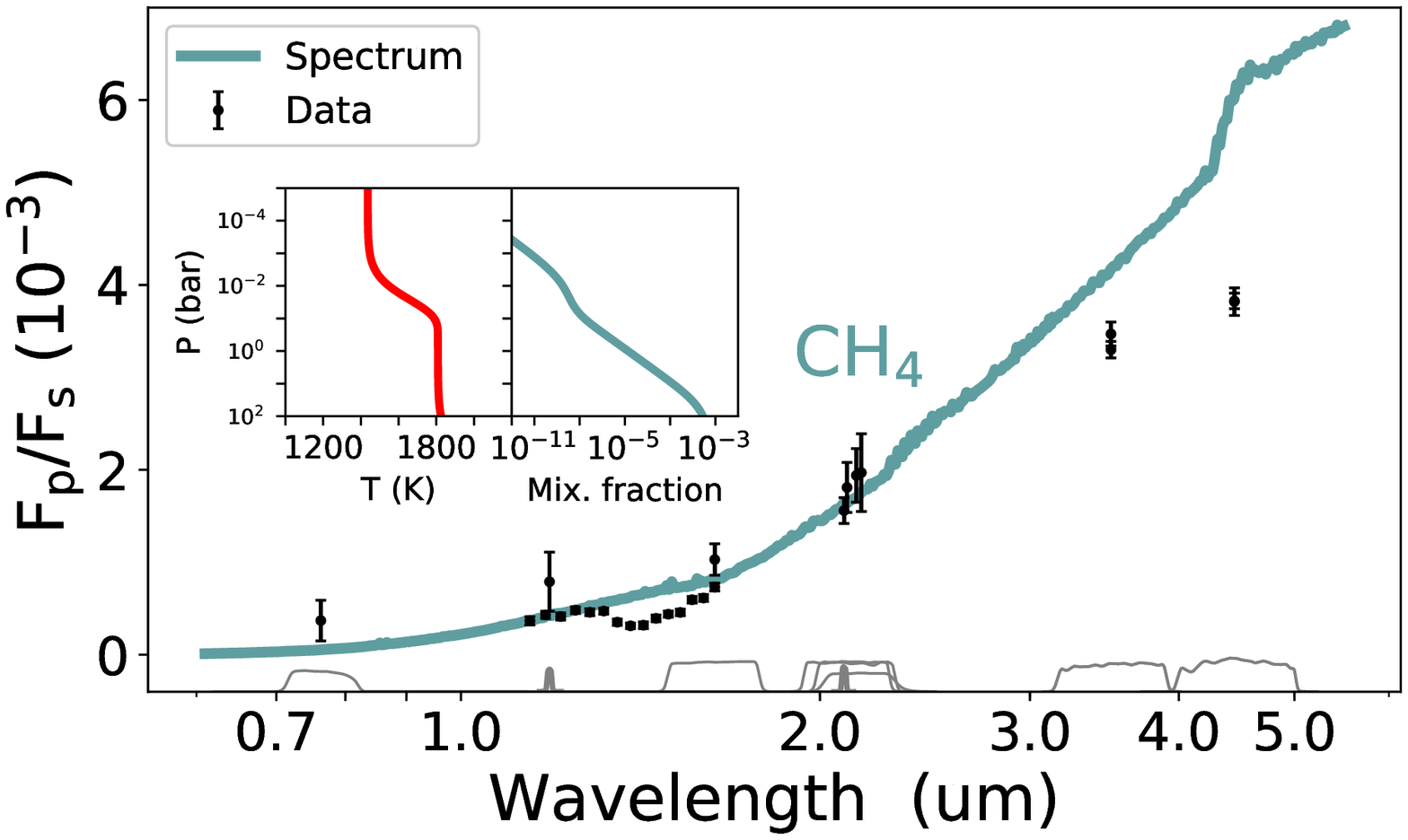}
\vspace{-2pt}

\includegraphics[width=.26\textwidth, height=3.5cm]{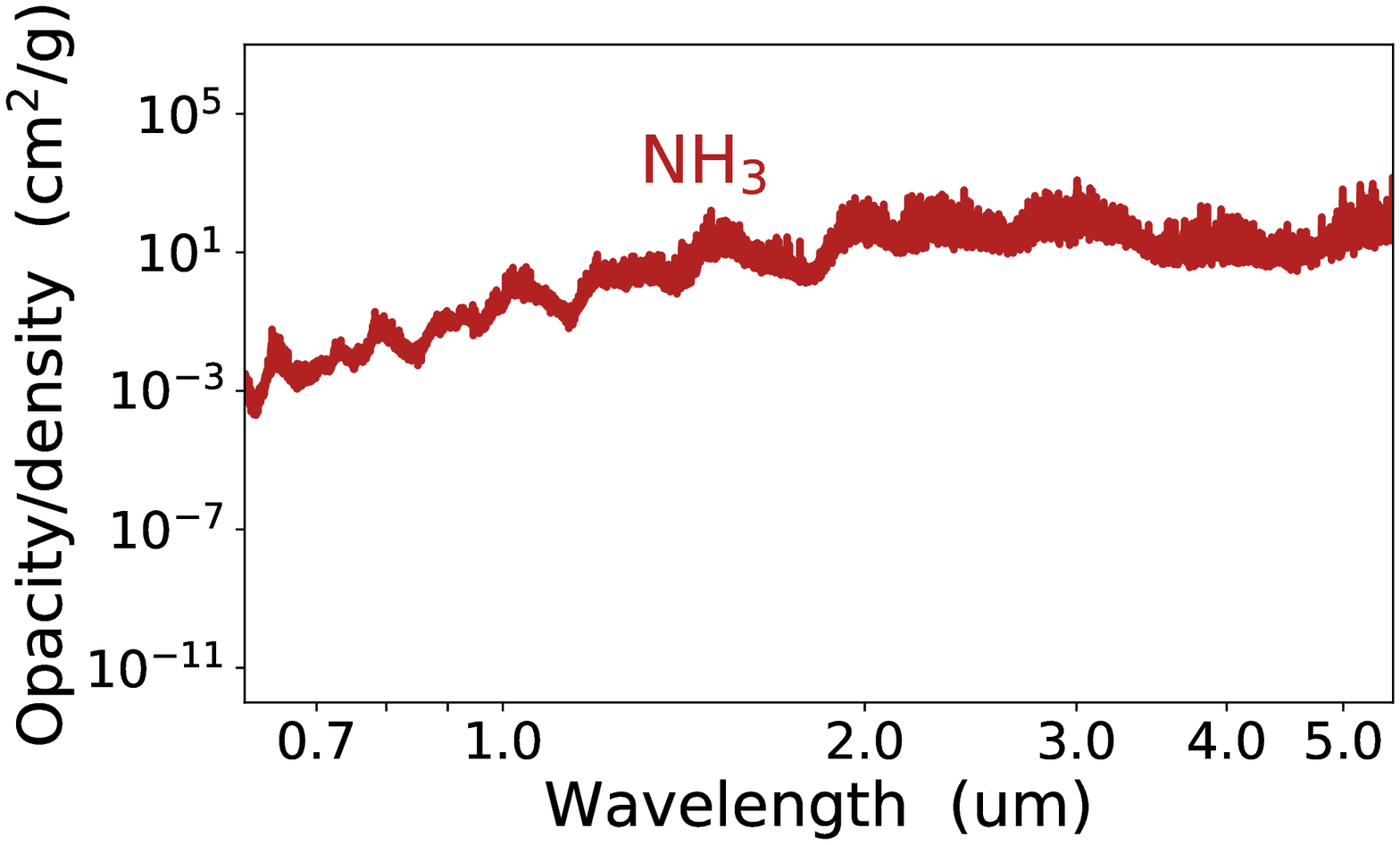}\hspace{-13pt}
\includegraphics[width=.26\textwidth, height=3.5cm]{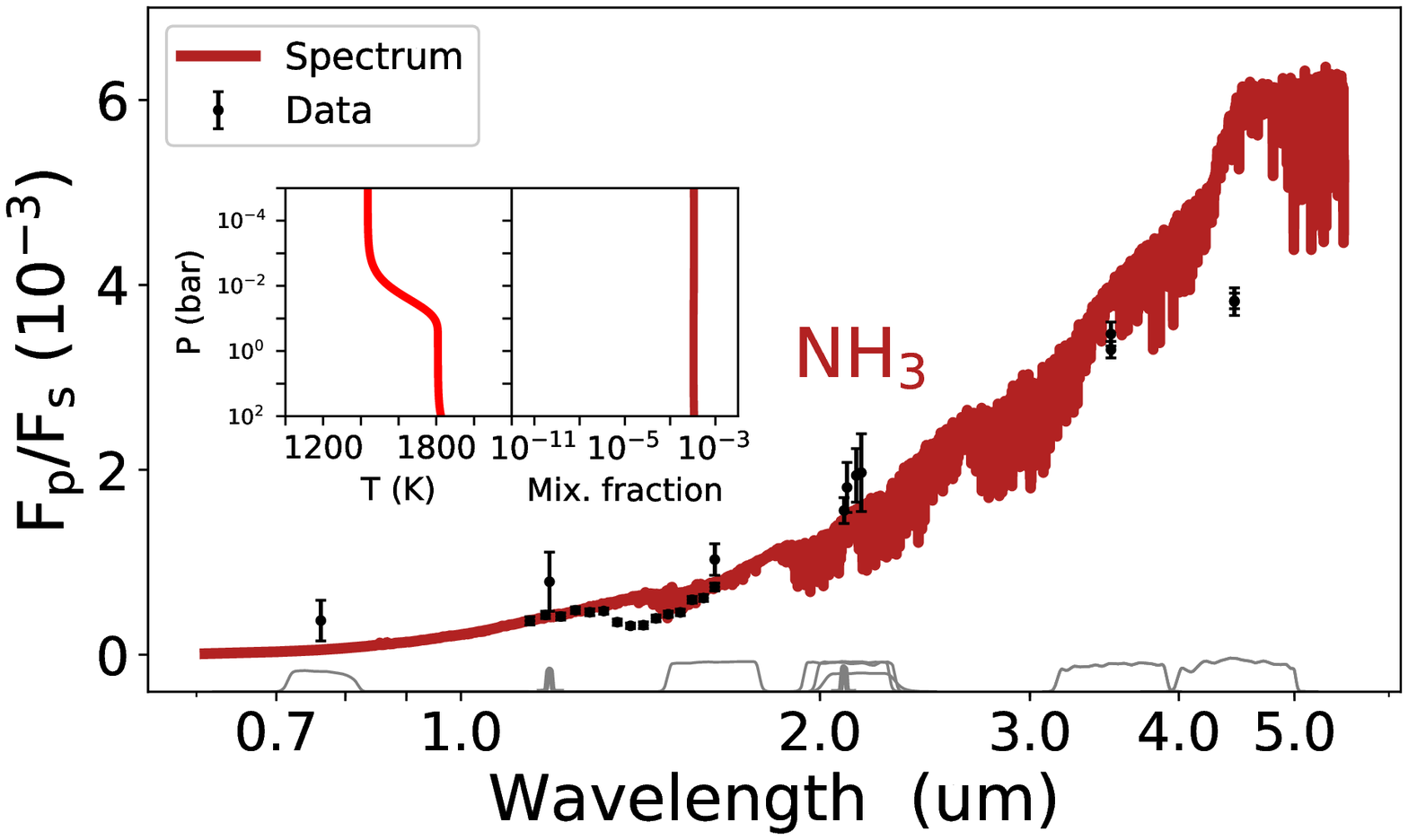}\hspace{-13pt}
\includegraphics[width=.26\textwidth, height=3.5cm]{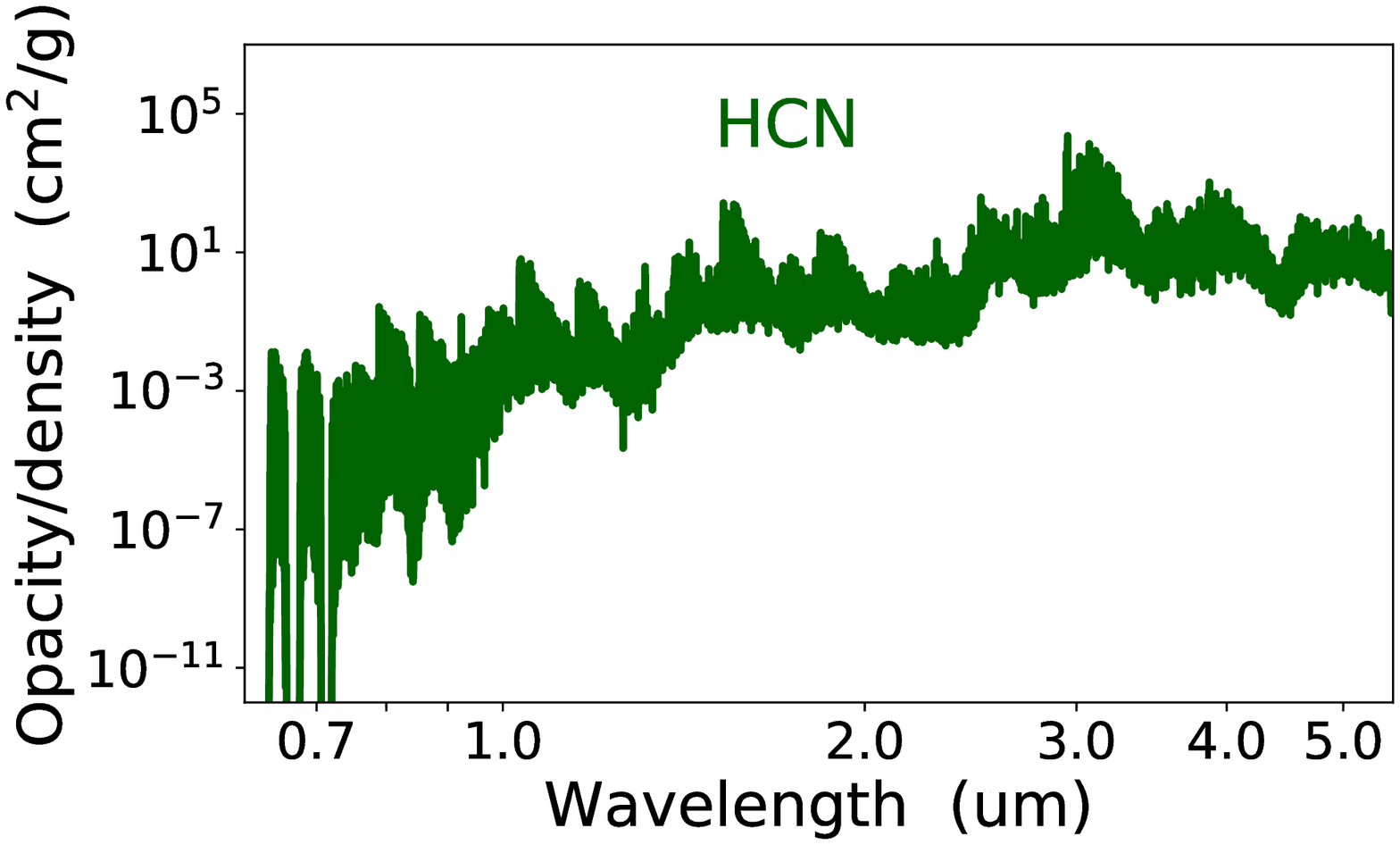}\hspace{-13pt}
\includegraphics[width=.26\textwidth, height=3.5cm]{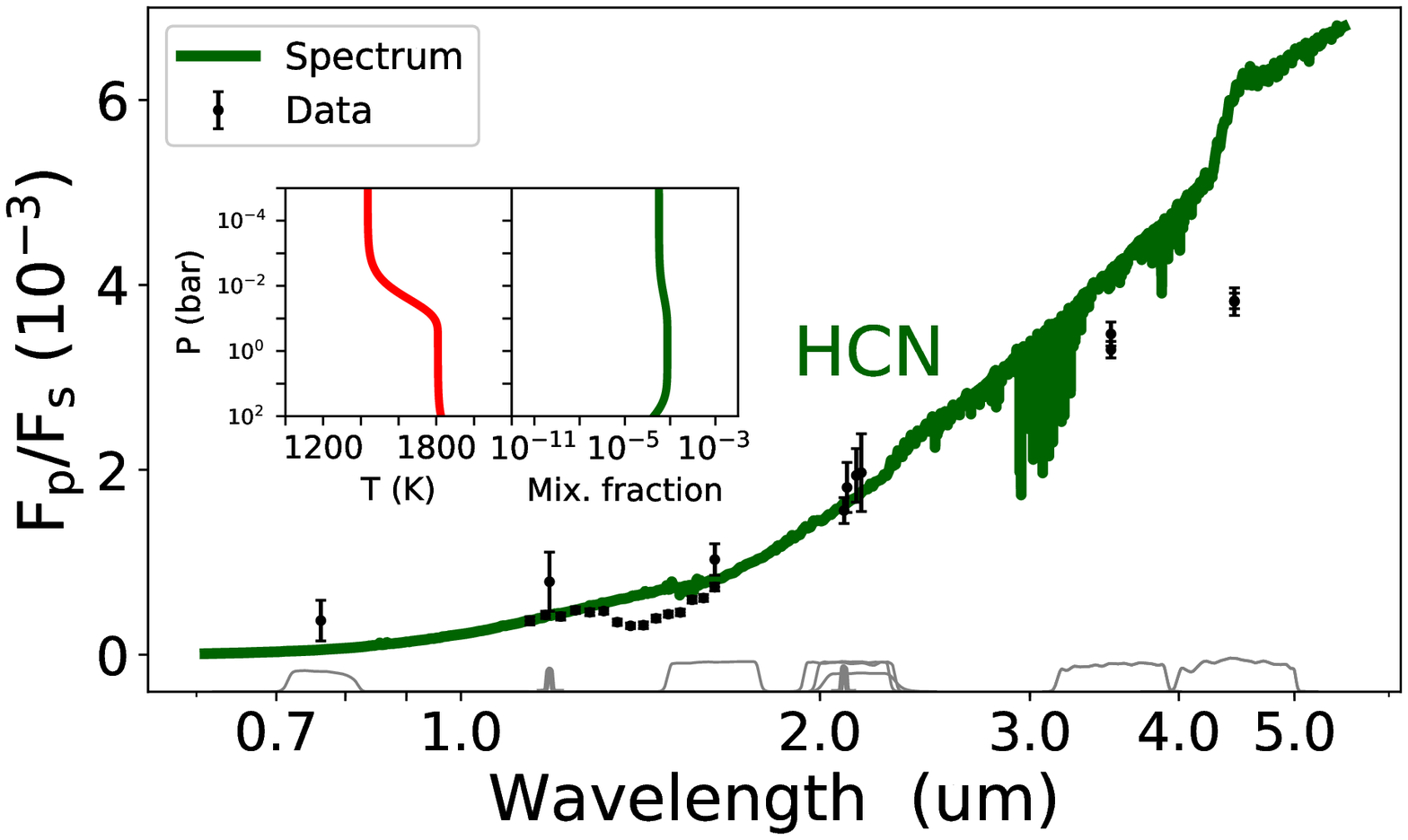}
\vspace{-2pt}

\includegraphics[width=.26\textwidth, height=3.5cm]{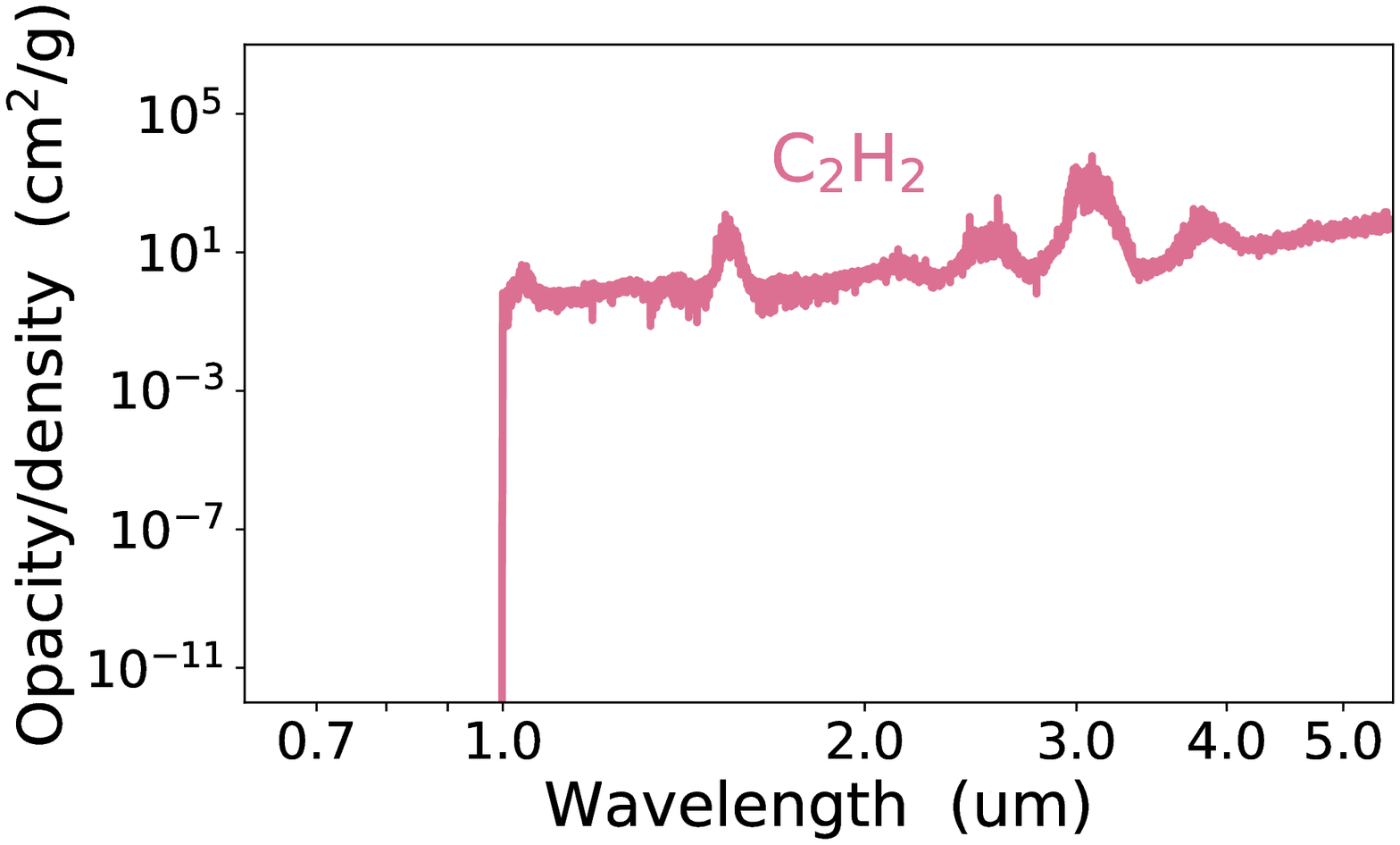}\hspace{-13pt}
\includegraphics[width=.26\textwidth, height=3.5cm]{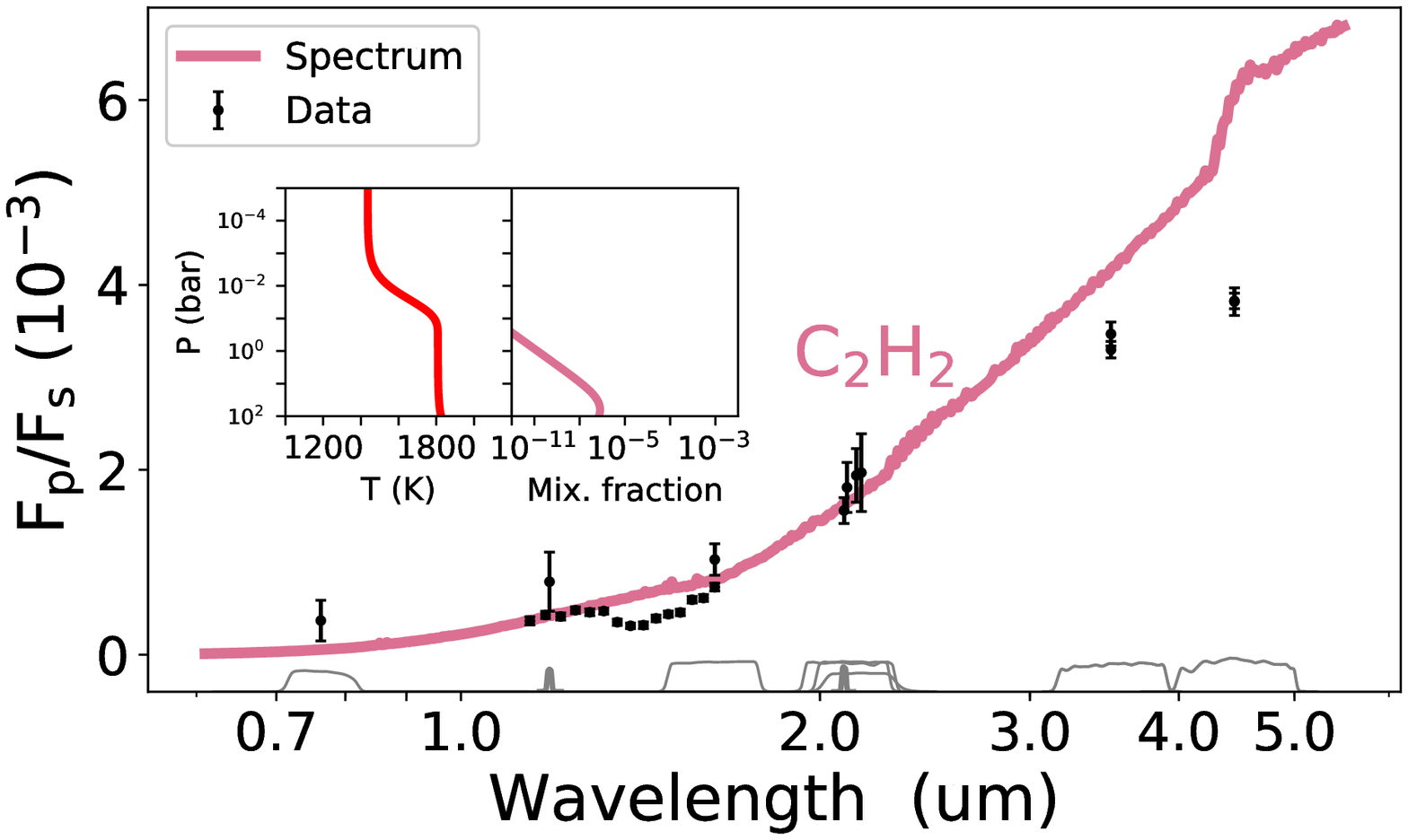}\hspace{-13pt}
\includegraphics[width=.26\textwidth, height=3.5cm]{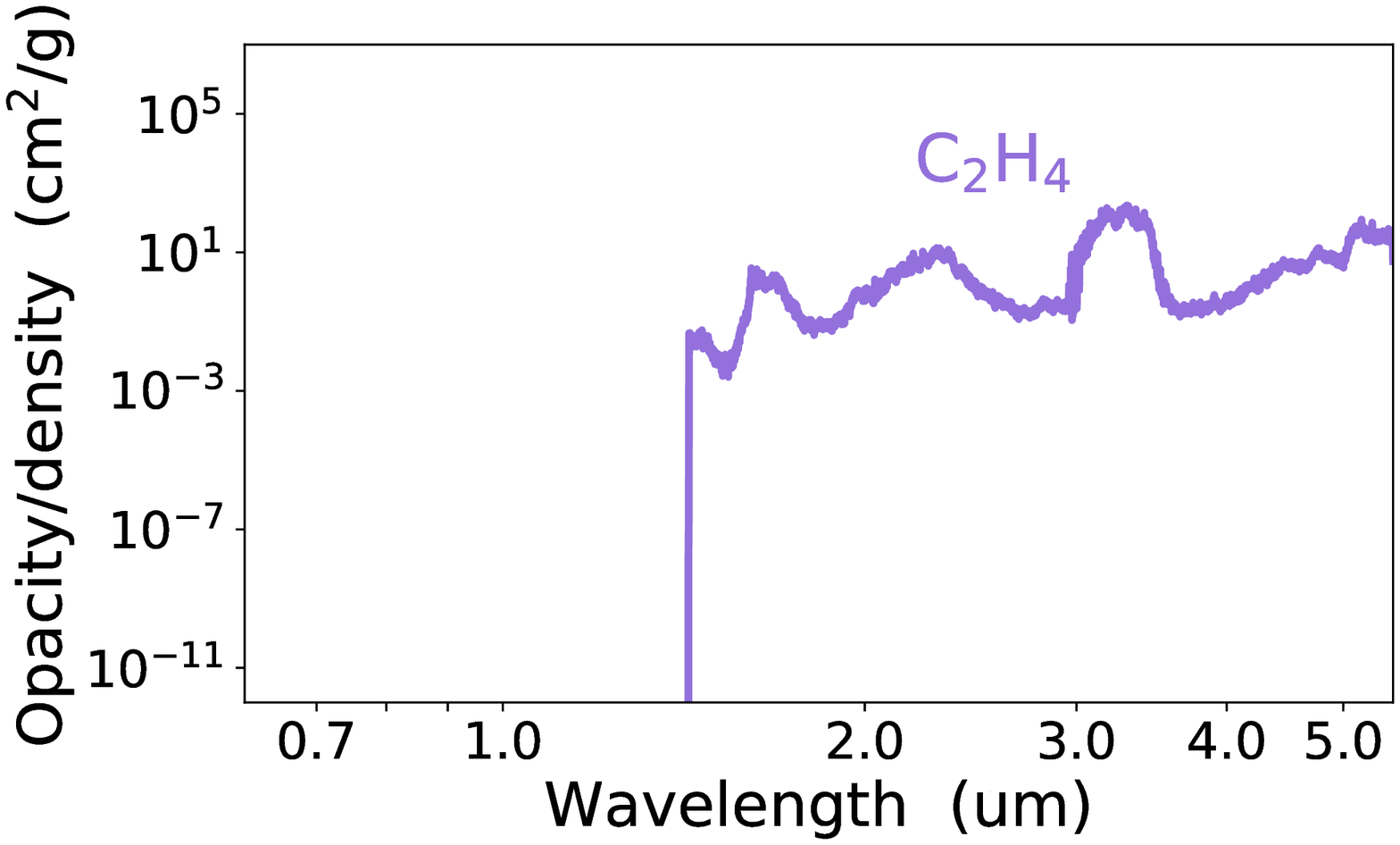}\hspace{-13pt}
\includegraphics[width=.26\textwidth, height=3.5cm]{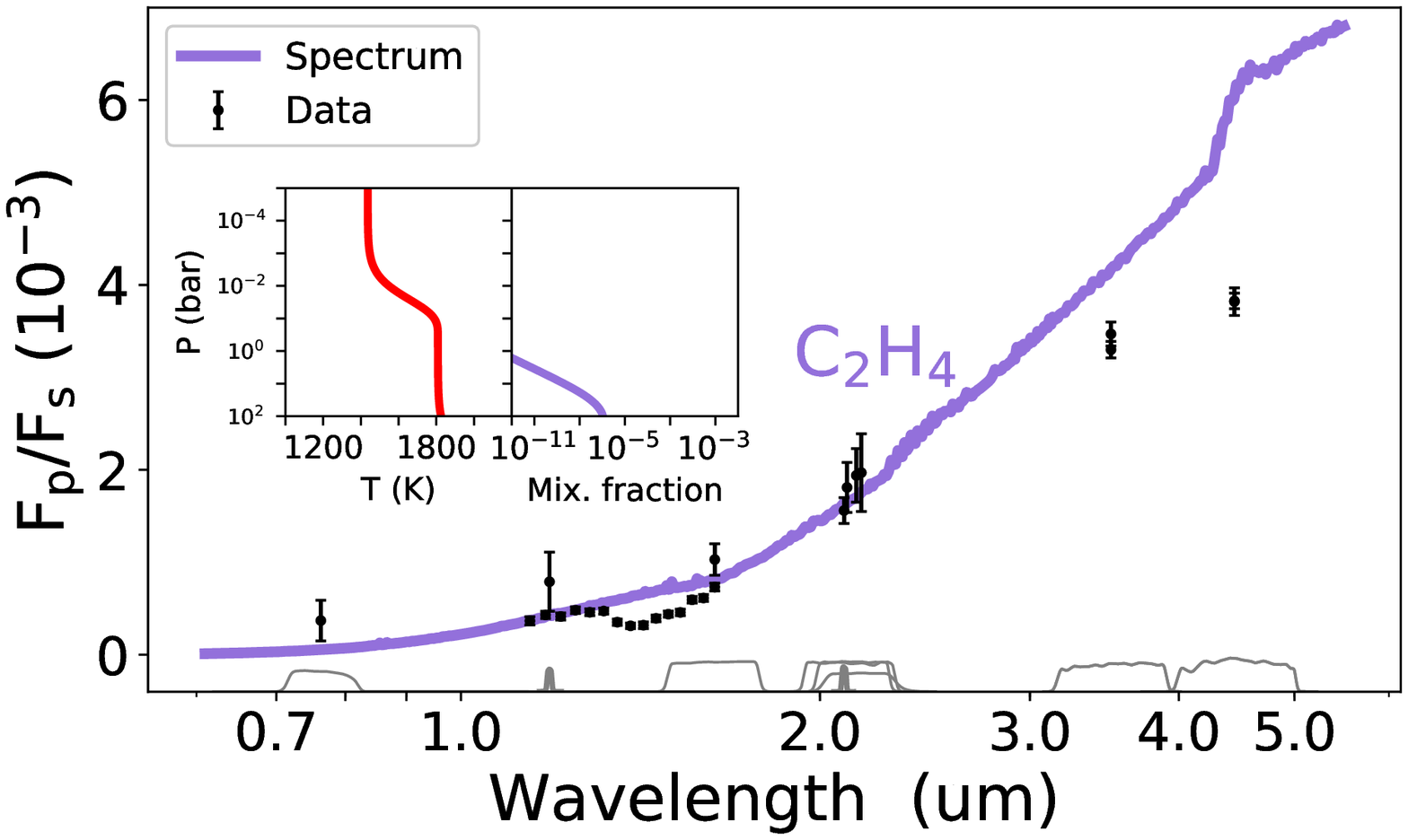}
\vspace{-2pt}

\includegraphics[width=.26\textwidth, height=3.5cm]{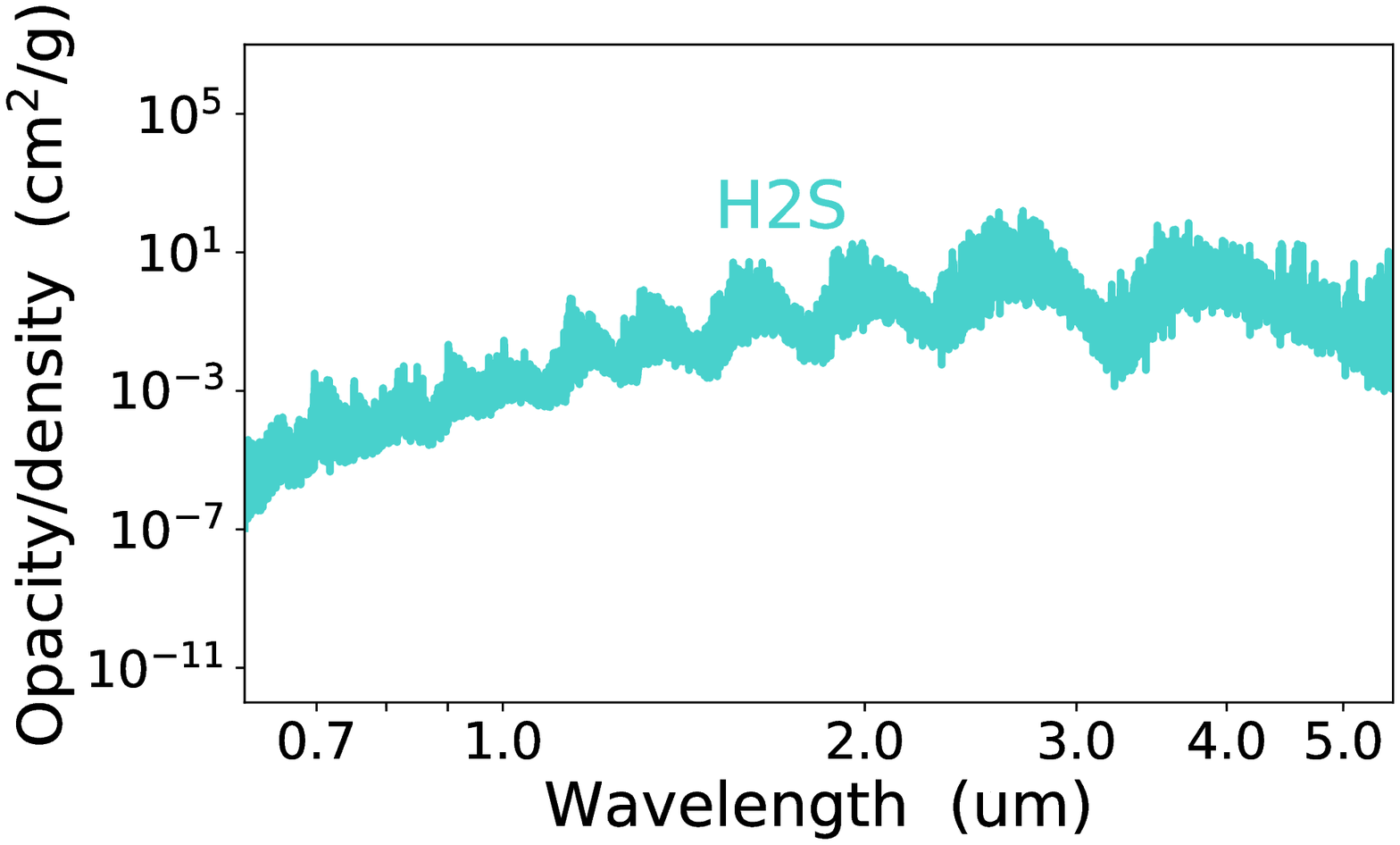}\hspace{-13pt}
\includegraphics[width=.26\textwidth, height=3.5cm]{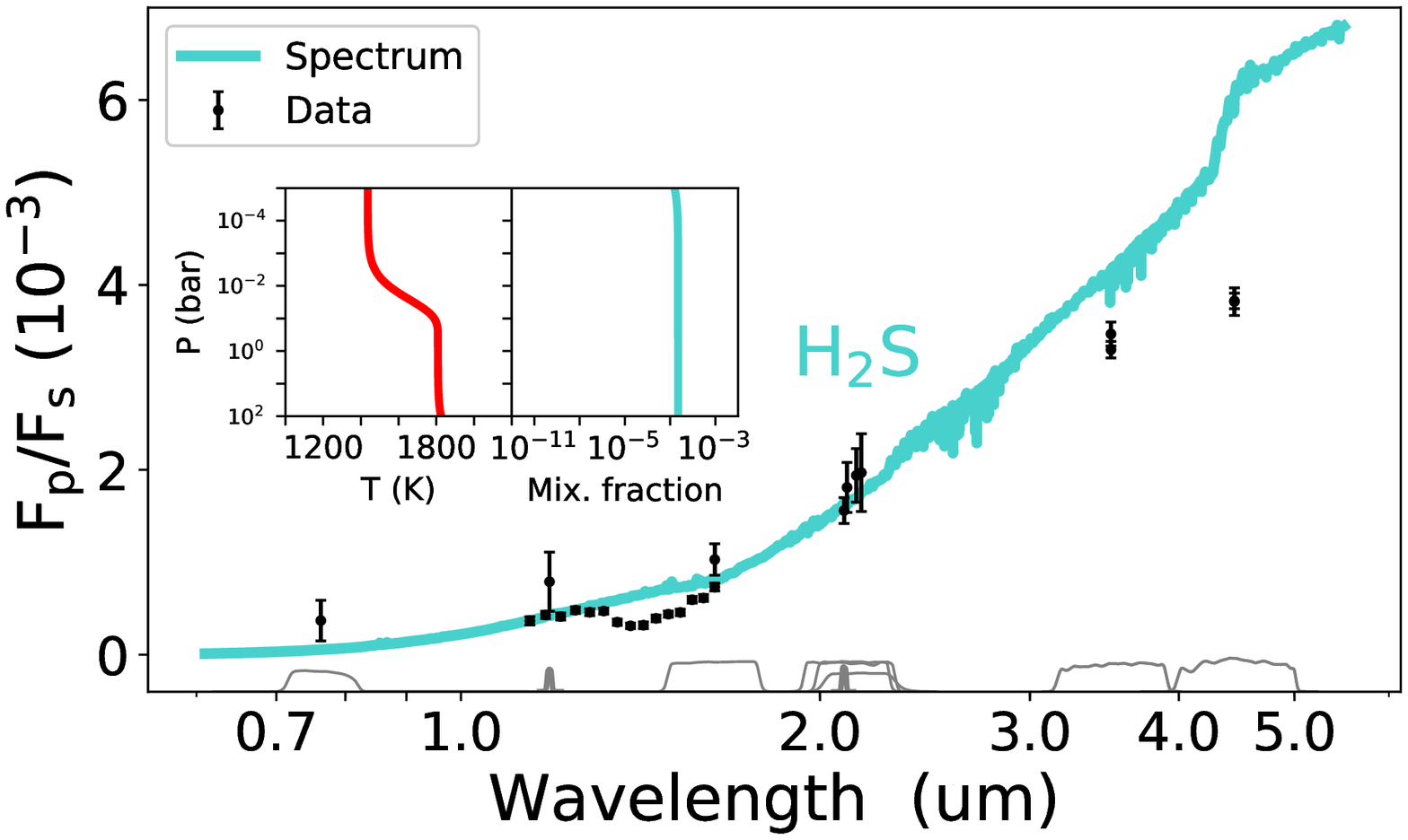}\hspace{-13pt}
\includegraphics[width=.26\textwidth, height=3.5cm]{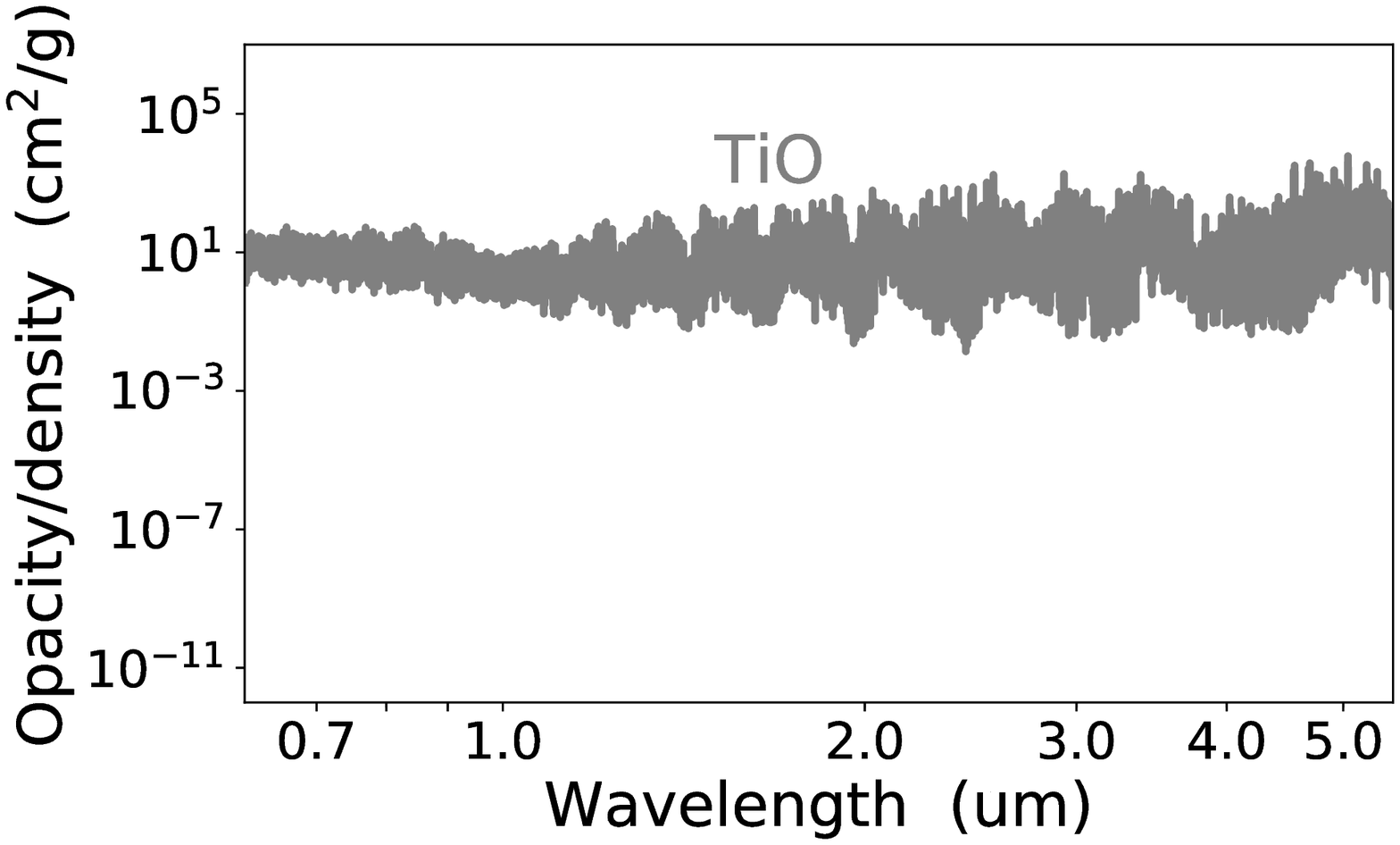}\hspace{-13pt}
\includegraphics[width=.26\textwidth, height=3.5cm]{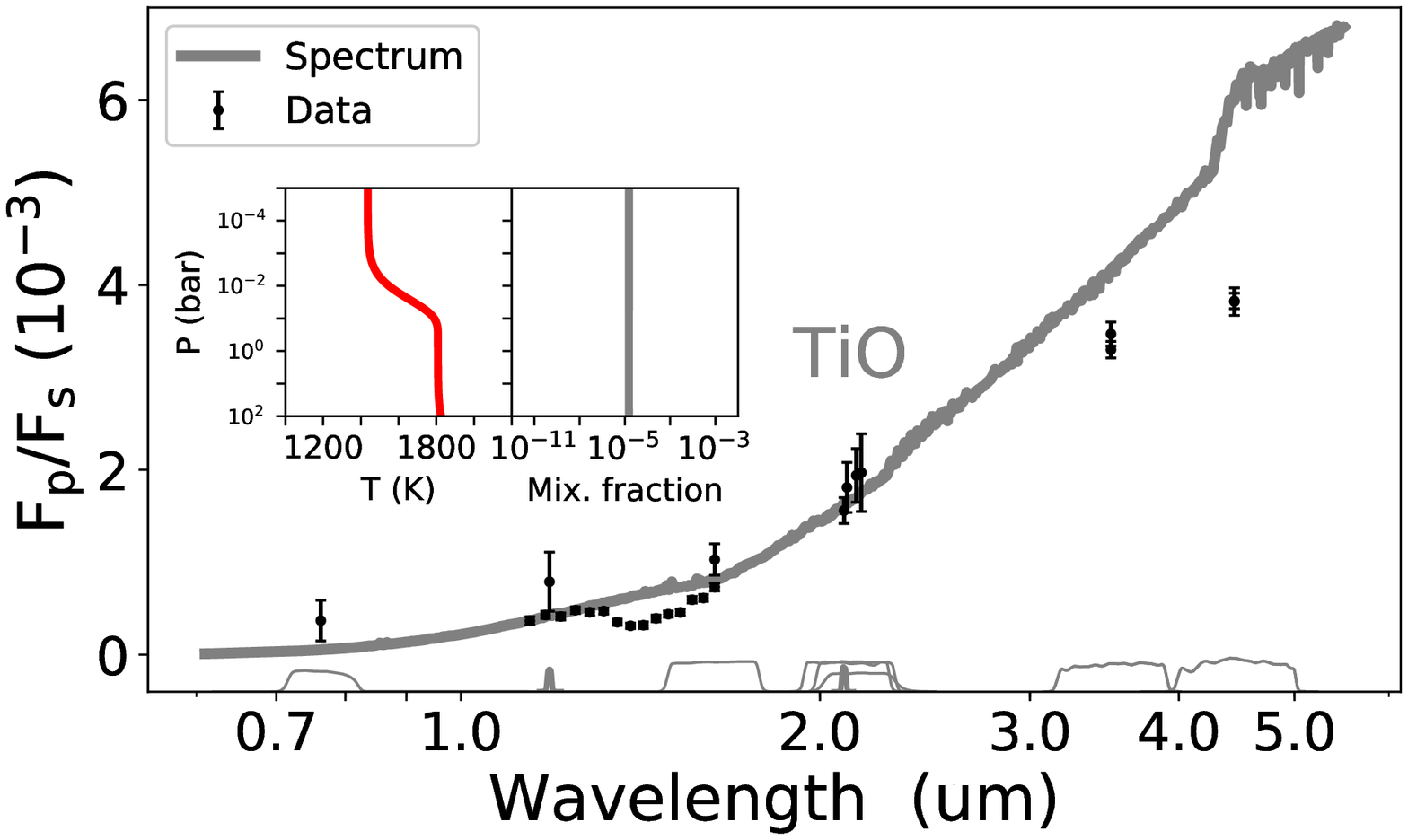}
\vspace{-2pt}

\includegraphics[width=.26\textwidth, height=3.5cm]{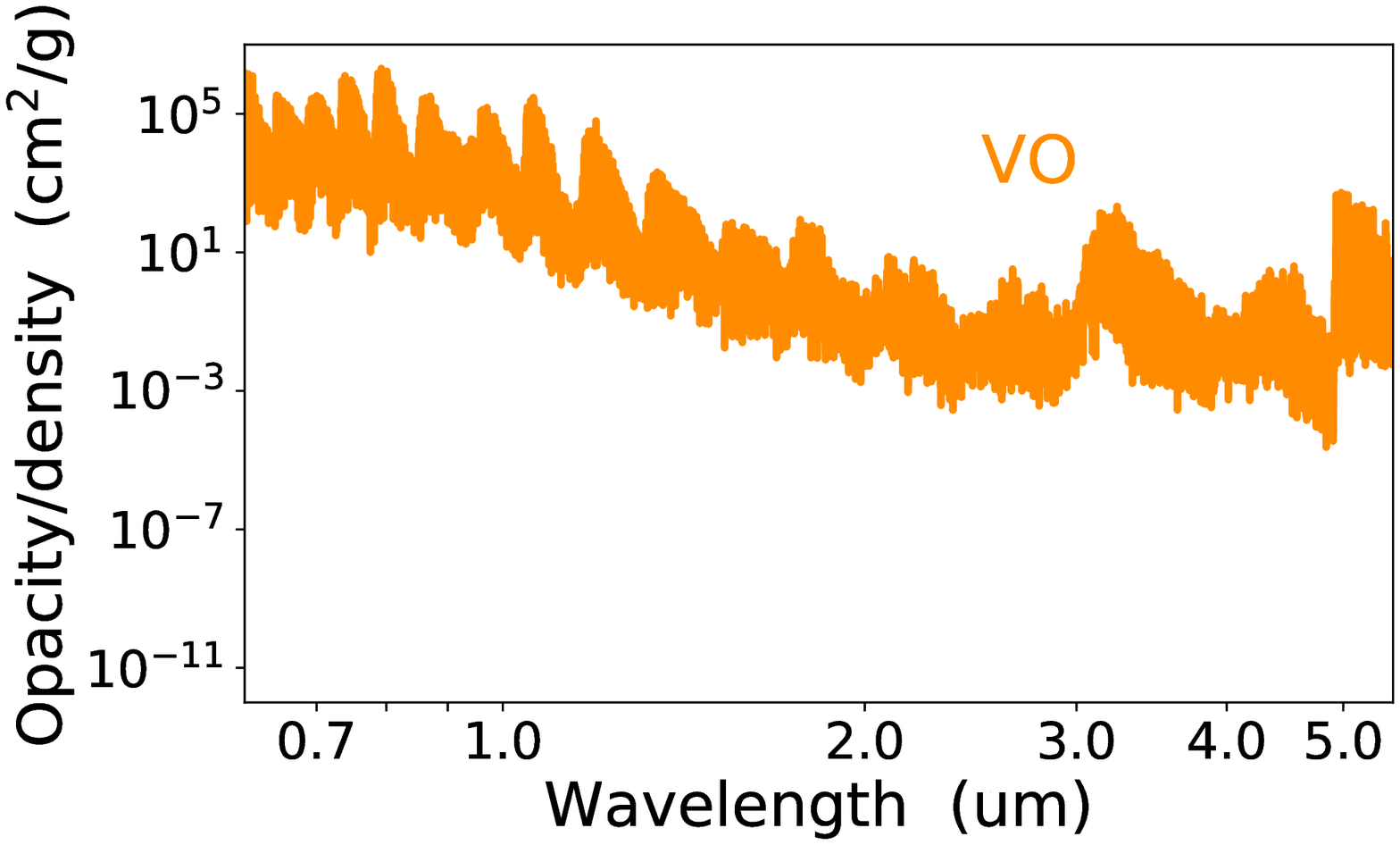}\hspace{-13pt}
\includegraphics[width=.26\textwidth, height=3.5cm]{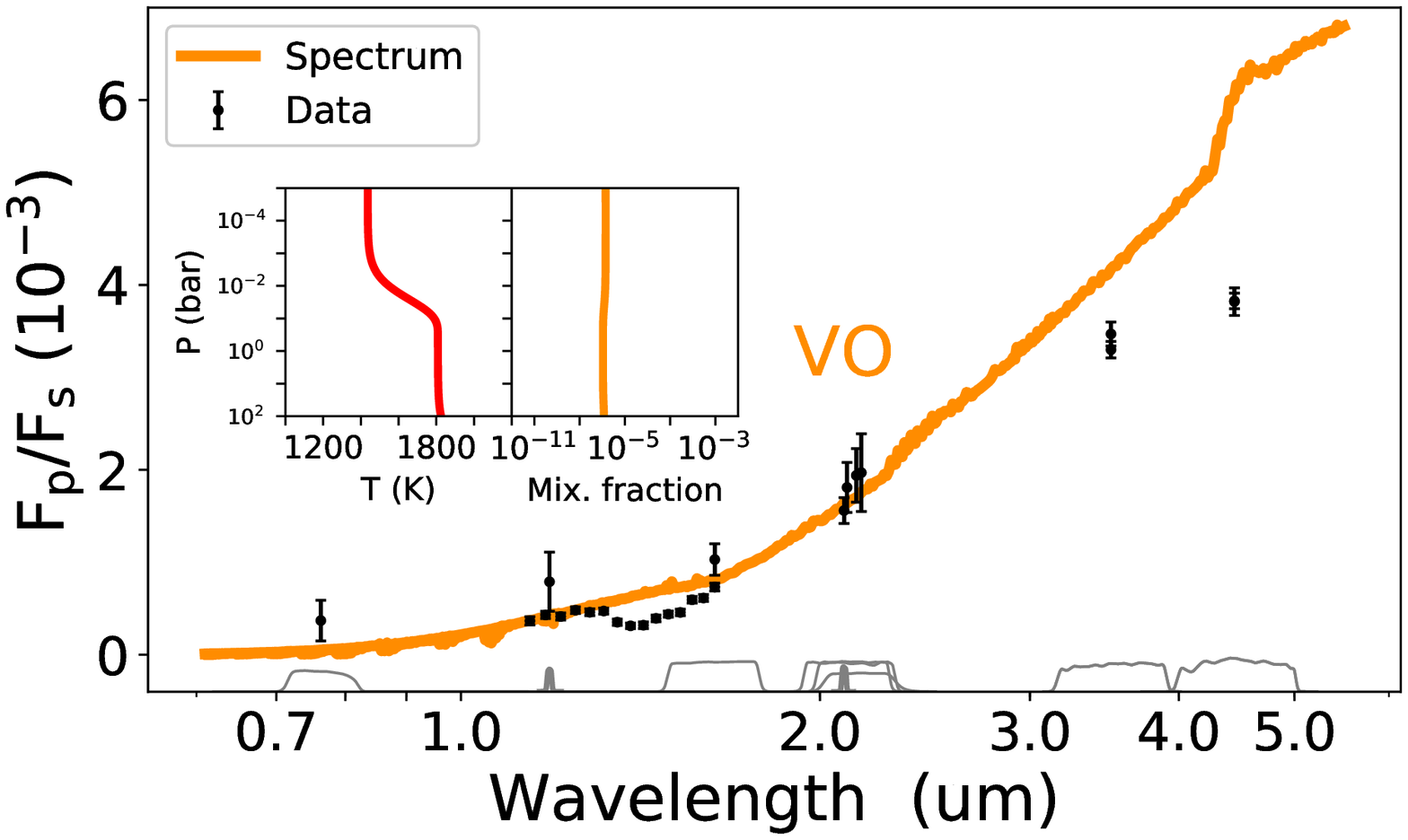}
\vspace{12pt}
\caption{Opacities of the species used in WASP-43b atmospheric retrieval, calculated at temperature of 1500 K and pressure of 1 bar, and their influence on the equilibrium spectra of WASP-43b. The spectrum model includes the corresponding molecular and CIA opacities (see Section \ref{sec:BARTsetup}). The gaps in the opacity figures come from the limited \math{y}-axis range (the values not shown are well below 10\sp{-12} cm\sp{2}/g). The \math{T(p)} profile and species abundances used to generate WASP-43b spectra are given in the inset figures. Black dots represent the data points with uncertainties. At the bottom of the figures, we show the bandpasses for each of the observations used in the analysis. From the individual equilibrium WASP-43b models, we see that H\sb{2}O, CO, NH\sb{3} and potentially HCN are the only absorbers that significantly influence the WASP-43b emission spectra in the wavelength range of the available observations, with H\sb{2}O being the most apparent.}
\vspace{15pt}
\label{fig:opacs}
\end{figure*}

\subsection{Retrieved models}
\label{sec:retMod}

It is well know in the retrieval community that the retrieval results can be biased depending on the atmospheric setup, assumed chemical species and their opacities included in the model, model selection, and physical and chemical processes assumed \citep[e.g., ][]{MadhusudhanEtal2011natWASP12batm, CrossfieldEatl2012ApJWASP12b-reavaluation, StevensonEtal2014-WASP12b, BarstowHeng2020, Barstow2020-modelSelection}. In an attempt to test this and the conclusions from, for example, \citet{hansen2014features} and \citet{Swain2013-WASP12b} on how the inclusion of additional opacity sources influences the best-fit model, we generated five atmospheric cases and compared them using statistical factors (Table \ref{tab:cases}). We present cases where we fit four major molecular species, H\sb{2}O, CO\sb{2}, CO, and CH\sb{4}, and seven molecular species, H\sb{2}O, CO\sb{2}, CO, CH\sb{4}, NH\sb{3}, HCN, and C\sb{2}H\sb{2}. We also include additional cases with non-fit opacities, assuming solar equilibrium composition. These exercises were performed to investigate whether the most often used basic approach for hot Jupiters, which includes only four major molecular species in the model and fits them in retrieval, is lacking some major spectral features from other relevant species; and to test how sensitive the data are on the inclusion of additional opacity sources, assuming equilibrium and non-equilibrium chemical composition.

\begin{table}[h!]
\centering
\caption{\label{tab:cases} Atmospheric Cases}
\begin{tabular}{l@{\hspace{25pt}}l@{\hspace{25pt}}l@{\hspace{25pt}}}
    \hline
    \hline
Case        & Fitted                    & Opacity\\
            & Species                   & Sources\\
    \hline
Case 1      & 4\tablenotemark{a}   & 4\tablenotemark{a}\\
Case 2      & 4\tablenotemark{a}   & 7\tablenotemark{b}\\
Case 3      & 4\tablenotemark{a}   & 11\tablenotemark{c}\\
Case 4      & 7\tablenotemark{b}   & 7\tablenotemark{b}\\
Case 5      & 7\tablenotemark{b}   & 11\tablenotemark{c}\\
    \hline
\end{tabular}
\tiny{
\begin{minipage}[t]{0.65\linewidth}
\tablenotetext{1}{H\sb{2}O, CO\sb{2}, CO, CH\sb{4}.}
\tablenotetext{2}{H\sb{2}O, CO\sb{2}, CO, CH\sb{4}, NH\sb{3}, HCN, C\sb{2}H\sb{2}.}
\tablenotetext{3}{H\sb{2}O, CO\sb{2}, CO, CH\sb{4}, NH\sb{3}, HCN, C\sb{2}H\sb{2}, C\sb{2}H\sb{4}, H\sb{2}S, TiO, VO.}
\end{minipage}
}
\end{table}

To assess different models quantitatively, we used two statistical factors: the reduced \math{\chi\sp{2}}, \math{\chi\sp{2}\sb{\rm red}} = \math{\frac{\chi\sp{2}}{N -k}}, where \math{N} is the number of data points, and \math{k} is the number of free parameters, and the Bayesian Information Criterion, BIC, \math{\chi\sp{2} + k\,ln(N)}. We consider BIC to be one of the most important factors. It allows us to compare goodness-of-fit for the models generated on the same dataset. Although in general, models with more free parameters improve the fit, BIC adds a penalty for any additional parameters in the system by  increasing its value. A lower BIC value indicates a better fit. We also considered the Bayesian evidence, another statistical criteria often used in literature \citep{Gregory2007-modelComparison, Trotta2008-BayesianEvidence, BennekeSeager2012-Retrieval}. The Bayesian evidence represents the ratio of marginal probabilities of two models and is calculated, in the first approximation, using BIC \citep[see Equations 20-22,][]{Raftery1995}. Since both statistical criteria lead to the same conclusions, we decided to use BIC factor in our subsequent analysis.

\subsection{Results - Four Fitted Species}
\label{sec:four}

\begin{figure*}[ht!]
\vspace{-5pt}
\centering
\subfigure{\includegraphics[height=5.9cm, clip=True]{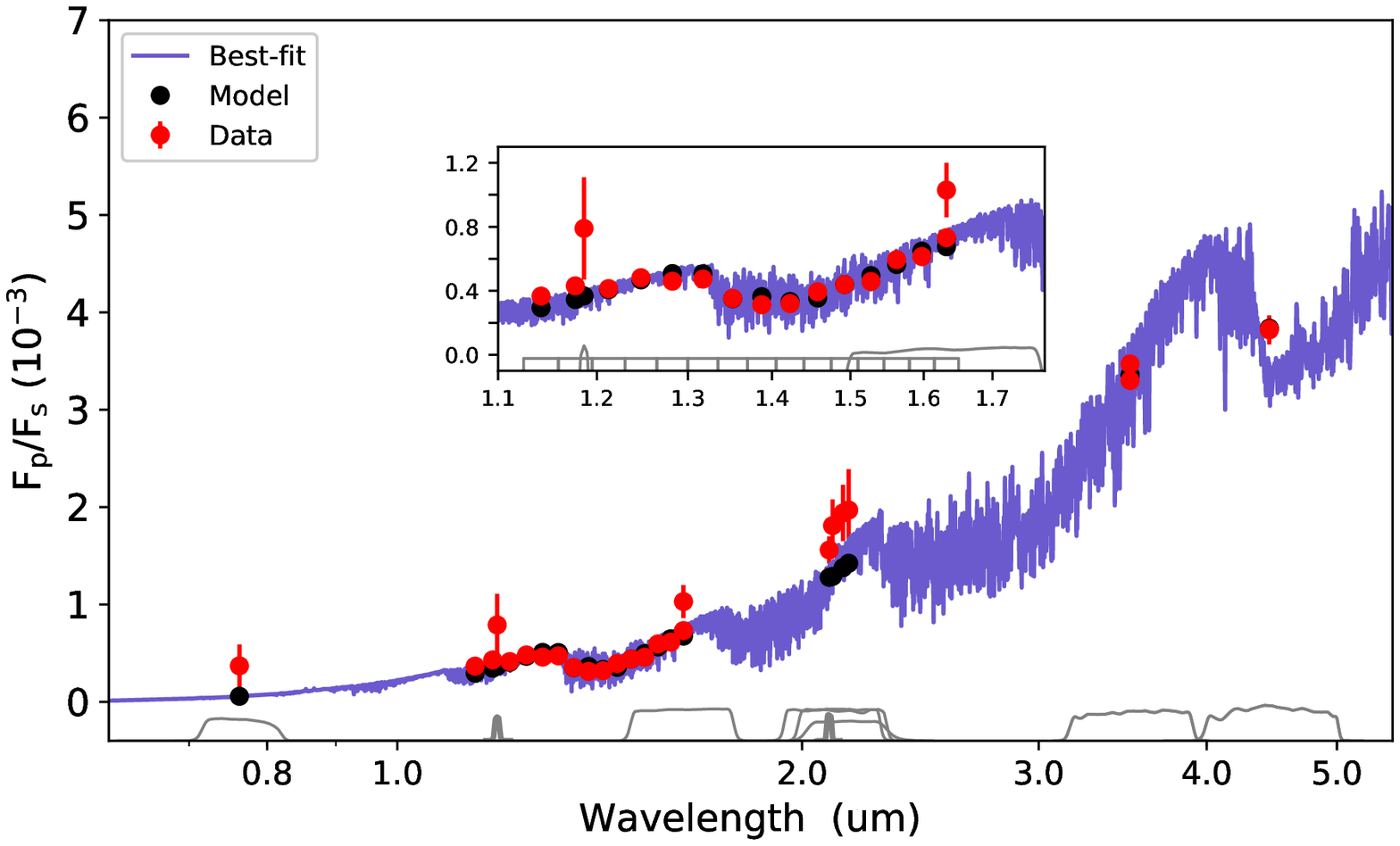}\hspace{-10pt}}
\subfigure{\raisebox{0.23cm}{\includegraphics[height=5.7cm, width=4.0cm]{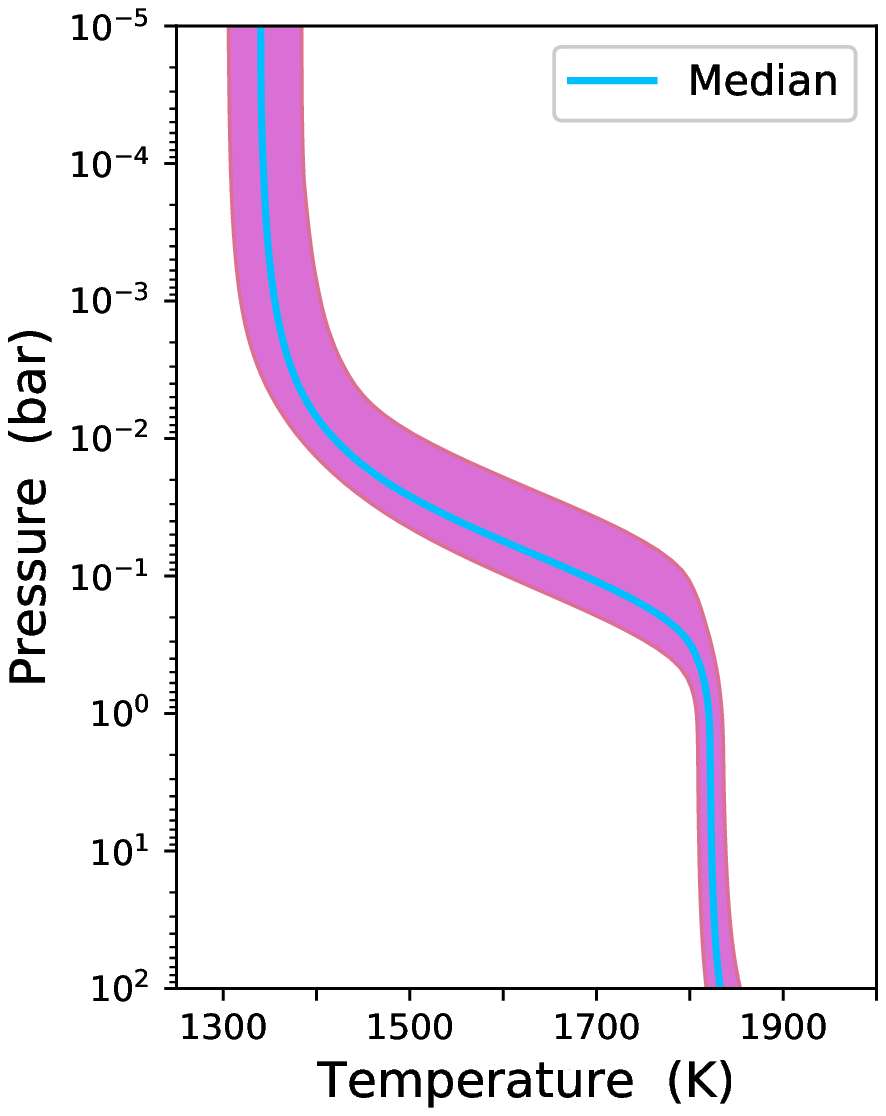}}}
\caption{Best-fit model and \math{T(p)} profile for Case 1, four fitted species. Left panel shows the best-fit spectrum. In red are the data points (eclipse depths) with uncertanties, in black are the integrated points of our model over the bandpasses of our observations shown at the bottom. The models are generated with the four major molecular species and their opacities. Right panel shows the median temperature and pressure profile with the 1\math{\sigma} confidence region. }
\label{fig:4mol}
\vspace{-5pt}
\end{figure*}

\begin{figure*}[ht!]
\vspace{-5pt}
\centering
\includegraphics[width=.22\textwidth, height=3.2cm]{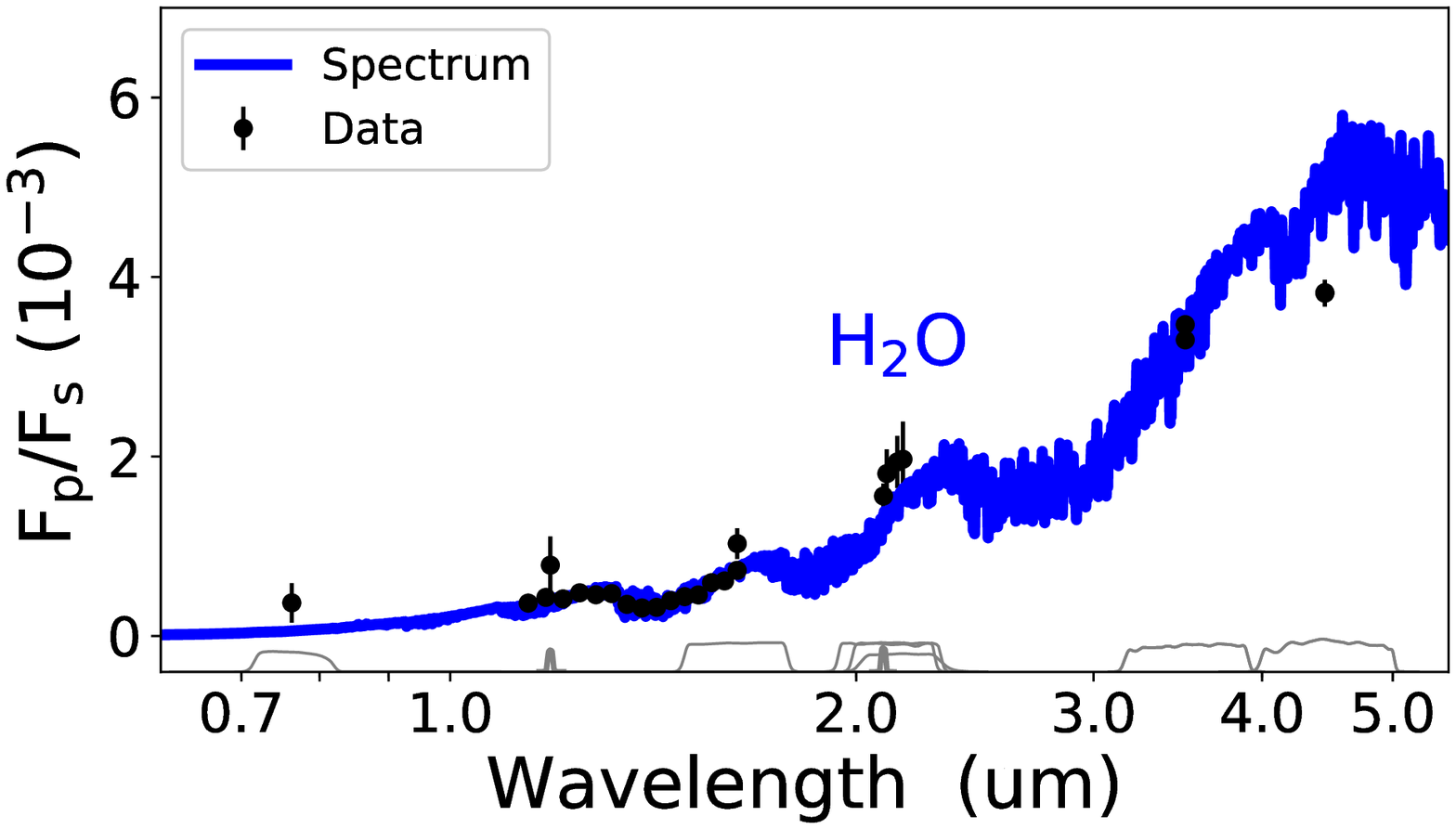}
\includegraphics[width=.22\textwidth, height=3.2cm]{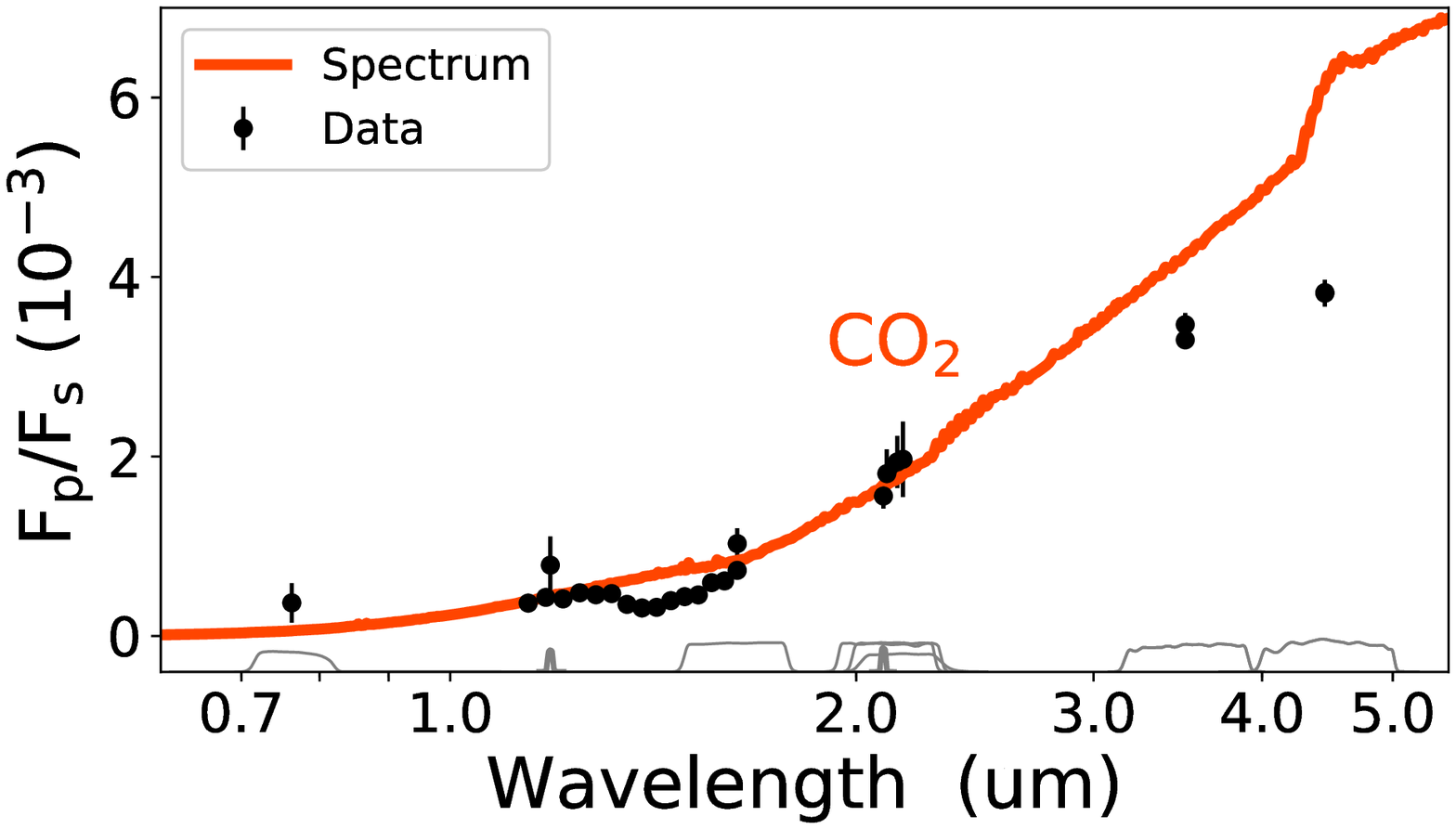}
\includegraphics[width=.22\textwidth, height=3.2cm]{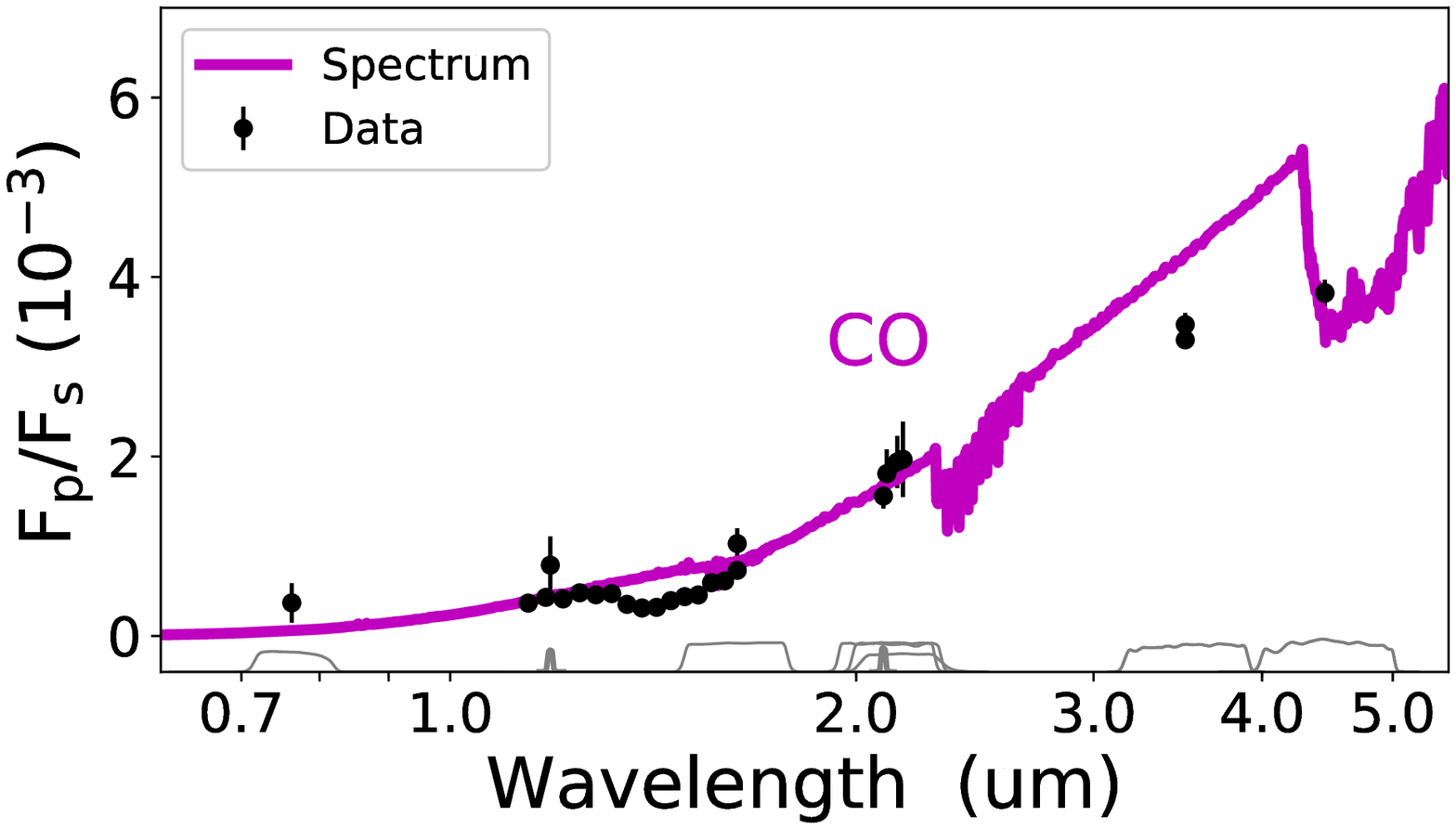}
\includegraphics[width=.22\textwidth, height=3.2cm]{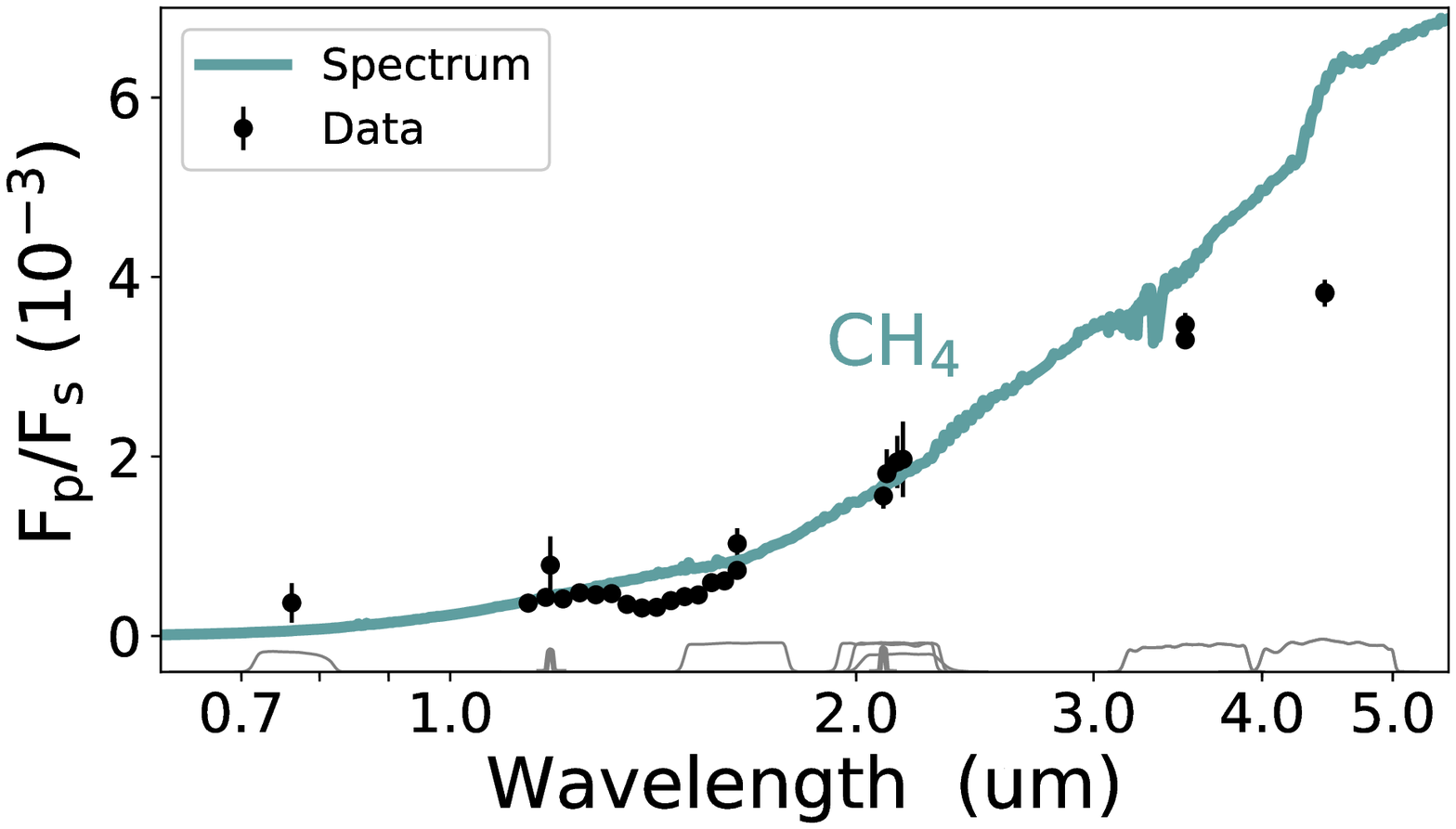}
\caption{Influence of each species on the best-fit model for Case 1, four fitted species.}
\label{fig:4mol-indSpecs}
\vspace{-5pt}
\end{figure*}

\begin{figure*}[ht!]
\vspace{-5pt}
\centering
\includegraphics[width=.57\textwidth, clip=True,trim=20 0 0 0]{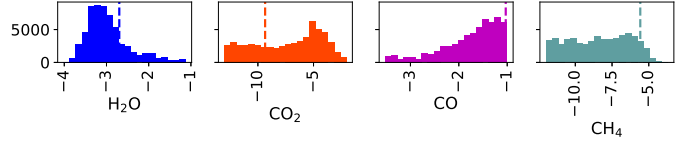}
\caption{Histograms for Case 1, four fitted species. Figure shows the species' abundances (\math{X}), expressed as \math{\log\sb{10}({X})} with their best-fit values shown as dashed vertical lines.}
\label{fig:4mol-hist}
\end{figure*}

\begin{figure}[h!]
\centering
\includegraphics[width=.5\textwidth, clip=True,trim=0 0 0 0]{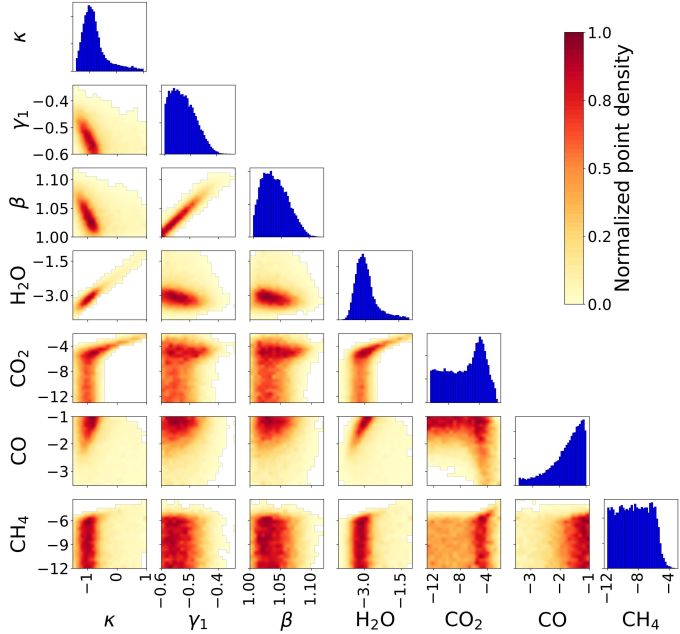}
\caption{Pairwise correlation plots for Case 1, four fitted species.}
\label{fig:4mol-pair}
\end{figure}

In this section, we describe cases where we fit four major molecular species,  H\sb{2}O, CO\sb{2}, CO, and CH\sb{4}. We first compare our results with \citet{LineEtal2014-Retrieval-I} and \citet{KreidbergEtal2014-WASP43b}, and then we investigate how the inclusion of additional species and their opacities affects the best-fit model. 

The initial \math{T(p)} parameters were set to the values that reproduce the best-fit temperature profile from \citet{LineEtal2014-Retrieval-I}. The initial species abundances were set to the equilibrium solar composition values at 0.1 bar level, assuming constant-with-altitude abundances, calculated using \textsc{TEA}. 

We generated three different cases (see Table \ref{tab:cases}):

\begin{enumerate}
\item We reproduced the setup from \citet{LineEtal2014-Retrieval-I} and retrieved the \math{T(p)} profile with four major molecular species. We, thus, included only the four major molecular species and their opacities in the mean molecular mass and opacity calculations. 

\item We tested the statistical significance when additional molecules and their opacity sources were included in the calculation, assuming thermodynamical equilibrium and solar composition. In this case, in addition to H\sb{2}O, CO\sb{2}, CO, and CH\sb{4}, we included nitrogen species NH\sb{3}, HCN, and the most abundant species occurring in the conditions when the atmospheric C/O ratio is larger than one, C\sb{2}H\sb{2}, testing if their spectral features are present in the model.

\item Same as Case 2 with the inclusion of the species that are considered to be responsible for thermal inversions in hot-Jupiter atmospheres, TiO, VO, H\sb{2}S, together with C\sb{2}H\sb{4}, the second most abundant species when C/O ratio is larger than one.
\end{enumerate}

\begin{table}[hb!]
\vspace{-20pt}
\footnotesize{
\caption{\label{table:4mol-fit} Goodness of Fit, Four Species}
\atabon\strut\hfill\begin{tabular}{lcccc}
    \hline
    \hline
                          & \math{\chi\sp{2}\sb{\rm red}}  & BIC\\
    \hline
Case 1, 4 opacities       & 2.0876                           & 62.4703\\
Case 2, 7 opacities       & 2.2203                           & 64.9932\\
Case 3, 11 opacities      & 2.2333                           & 65.2399\\
    \hline
\end{tabular}\hfill\strut\ataboff
}
\end{table}

\begin{figure*}[ht!]
\vspace{-10pt}
\centering
\subfigure{\includegraphics[height=5.9cm, clip=True]{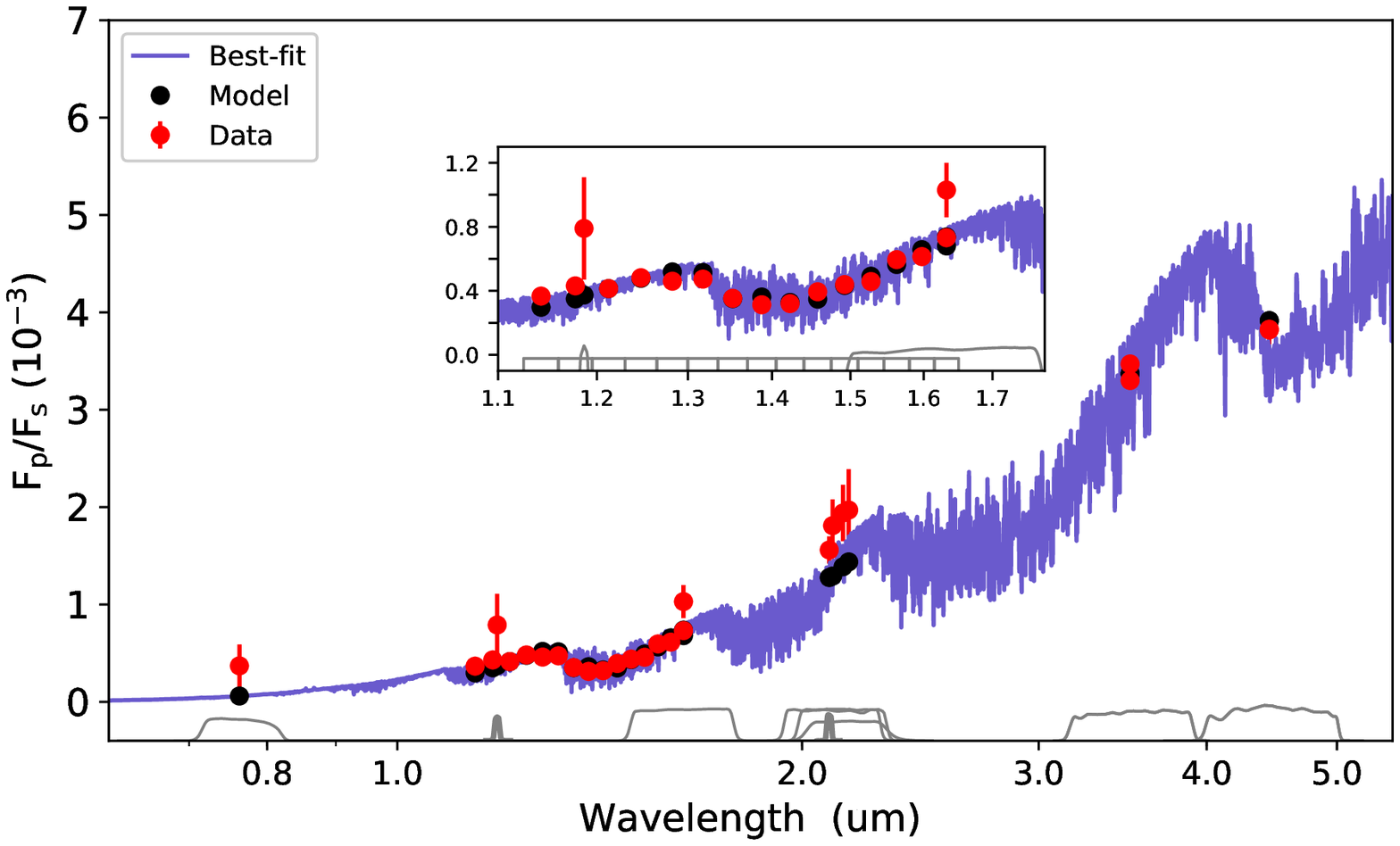}\hspace{-10pt}}
\subfigure{\raisebox{0.23cm}{\includegraphics[height=5.7cm, width=4.0cm]{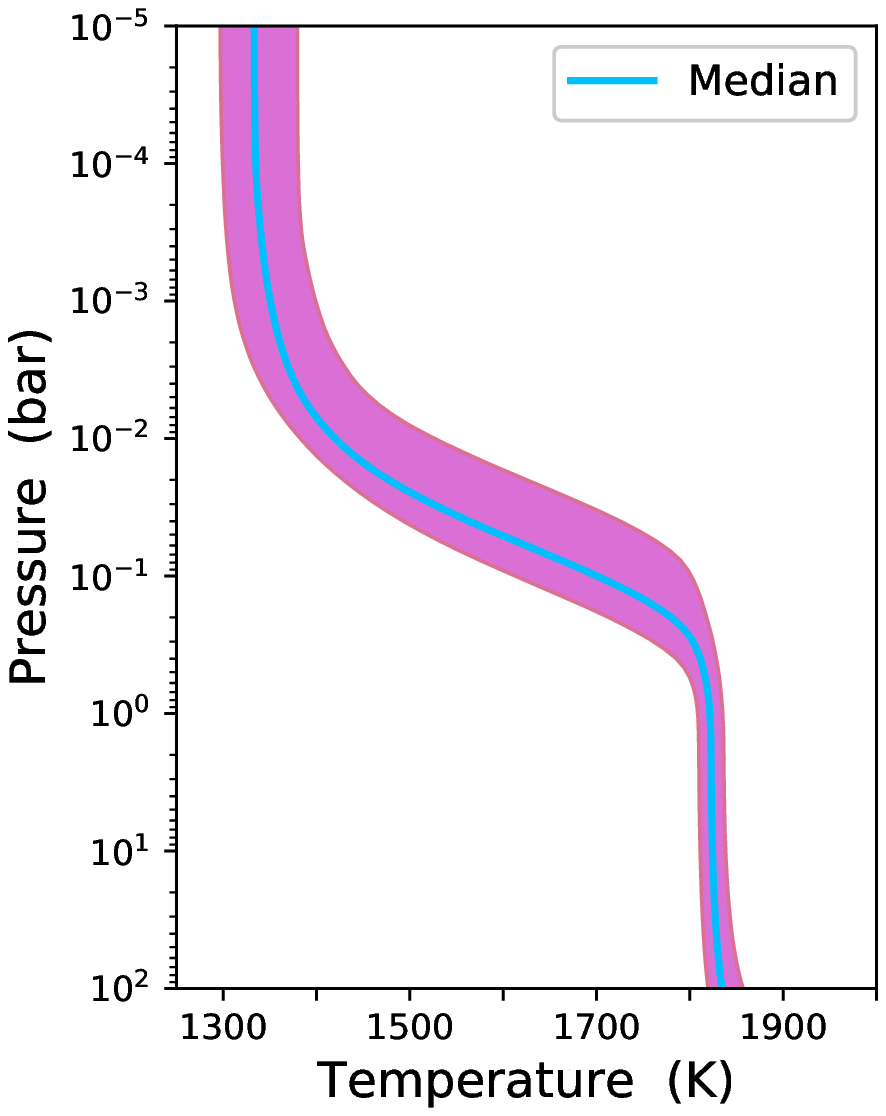}}}
\vspace{-5pt}
\caption{Best-fit model and \math{T(p)} profile for Case 4, seven fitted species. Left panel shows the best-fit spectrum. In red are the data points (eclipse depths) with uncertanties, in black are the integrated points of our model over the bandpasses shown in gray. The models are generated with the seven molecular species and their opacities. Right panel shows the median temperature and pressure profile with 1\math{\sigma} and 2\math{\sigma} confidence regions. }
\label{fig:7mol}
\end{figure*}

\begin{figure*}[ht!]
\vspace{-10pt}
\centering
\includegraphics[width=.22\textwidth, height=3.2cm]{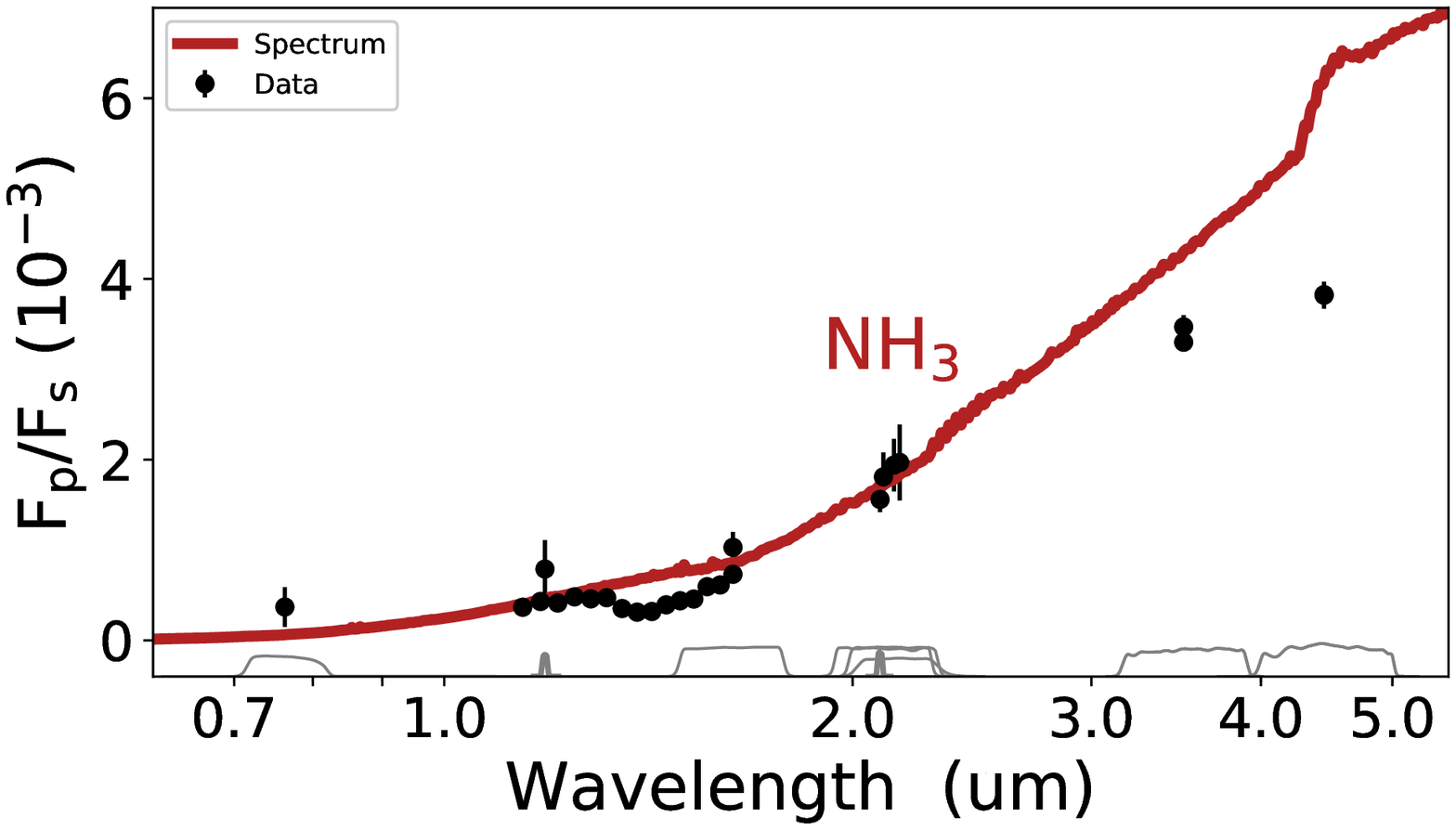}
\includegraphics[width=.22\textwidth, height=3.2cm]{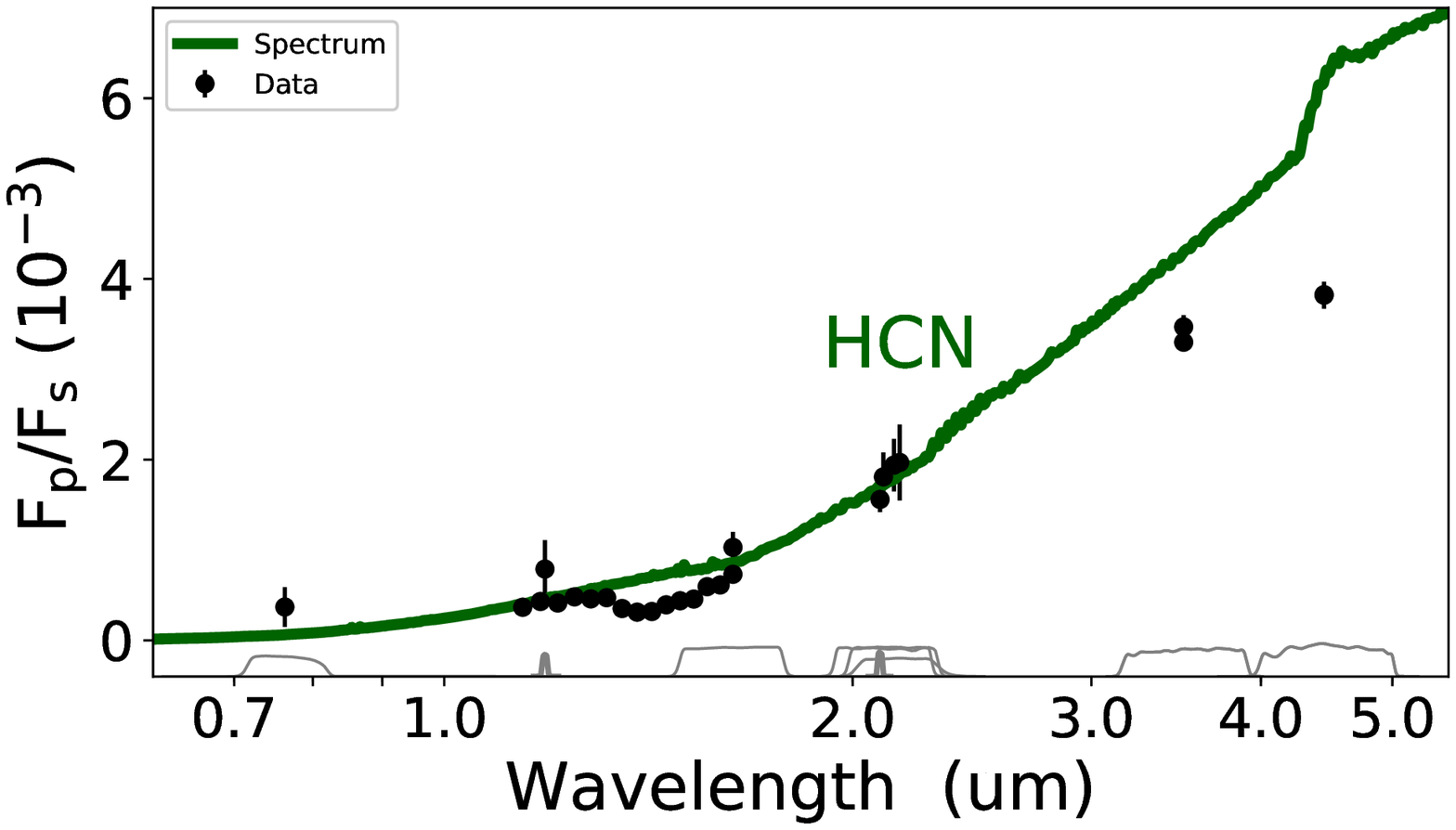}
\includegraphics[width=.22\textwidth, height=3.2cm]{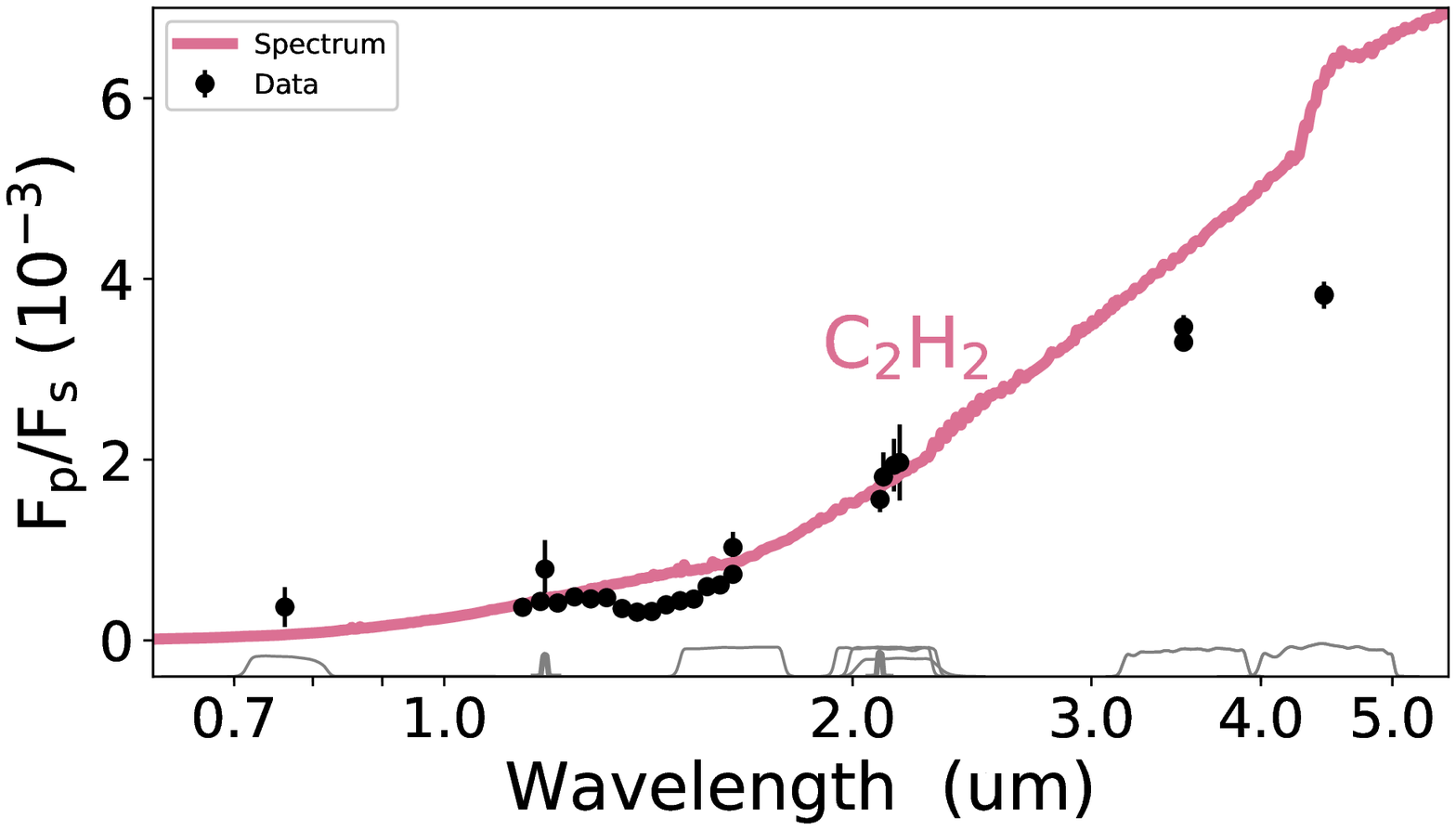}
\caption{Influence of the NH\sb{3}, HCN, and C\sb{2}H\sb{2} species on the best-fit model for Case 4, seven fitted species. H\sb{2}O, CO\sb{2}, CO, CH\sb{4} have similar influence on the spectra as seen in Figure \ref{fig:4mol-indSpecs}.}
\label{fig:7mol-indSpecs}
\vspace{-8pt}
\end{figure*}

\begin{figure*}[ht!]
\centering
\includegraphics[width=.7\textwidth, clip=True,trim=20 0 0 0]{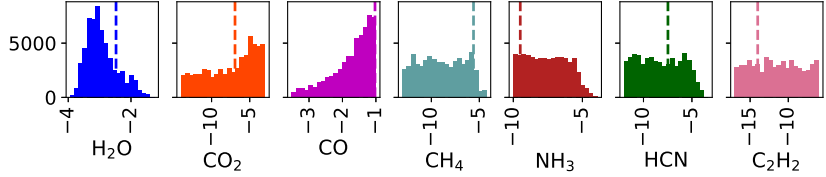}
\caption{Histograms for Case 4, seven fitted species. Figure shows the species' abundances (\math{X}), expressed as \math{\log\sb{10}(X)} with their best-fit values shown as dashed lines.}
\label{fig:7mol-hist}
\end{figure*}

\begin{figure}[ht!]
\vspace{-3pt}
\centering
\includegraphics[width=.5\textwidth, clip=True,trim=0 0 0 0]{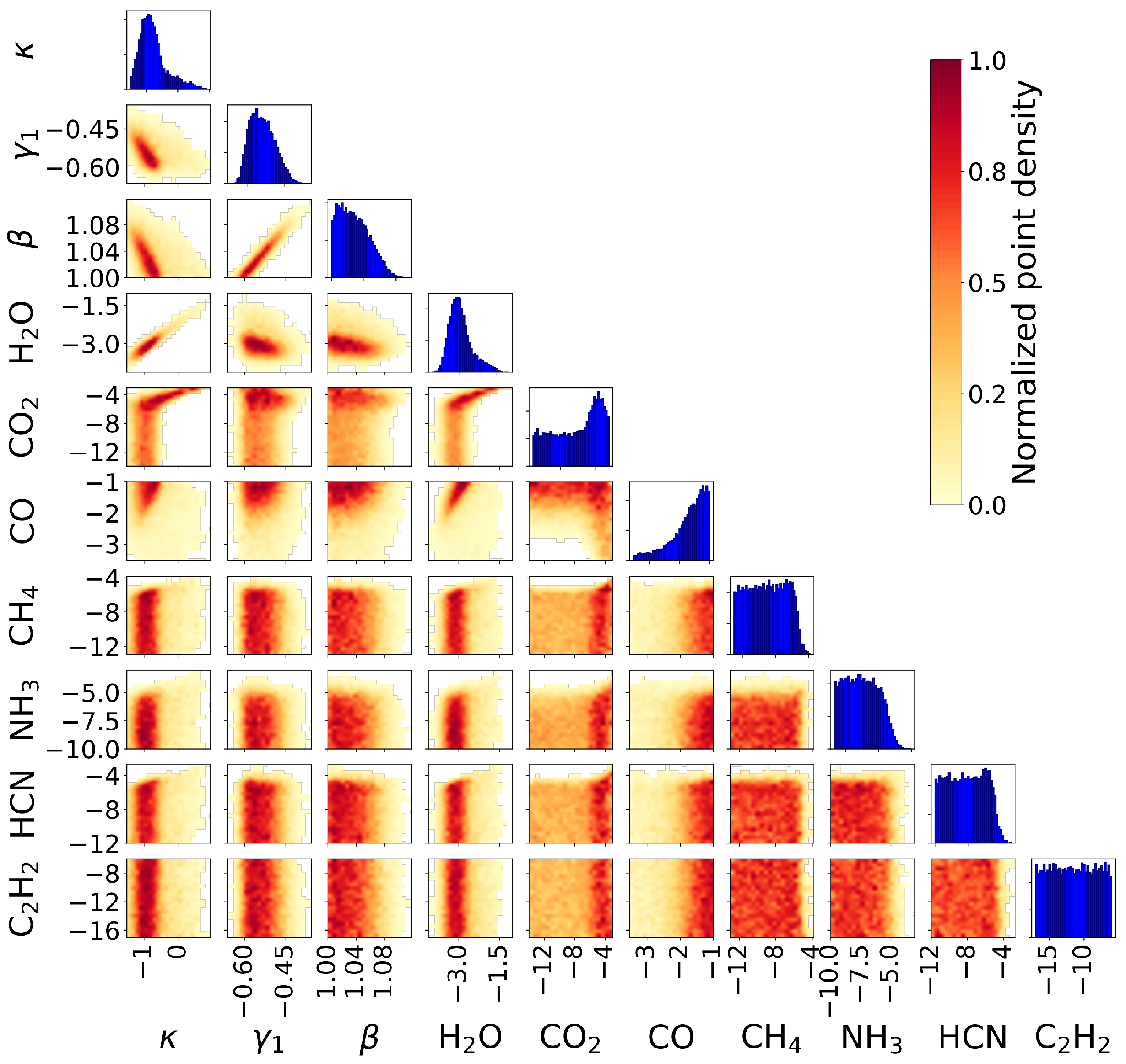}
\caption{Pairwise correlation plots for Case 4, seven fitted species.}
\label{fig:7mol-pair}
\end{figure}

Table \ref{table:4mol-fit} lists \math{\chi\sp{2}\sb{\rm red}} and BIC values for Cases 1, 2, and 3. According to BIC, Case 1 is slightly favored compared to Case 2 by a probability ratio (\math{e\sp{\Delta{BIC}/2}}) of 3.5 and to Case 3 by a probability ratio of 4.0. Figure \ref{fig:4mol} shows Case 1 best-fit spectrum and \math{T(p)} model. In the inset, we give the wavelength range covered by the {\em HST} data. Figure \ref{fig:4mol-indSpecs} shows the influence of the individual species on the best-fit model for Case 1, Figure \ref{fig:4mol-hist} the histograms of the posterior distribution of the retrieved molecular species with their best-fit values, while Figure \ref{fig:4mol-pair} shows the pairwise correlation plots. The best-fit model, \math{T(p)} profile, histograms, pairwise correlation plots and influence of each species on the best-fit spectra for Case 2 and 3 are almost identical to Case 1. Table \ref{table:4mol-fit} compares the goodness of fit for all three cases. According to BIC, the inclusion of additional opacity sources does not improve the fit, as their effect on WASP-43b best-fit model is negligible.

As seen, water features dominate the spectrum, and thus water is constrained the best in this analysis. There is some evidence for CO\sb{2}, showing a soft upper limit, and we also see a more pronounced upper limit of CH\sb{4}. CO appears abundant, hitting the set limit and peaking just below it. 

These conclusions agree well with the results from \citet{LineEtal2014-Retrieval-I}. In particular, our retrieved \math{T(p)} profile is fully consistent with theirs.

\subsection{Results - Seven Fitted Species}
\label{sec:seven}

We continued our analysis by fitting seven molecular species:  H\sb{2}O, CO\sb{2}, CO, CH\sb{4}, NH\sb{3}, HCN, and C\sb{2}H\sb{2}. Again, we tested the effect of including additional opacity sources, and we generated two different cases:

\begin{enumerate}
\item [4.]  Using the same initial \math{T(p)} profile and vertically-uniform species abundances as in our previous cases, we modeled the atmosphere of WASP-43b including the opacity sources for all seven molecules. 
\item [5.]  Same as Case 4 with the inclusion of all 11 species and their opacities in the calculation (C\sb{2}H\sb{4}, H\sb{2}S, TiO, and VO in addition to the species from Case 4), assuming equilibrium, solar composition.
\end{enumerate}

\begin{table}[h!]
\vspace{-10pt}
\footnotesize{
\caption{\label{table:7mol-fit} Goodness of Fit, Seven Species}
\atabon\strut\hfill\begin{tabular}{lccc}
    \hline
    \hline
                          & \math{\chi\sp{2}\sb{\rm red}}  & BIC \\
    \hline
Case 4, 7 opacities       & 2.4754                           & 72.1866\\
Case 5, 11 opacities      & 2.5070                           & 72.6932\\
    \hline
\end{tabular}\hfill\strut\ataboff
}
\end{table}

Figure \ref{fig:7mol} shows the best-fit spectrum and \math{T(p)} for Case 4. Figure \ref{fig:7mol-indSpecs} shows the influence of the individual species on the best-fit model, Figure \ref{fig:7mol-hist} the histograms of the posterior distribution of the retrieved molecular species with their best-fit values, while Figure \ref{fig:7mol-pair} shows the pairwise correlation plots. As in Section \ref{sec:four}, the most influence on the spectrum comes from the water spectral features, thus, water produced the most precise abundance constraint (Figure \ref{fig:7mol-hist}). The additional species retrieved, NH\sb{3}, HCN, and C\sb{2}H\sb{2}, show no influence on the WASP-43b spectrum (Figure \ref{fig:7mol-indSpecs}). 

According to BIC values (Table \ref{table:7mol-fit}), Case 4 is favored over Case 5 for a probability ratio of 6.3. However, Case 1 is favored to Case 4 for a probability ratio of 129. The inclusion of any additional opacity sources does not improve the fit. We also see that BIC values are generally larger when we fit 7 species (Table \ref{table:7mol-fit}) than when we fit 4 species. In Table \ref{table:results}, we list the best-fit species abundances for Case 1 and Case 4, and their corresponding confidence regions. Table \ref{table:accuracy} lists Case 1 and Case 4 posterior accuracy, i.e., 1\math{\sigma} errors on our credible regions calculated following \citet[][Section 5 and Appendix C]{HarringtonEtal2021-BART_I}.  

\newpage
\subsection{WASP-43\lowercase{b} Contribution Functions}
\label{sec:cfWASP43b}

Figure \ref{fig:cf} shows the \math{T(p)} profile and normalized contribution functions for {\em HST, Spitzer}, and ground-based observations, for the best-fit model from Section \ref{sec:four}, Case 1, our lowest BIC model. The right panel shows the pressures where the maximum optical depth is reached. We see that the best-fit atmospheric model probes mostly the thermal structure around 1 bar. The observations done by {\em Spitzer}'s channel 2 (4.5 {\microns}) probe lower pressures, around 0.1 bar, explained by the presence of several opacity sources in this bandpass. Figure \ref{fig:4mol-indSpecs} shows that most of the spectral features from CO are concentrated in this region, in addition to H\sb{2}O features. 

\section{Summary and Conclusions}
\label{sec:conc}

This paper is one of three papers that presents a novel retrieval framework, the Bayesian Atmospheric Radiative Transfer (\textsc{BART}) code. \textsc{BART} is an open-source, open-development, Bayesian, thermochemical, radiative-transfer code under a reproducible-research license available at \href{https://github.com/exosports/BART}{https://github.com/exosports/BART}. In this paper, we presented the implementation and the underlying theory of the initialization routines, \textsc{TEA} module, atmospheric profile generator, best-fit routines, and the contribution functions module. Other modules and packages are described in \citet{CubillosEtal2021-BART_II}, while \citet{HarringtonEtal2021-BART_I} describes the overall framework, tests, and Bayesian best practices. 

We also present forward and retrieval analyses of space- and ground-based secondary eclipse data, in an attempt to compare our results with some previous analyses and validate our framework, and to investigate several new scenarios. We find that the temperature decreases with increasing altitude for all cases, in agreement with previous studies from \citet{StevensonEtal2014-WASP43b, LineEtal2014-Retrieval-I} and \citet{BlecicEtal2014-WASP43b}. The data are best fit with the model including only four major molecular species (H\sb{2}O, CO\sb{2}, CO, and CH\sb{4}). We do not find the signatures of HCN, C\sb{2}H\sb{2}, and C\sb{2}H\sb{4} in any of our models, inferring that it is unlikely that the atmosphere of WASP-43b has high C/O ratio or high metallicity. We also find that the inclusion of TiO, VO, and H\sb{2}S opacities do not improve the fit, suggesting that there is no indication of strong absorbers in the visible that could lead to thermal inversions in WASP-43b. Our results are in agreement with the conclusions made by \citet{LineEtal2014-Retrieval-I} and \citet{KreidbergEtal2014-WASP43b}. Overall, the inclusion of additional opacity sources does not change the shape of the best-fit spectrum, only marginally influences the best-fit species abundances, and does not improve the fit. According to BIC, the atmospheric model with only four major opacity sources is the best match to the data. 

\begin{table*}[t!]
\footnotesize{
\caption{\label{table:results} Log of Retrieved Species Abundances}
\atabon\strut\hfill\begin{tabular}{lcccccccc}
    \hline
    \hline
Case         &                & \math{\rm log_{10}}(H\sb{2}O) & \math{\rm log_{10}}(CO\sb{2}) & \math{\rm log_{10}}(CO) & \math{\rm log_{10}}(CH\sb{4}) & \math{\rm log_{10}}(NH\sb{3}) & \math{\rm log_{10}}(HCN) & \math{\rm log_{10}}(C\sb{2}H\sb{2}) \\
    \hline
Case 1       & Best Fit          &    -2.7        &    -9.3         &    -1.0        &    -5.6         & / & / & / \\
4 opacities  & 68.27\% region  & [-3.5, -2.8] & [-12.6, -3.4] & [-1.9, -1.0] & [-11.7, -5.8] & / & / & / \\
    \hline
Case 4       & Best Fit          &    -2.5      &    -6.9       &    -1.0      &    -5.6       &    -9.5      &    -7.5       &    -14.0    \\
7 opacities  & 68.27\% region  & [-3.5, -2.7] & [-13.6, -3.1] & [-1.9, -1.0] & [-12.2, -5.8] & [-9.6, -6.3] & [-11.5, -5.2] & [-16.2, -6.6]\\
    \hline
\end{tabular}\hfill\strut\ataboff
}
\end{table*}

\atabon\begin{table*}[!t]
\vspace{-10pt}
\centering
\caption{Retrievals: Posterior Accuracy}
\label{table:accuracy}
\begin{tabular}{lcccccl}
\hline\hline
Case &  \multicolumn{1}{p{2cm}}{\centering Number of iterations} & ESS\sp{1} & \multicolumn{1}{p{2cm}}{\centering 68.27\% Credible Region Error} & \multicolumn{1}{p{2cm}}{\centering 95.45\% Credible Region Error} & \multicolumn{1}{p{2cm}}{\centering 99.73\% Credible Region Error} \\
\hline
{\tt Case 1}   & 6.0 \math{\times} 10\sp{5} & 668  &  1.79\% & 0.80\% & 0.20\% \\
{\tt Case 4}   & 6.0 \math{\times} 10\sp{5} & 887  &  1.56\% & 0.70\% & 0.17\% \\
\hline
\end{tabular}
\begin{minipage}[t]{0.66\linewidth}
\tablenotetext{1}{ESS - steps per effectively independent sample, see \citet[][Section 5 and Appendix C]{HarringtonEtal2021-BART_I}}
\end{minipage}
\end{table*}\ataboff

\begin{figure*}[t!]
\centering
\subfigure{\includegraphics[width=0.60\textwidth]{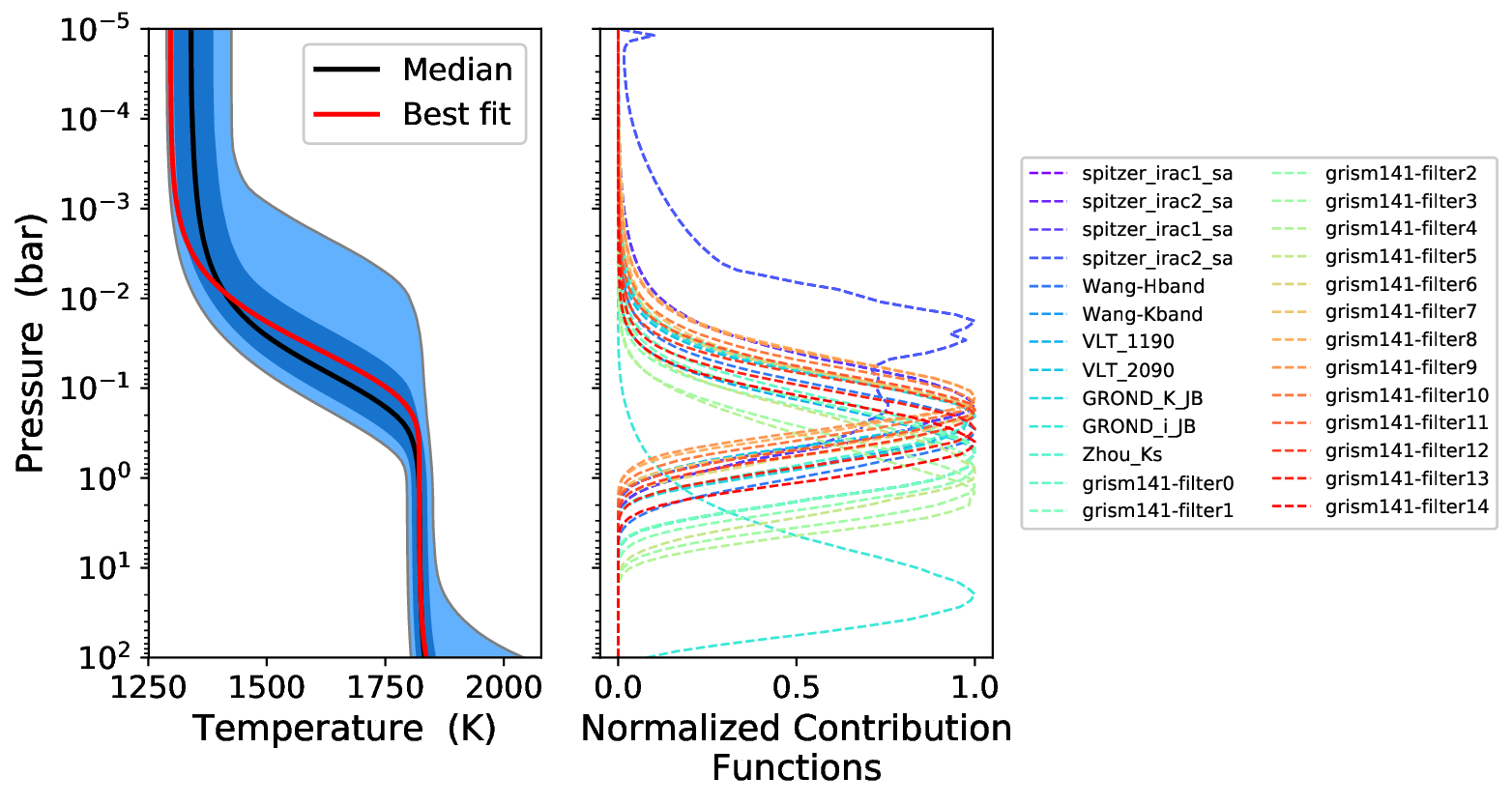}}
\subfigure{\raisebox{1.7cm}{\includegraphics[width=0.30\textwidth, clip=True]{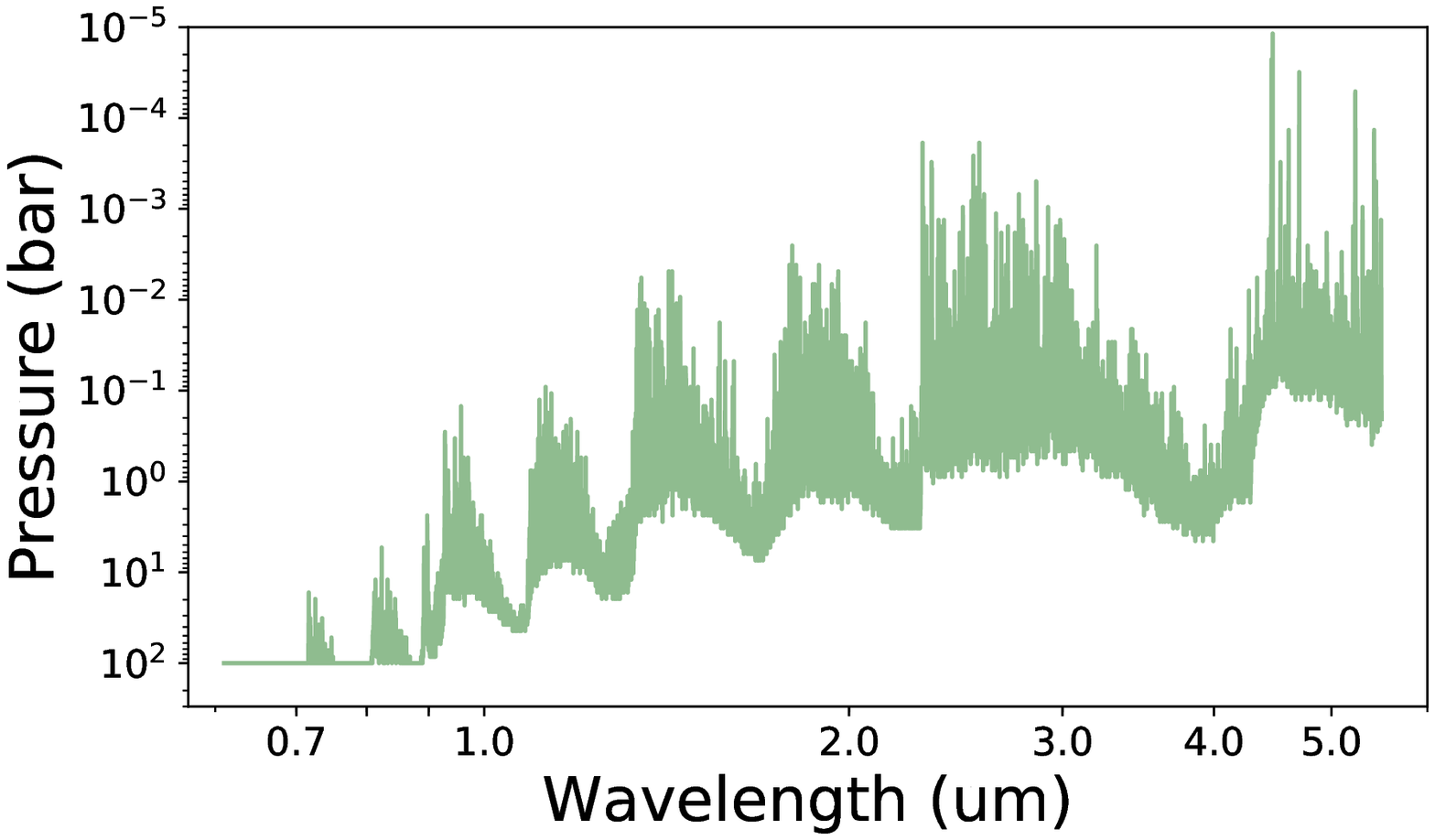}}}
\caption{{\bf Left:} Contribution functions for the lowest BIC atmospheric model, Case 1, Section \ref{sec:four}. First panel shows the the median temperature and pressure profile with 1\math{\sigma} and 2\math{\sigma} confidence regions, while the second panel shows the individual normalized contributions function of each observation.  {\bf Right:} Pressures where the maximum optical depth is reached for each wavelength.}
\label{fig:cf}
\end{figure*}

The spectrum is dominated primarily by H\sb{2}O, and to some extent by CO at specific wavelengths, showing spectral features in the bandpasses of our observations, with water being the most dominant. CO\sb{2}, CH\sb{4}, NH\sb{3} and HCN are detected giving an upper limit, while C\sb{2}H\sb{2} is unconstrained. We also see a degeneracy between CO and CO\sb{2}, with CO being preferred over CO\sb{2}. Both species have spectral features around 4.5 {\microns}, but the broad band and sparse coverage does not allow us to entirely distinguish between them. Finally, we calculate the water abundance using our best-fit model. Assuming the same solar water abundance of 6.1\math{\times}10\sp{-4} as \citet{KreidbergEtal2014-WASP43b}, we constrain the water abundance on the dayside of WASP-43b to 2-6\math{\times}solar, similar to the conclusions made by \citet{BlecicEtal2014-WASP43b} and \citet{KreidbergEtal2014-WASP43b}. 

The reproducible-research compendium for this paper including all codes, inputs, and outputs is available at: \href{https://github.com/dzesmin/RRC-BlecicEtal-2019-ApJ-BART3}{https://github.com/dzesmin/RRC-BlecicEtal2021-ApJ-BART3} (and will be uploaded on Zenodo upon acceptance).

\acknowledgments This project was completed with the support of the
NASA Earth and Space Science Fellowship Program, grant NNX12AL83H, 
held by Jasmina Blecic and NASA ROSES-2016/Exoplanets Research Program, grant NNX17AC03G
held by Ian Dobbs-Dixon and Jasmina Blecic, and by NASA Planetary Atmospheres grant NNX12AI69G, NASA Astrophysics Data Analysis Program grant NNX13AF38G, and NASA Exoplanets Research Program grant NNX17AB62G, held by Joseph Harrington. Part of this work is based on observations made with the {\em Spitzer Space Telescope}, which is operated by the Jet Propulsion Laboratory, California Institute of Technology under a contract with NASA. We would like to thank Kevin B. Stevenson for providing the 3.6 and 4.5 {\micron} {\em Spitzer}\/ data, Guo Chen and George Zhou for the transmission response functions, and Jonathan Fortney for a useful discussion. We also thank contributors to
SciPy, NumPy, Matplotlib, and the Python Programming Language; the
open-source development website GitHub.com; and other contributors to
the free and open-source community. \\

\bibliography{BART-III}

\end{document}